\documentclass[preprint]{aastex}
\usepackage{lscape}
\usepackage{ulem}
\usepackage{subfloat}
\renewcommand*\thesubfloatfigure{\themainfigure~(\alph{subfloatfigure})}

\def\gsim{\mathrel{\raise0.35ex\hbox{$\scriptstyle >$}\kern-0.6em
\lower0.40ex\hbox{{$\scriptstyle \sim$}}}}
\def\lsim{\mathrel{\raise0.35ex\hbox{$\scriptstyle <$}\kern-0.6em
\lower0.40ex\hbox{{$\scriptstyle \sim$}}}}

\begin{document}

\title{Photometric Response Functions of the SDSS Imager}

\author{Mamoru Doi$^{a,b}$, Masayuki Tanaka$^c$, Masataka Fukugita$^{b,d,e}$, James E. Gunn$^f$, Naoki Yasuda$^{b}$, 
\v{Z}eljko Ivezi\'{c}$^{g}$, Jon Brinkmann$^h$, Ernst de Haars$^f$,
Scott J. Kleinman$^i$, Jurek Krzesinski$^j$,  and R. French Leger$^h$}

\affil{$^{a}$Institute of Astronomy, University of Tokyo, Mitaka, Tokyo 1810015, Japan\\
$^{b}$Institute for the Physics and Mathematics of the Universe, University of Tokyo, Kashiwa 2778582, Japan\\
$^c$European Southern Observatory, Karl Schwarzschild Stra\ss e 2, 
Garching bei M\"unchen 85748, Deutschland\\
$^d$Institute for Advanced Study, Princeton, NJ08540, U. S. A.\\
$^e$Institute for Cosmic Ray Research, University of Tokyo, Kashiwa 2778582, Japan\\
$^f$Princeton University Observatory, Princeton, NJ 08540, U. S. A.\\
$^g$Department of Astronomy, University of Washington, Box 351580, Seattle, WA 98195, U. S. A.\\
$^h$Apache Point Observatory,@2001 Apache Point Road, P.O. Box 59, Sunsport, NM88349-0059, U. S. A.\\
$^i$Gemini Observatory, 670 N. A'ohoku Place, Hilo Hawaii, 96720, U. S. A.\\
$^j$Mt. Suhora Observatory, Cracow Pedagogical University, ul. Podchorazych 2, 
30-084 Cracow, Poland\\
}

\begin{abstract}

The monochromatic illumination system is constructed to carry out {\it
in~situ} measurements of the response function of the mosaicked CCD
imager used in the Sloan Digital Sky Survey (SDSS).  The system is
outlined and the results of the measurements, mostly during the first
6 years of the SDSS, are described.  We present the reference response
functions for the five colour passbands derived from these
measurements, and discuss column to column variations and variations
in time, and also their effects on photometry. We also discuss the
effect arising from various, slightly different response functions of
the associated detector systems that were used to give SDSS
photometry. We show that the calibration procedures of SDSS remove
these variations reasonably well with the resulting final errors from
variant response functions being unlikely to be larger than 0.01 mag
for 
$g$, $r$, $i$, and $z$ bands
over the entire duration of the survey.  The
considerable aging effect is uncovered in the u band, the response
function showing a 30\% decrease in the throughput in the short
wavelength side during the survey years, which potentially causes a
systematic error in photometry.  The aging effect is consistent with
variation of the instrumental sensitivity in $u$-band, which is
calibrated out. The expected colour variation is consistent with
measured colour variation in the catalog of repeated photometry. The
colour variation is $\Delta (u-g) \sim 0.01$ for most stars, and at
most $\Delta (u-g) \sim 0.02$ mag for those with extreme colours. We
verified in the final catalogue that no systematic variations in
excess of 0.01 mag are detected in the photometry which can be
ascribed to aging and/or seasonal effects except for the secular $u-g$
colour variation for stars with extreme colours.

\end{abstract}

\section{Introduction}

A unique and unprecedented feature of the Sloan Digital Sky Survey
(SDSS: York et al. 2000) is the wide field CCD imager (Gunn et
al. 1998) that enables us to image 1.52 square degrees (the physical
area size is 725 cm$^2$) of the sky at a time. The prime concern is
the accurate characterisation of the imaging efficiency across the
five colour passbands of the SDSS.  For this purpose we have
constructed a monochromatic illumination system and installed it 
permanently
in the imager enclosure. We have carried out detection efficiency
measurements several times during the SDSS operation. This
enables us to study the variation of the response function over the
survey period.  We have then compared photometry with the designed
and/or measured characteristics of the imager in the laboratory, 
and with photometry of the associated
telescopes with SDSS filters to qualify SDSS photometry. 

This paper describes the monochromatic illumination system and the
detector efficiency measurement, and presents the
response function for the SDSS main imager which is 
used as the reference.  We discuss
the effect that is expected on SDSS photometry from seasonal, 
secular and chip to chip
variations of the response efficiency,
and compare it with the actual data acquired
by the survey after the routine photometric calibration. We also
study response functions of the associated telescopes used in
photometric calibrations for SDSS and study the effect on SDSS
photometry from their slightly variant system responses.  We believe
that this helps us understand SDSS photometry better as well as serves as a
useful guide for designing comparable systems in the instrumentation
planned in the future.

The photometric system comprises five colour bands, $u$, $g$, $r$, $i$,
and $z$, that divide the entire range from the atmospheric ultraviolet
cutoff at 3000~\AA\ to the sensitivity limit of silicon CCDs
at~11000~\AA\ into five non-overlapping passbands,
maximising the band width of each band to enhance the detection
efficiency (Fukugita et al. 1996, hereinafter F96).  The blue side of
the $g$, $r$ and $i$ passbands is cut off by colloidal colour glass elements
(GG400, OG550 and RG695, respectively) and their red side 
by short-pass interference multilayer (30$-$45 layers) coatings made of
TiO$_2$ and SiO$_2$.  The characteristics of the $u$ filter is
determined by ionically coloured glass, BG38 and UG11, for the blue and
red sides, respectively, while an intereference coating made of  
Ta$_2$O$_5$ and SiO$_2$ is applied to suppress the redleak that would
otherwise appear at 6600$-$8000\AA.  For the $z$ passband, the
blue side is cut off by colloidal colour glass RG830, and the red side is open,
naturally cut off by the silicon CCDs.

CCD detectors for the main imager are thinned back-illuminated devices
except for those for the $z$ band which use thick front-illuminated
devices, all procured from Scientific Imaging Technologies, Inc (SITe).
An ultraviolet-enhancing antireflection 
coating was applied to the CCDs
used for the $u$ band.  The filters, directly cemented on the second quartz
corrector of the Richey-Chr\'etien telescope, are placed just before
the CCDs.  The containers that house the filters and CCDs are vacuumised
all the time for the duration of the survey except for necessary maintenance
periods. The CCD that is used at the
50-cm Photometric Telescope (hereafter PT), which has been used to set
the zero point of photometry and to monitor atmospheric extinction,
and the CCD at the USNO 1m telescope system, which was used to preset
brightness of standard stars, are also UV-enhancing coated, thinned
back-illuminated devices similar to the $u$ CCDs in the main camera.

We have measured the transmission of filters in the laboratory before
their installation to the imager. The transmission was verified to be
sufficiently close to the design, allowing for some variation in the
cutoff wavelength up to 10$-$20\AA~ that arise from fluctuations in the
commercially available colour glass elements (Schott, Mainz) and
different batches of coatings (Asahi Spectra Co., Tokyo).  The
response of the CCD was measured at SITe but only at a room temperature.
This curve was then modified at long wavelengths to fit the data
obtained at operating temperature (about $-$80C) in our laboratory,
using, however, coarsely sampled measurements.  Synthesized response
curves for the specific device used for PT were published in F96,
which defines the original photometric system of SDSS.  The response
curves used for the survey with the 2.5m telescope camera,
therefore, are expected to differ by some extent from the one that
defines the SDSS photometric system.  The camera itself has six
assemblies (called Camcols) of 5 detectors, one for each colour (hence 30 CCD's and filters), and the
six individual systems are, of course, not exactly the same as
each other.

We also anticipated that the system response may be
subject to some seasonal variation (the CCDs are cooled, but the filters 
run approximately at the ambient temperature of the telescope enclosure)
and possibly to aging effects.
We have designed and constructed a monochromatic
illumination system to characterize the wavelength response of all of
the camera CCDs/filters, and
have occasionally measured the system response during the
duration of the survey to monitor the seasonal and secular effects.

After we began the operation of the 2.5m telescope (Gunn et al. 2006)
we found that the response functions deviated significantly due to an
effect that was completely unexpected in the beginning.  The filters
are in vacuum, with the coated surface exposed. In vacuum, water
molecules, which have been adsorbed into the voids in the relatively
low-density evaporated films used in the filters, migrate away,
lowering the refractive index of the films and moving the cutoff
wavelength of the $g$, $r$, and $i$ filters blueward by about a 2.3
percent (120\AA, 160\AA, and 175\AA, respectively). The same effect
would also modify the redleak suppression of the $u$
filters. Laboratory measurements of this effect, motivated by the
early measurements of the imager using the illumination system, are
reported in Fukugita \& Shimasaku (2009) (hereafter FS09). The result
of our early measurement was given in Fan et al. (2001).

The preparation for the calibration work has been somewhat patchy, due
to a pressing time schedule when the survey began and to a number of
problems that surfaced in the early stage of observations. 
The basic system was
defined in F96 for the combination of the filters and the CCD that
were supposed to be used for the Photometric Telescope, 
originally intended to be used both to define magnitudes of the
standard stars and to carry out the photometric calibration for the
2.5m telescope, together with a  daily monitor of atmospheric
extinction.  However, due to technical problems and a subsequent delay in
installing the PT (called the Monitor Telescope at that
time) photometry of the standard stars was actually carried out using the
system at the USNO 1m telescope in Flagstaff with a set of filters
that were slightly different, due to manufacturing variations, 
from those for PT. 
The zero points for the USNO system were adjusted to the F subdwarf spectrophotometric
standard BD+17$^\circ$4708 in the PT system, as given by F96.

The 50cm Photometric Telescope eventually installed at Apache Point
Observatory (APO) has observed both standard stars, whose magnitudes
were adopted from observations at USNO, and stars in secondary patches
which were simultaneously measured by the 2.5m main telescope and thus
used as transfer fields.  The brightness of stars in the secondary
patches was measured with respect to the USNO brightness for the set
of standard stars that define the zero point of SDSS photometry (Smith
et al. 2002; Tucker et al. 2006), originally denoted as primed
magnitudes (FS96). Further complications were caused by the fact that
the set of filters used at PT was replaced in the middle of the survey
(18 August 2001, Modified Julian Date (hereafter MJD) 52140) with a
new set which was fabricated using a more modern technique of
ion-assisted deposition coating. This is forced by the fact that we
discovered that the old filters made with traditional evaporated
coatings display changes with temperature and humidity, smaller than
the changes observed going to a fully evacuated environment, but still
substantial. These changes are almost certainly associated again with
the adsorption of water in the low-density evaporated films. The
ion-assisted films are much denser and do not show the effects, as we
also confirmed in laboratory experiments. The filters with
evaporation-deposition films show a temperature dependence in excess
of what was anticipated from the temperature dependence of colour
glass\footnote{ The red cutoff wavelength shifted by an unexpected
amount, 42\AA~for $g$, 58\AA~for $r$ and 77\AA~for $i$ redwards when
temperature increases by 20 deg (FS09).  They are about 5 times worse
than the temperature dependence for the colour glass used for the blue
side cutoff.  These shifts were much more than anticipated and would
cause a few hundredth of magnitude shift.  As mentioned in the text
above, a very dry condition would affect the coating surface and hence
the character of transmission.  The filter at PT is purged with very
dry air.  It is so dry that the transmission of filters has shifted
halfway between the air and the vacuum.  We realised that it is
difficult to control the exact condition; this forced us to decide
that we replace the filters with ones made with the new technique. We
confirmed that this effect together with the temperature dependence
disappears with filters fabricated by the ion-assisted deposition
coating technique.  }.  In vacuo, as in the 2.5m imager, the
evaporated films show {\it no~ temperature} effects, however.

This instability was also noted by Tucker et al. (2006) during their
work on photometric standard stars at the PT, who attempted to remove
the effect by using variable color terms (`b-terms') so as not to
affect the final results. The determination of these transformations
and tracking them in time, however, is difficult and results in
somewhat ambiguous photometric results.  We thus decided to replace
the PT filter set with a new one made with ion-assisted deposition
coating, which makes the filter characteristics stable against
environmental conditions. We designed the transmission of PT filters
to fall between the one in vacuo and the one in air so that it can be
close to the PT system in the dry air-purged environment with which
some amount of the work had already been done.

We thus have used several systems that have somewhat different
response functions to set up SDSS photometry.  It is, therefore,
important to examine that all relevant photometric response functions 
are sufficiently close to each other so that the resulting photometry
defines a system at least with tolerable internal errors.  This
compels us to measure the response functions of PT (before and after
the filter replacement), USNO and all 30 devices of the 2.5m telescope
imager, in addition to that used to define the original photometric
system.  We note that the fiducial response functions of the 2.5m
telescope imager have been made public through the world wide web at
the SDSS site (SDSS public web site 2001)\footnote{
http://www.sdss.org/dr7/instruments/imager/filters/u.dat,...,z.dat}
using our measurements in December 2000.  In this paper we present the
newly estimated response function based on the entire measurements,
but primarily resorting to those measured in the autumn of 
2004 with more careful
wavelength calibrations.  The response function for the $u$ band we
shall present is significantly different from the one made public
because of an unexpected large aging effect, but the other reference
response functions we present differ little from the ones in the
public web site.  Even with the response functions as much varied as
in the $u$ band, the effects on calibrated photometry in the SDSS
output turn out to be tolerably small.  We note in passing that though we
will do all our calculations as if the SDSS photometry is an AB
system, we will not address the very difficult problem of determining
the AB zero-point corrections, which could well be as large as a few
hundredth of a magnitude.

In Section 2 we present the design of the monochromatic illumination
system.  In section 3 we present the results of the response function
measurements, and present the new reference response
function constructed therefrom.  In section 4 we consider
the effects expected from the variation of response functions on
photometry. We also discuss the response function of associated telescopes
used to give SDSS photometry and errors in photometry when these
slightly different systems are used. We also study the actual data, 
which are published in
SDSS catalogues (Abazajian et al. 2003; 2004; 2005; Adelman-McCarthy
et al. 2006; 2007; 2008) after processing and calibrations by the
photometric pipelines with the aid of the monitor telescope pipeline
(Stoughton et al. 2002; Tucker et al. 2006).

\section{Monochromatic illumination system}

The illumination system consists basically of a lamp, a monochrometer
consisting of a grating, order sorting filters and slits, an
integrating sphere and a photodiode to measure the flux of the
illumination. This system can be moved accurately over the focal plane
to illuminate any CCD in the camera, and also to illuminate a
calibrated reference diode. The output of the integrating sphere is a
few inches above the quartz corrector element which also serves as the
mechanical substrate upon which the imager dewars are mounted, about
10 inches above the focal plane (Gunn et al. 1998).  The X direction
(the direction across different camera columns for the same colour
band) is controlled by a ball-screw stepping motor, while the Y
direction (the direction across five different colour filters in a
single dewar) is manually controlled.  Thus six CCDs with a given
colour band can be reached by remote control, but measuring different
colour bands requires manual intervention. Since the (manual) slit
width must sometimes be changed when going from one colour band to
another anyway, this is not onerous.  The illumination covers 250 nm
to 1200 nm with the resolution $R=\lambda/\Delta\lambda=50-200$ for a
0.5$-$2 mm slit width.  The lamp is a Philips quartz-halogen
tungsten-filament lamp of 150W 
with the filament size 3mm$\times$5mm.  The flux is 5000 lm and the
colour temperature is 3400K at 24 V. Its life time, 50 hours at 24 V,
is increased to 500 hours when run at 20 V, at which voltage the
colour temperature is 3200K.  

The light is condensed with a mirror
system, and is then guided to a slit of the monochrometer with
adjustable width 1-20 mm by a triangular mask.  The monochrometer
(JASCO CT-10) consists of four mirrors which make a collimated beam
with a diameter of about 30 mm. The mirror focal length is 100 mm,
with an approximate focal ratio of F/3. A grating of 28 mm$\times$28
mm with 1200 lines/mm is used to disperse the light with the spectral
dispersion about 6.7 nm/mm at 600 nm.  The grating is controlled by a
stepping motor.  The beam is guided through the outlet slit, whose
width is set equal to that of the input slit, to the order-sorting
filter, consisting of three filters, U-330 for 250$-$390 nm, L-37 for
390$-$680 nm and R-64 for 680$-$1200 nm. The output light illuminates
an integrating sphere (ORIEL) 8 inches in diameter, and a shutter is
placed before the integrating sphere.  The integrating sphere makes
the illumination over the surface of a CCD as uniform as we can.  The
output is through a circular hole of 2 inches in diameter.  The
illuminating hole is placed vertically so that distance to CCD mimics
the F/5 beam of the telescope.  A 61.7mm thick aluminium block 
with a cylindrical 
hole of 2.5 inch diameter is attached at the illuminating hole, which works
as a light guide.
The inside of the cylinder is anodised so that it works as a light baffle.
Figure 1 shows the entire unit.

The light flux is monitored by a photodiode (Hamamatsu Photonics,
S2281) whose detection efficiency is accurately known.  
The sensitivity of the photodiode was measured by
Hamamatsu Photonics from 200 to 1080 nm at a 10 -- 20 nm interval.  We
may obtain the idea about the error of the sensitivity by replacing
the photodiode with alternative photodiode, which leads us to convince
that it is no more than $\approx$ 0.5\%.  The absolute sensitivity of
the photodiode is verified by Japan Quality Assurance Organisation,
showing that the accuracy is better than 0.6 -- 0.7\% at 488.0 nm and
at 835.1 nm.  This photodiode is placed at the baffle of the cylinder
hole of the block beneath the 
integration sphere. This is used to monitor the
illumination flux, i.e., the efficiency of the CCD including the filter
transmission is measured relative to the flux received at this
photodiode.

The photodiode signals are converted into voltage with an AD743
operational amplifier, and data acquisition is done with National
Instruments DAQ-700 at 1000 data points per exposure. The illumination
apparatus and the photodiode data acquisition are controlled with
National Instruments PCMCIA-GPIB.  The entire measurement sequences
are programmed and controlled by a LINUX computer in the SDSS control
room, except for changing the manual stage positions, choosing a slit
width and a lamp voltage.

The parameters of the monochromator are set as follows: the slit
width is 0.5 mm for g, r, and i-band, 0.75 mm for u-band, and 1.0 mm
for z-band. which corresponds to the spectral resolution of 5.6 nm,
3.6 nm, 3.3 nm, 3.0nm, and 5.5 nm for u, g, r, i, and z-band
respectively.  The QTH lamp works at 20.0 Volt for g, r, i, and
z-band, and 22.5 Volt for u-band. The exposure time is 2 sec to 100
sec depending on the efficiency of the passband at the wavelength
being measured so that the CCD output is 3,000 -- 30,000 electrons per
pixel. Typical errors of repeated measurements were found to be less
than $\sim0.5\%$ in efficiency.  

We were not prepared to measure accurately the absolute detection
efficiency at each measurement.  We occasionally placed another
photodiode at the distance the same as that to CCD to estimate the
absolute detection efficiency. The absolute normalisation of the
detection efficiency we quote in the next section is obtained from
such measurements, unless otherwise stated. We expect systematic
errors of the order no worse than 10\% in the absolute normalisation,
which is mainly due to geometrical uncertainty of the hardware
setting. Relative efficiency was estimated with the photodiode placed
at the integrating sphere. We scaled each measurement to that with the
absolute measurement by adjusting it with an appropriate normalization
factor. 

Another important issue is the wavelength calibration. We measured the
wavelength of the monochromatic illumination repeatedly with a
portable fibre spectrometer, Asahi Spectra HSU-100S, which was
calibrated against lines of a Hg+Ar lamp for the range 250$-$1000 nm.
As a whole we estimate that the wavelength accuracy of the
monochromatic illumination is kept to an error less than $\pm$3\AA, at
least for the measurements with wavelength calibration carried out in
July 2004 and after.  In earlier measurements we did not use the
portable spectrometer and the monochrometer was directly calibrated
with H$\alpha$ and H$\beta$ emissions from a hydrogen lamp; the
position of lamp could not be accurately controlled and was likely
somewhat different from the original one for the quartz-halogen lamp
and hence the wavelength calibration was not as accurate as it was
with later measurements. In this report we correct early measurements
using 
ones measured after July 2004 by matching the red side
cutoff, whose wavelength has been sufficiently stable in the vacuum
chamber. 

We note that the deviations of the incident angle of the light from exactly perpendicular
causes a shift of cutoff wavelengths of interference filter, as
\begin{equation}
\Delta\lambda = \sqrt{1-c_i\sin^2\theta}
\end{equation}
where $c_i=0.62, 0.57, 0.58$ for $g$, $r$, and $i$ filters;
$\theta$ is the angle from perpendicular. This gives $\approx 0.2$\%
in the wavelength, or 8$-$12\AA~ depending on the passband for
$\theta=4$ deg, corresponding approximately to the average over an F/5
beam with the secondary obscuration. Hence the real transmission of
the 
red-side 
cutoff in $g$, $r$, and $i$ band may systematically be
shorter by 8$-$12 \AA ~than with the parallel beam. Our illumination
system produces an F/5 beam, however, without the secondary
obscuration which gives $\approx 0.1$\% (4$-$6 \AA) shifts.  Hence, we
still expect 4$-$6 \AA shift for the 
red 
cutoff in $g$, $r$, and $i$
band responses.  This effect is not large, and is wavelength dependent
within each band pass. We do not correct for this effect in the
present work, but include this into systematic errors. 

The illumination of the monochrometer covers roughly the whole
size of the CCD, on which the flux varies only by $<8$\%.  Some
fringing pattern is visible in the red end of the $i$ band
illumination, but it is at most 2\%, and the effect can be small
enough when an average is taken over some extended pixels.  We set a
region with the circular aperture of 100 pixel radius near the centre
for our measurements. The repeated measurements with exposures using
the same setting give the median counts that agree within 0.5\%. (They
are however mostly due to small flux variation of the QTH lamp, and
are mostly absorbed when compared against the flux at the
photodiode.) 

Along with the measurement of the response functions, we
occasionally measured gains of the detectors.  A few pairs of the same
exposures were taken with the same set up of the monochrometer as the
response function measurement, and the images were subtracted between
the two and the standard deviations in counts were measured.  The
input fluxes in counts and the deviations then give the detector gain.
The raw data for the response functions derived above were multiplied
by the gains to obtain the quantum efficiency.

Figure 2 shows an example of the measurement of the quantum efficiency
of a UV-enhancing coated thinned CCD without a filter, the specific
device used for the Photometric Telescope.  The solid points are
obtained with the present monochrometer illumination system with the
CCD cooled to operating temperature and the open points represent the
data which were measured at SITe at room temperature but were warped
to trace the broadband measurements at a working temperature in our
laboratory for the same CCD.  We see agreement between the two at
1$-$2\% accuracy, although the measurement in the UV is somewhat
noisy. The latter is the curve we used in the characterisation of the
SDSS response function in F96.

\section{The measurement}

We have carried out measurements at 11 epochs before 2006 (and one in
2008); the journal is given in Table 1. The measurement is time
consuming, and full measurements were not done at all the epochs,
because of time constraints.  The results are summarised in Fig. 3
(u), Fig. 4 (g), Fig. 5 (r), Fig. 6 (i) and Fig. 7 (z). The response
functions for the 2.5m telescope main imager that we will take as the
reference are shown by thick solid curves.  They are obtained by
averaging our measurements over 6 columns of detectors, obtained in
October and November 2004, as described below.  When some measurements
for specific columns are missing we supplement the data from other
measurements, after applying a correction for temperature effects.  In
order to take into account the reflection losses due to the primary
and the secondary mirrors of the telescope, we included reflection
losses from two fresh aluminium surfaces, which modify the
efficiencies by 20$-$30 percent while change the shapes of the curves
only slightly.  
The attenuation due to the first corrector lens
is significantly smaller than 1\% at all the wavelengths, and hence it
is not included in our analysis.  The monochromator measurements
presneted in this paper include the attenuation due to the secondary
corrector.

The thin solid curves indicate the response function that has been
taken as the standard from the beginning of the survey as described in
F96. It was obtained by the synthesis of the measured transmission and
the quantum efficiency of the relevant filters and CCD. These specific
filters and CCD have been used at PT before the filter replacement.
The curves also include reflection of two aluminium surfaces. This
represents the response functions for PT with the original filter set,
allowing, however, for variations caused by environmental effects.

There are two more curves drawn. The thin dashed curves are
synthesised response functions expected for the USNO 1m telescope
system. Since the actual quantum efficiency was not measured for the
CCD used in the USNO system, the quantum efficiency of the PT CCD was used,
for the USNO system uses a CCD with surface treatment identical to
that used at PT.  The other curves, drawn dotted, are the response
functions for the PT with the new set of filters.

For the $u$ band (Figure 3) a large departure of the reference
response function for the 2.5m telescope imager from the others is
apparent at shorter wavelengths $\lambda<3650$\AA. This is ascribed to
the aging effect we will discuss below.

Another conspicuous set of deviations of the 2.5m reference response
curves is seen around the red edges with the $g,r$ and $i$ bands. We
see that the red edges of the 2.5m reference response curves are
significantly blueward of the other curves.  This shift is caused by
coating films placed in the vacuum environment, as mentioned
earlier.  The shifts amount to 120 ($g$), 160 ($r$) and 175\AA ($i$),
which are also confirmed in laboratory experiments (FS09). Some
variations up to $(40-50)$\AA~ are also seen at the blue edges, which
are mostly due to fluctuations of characteristics of colour glass, and
to a minor extent to the temperature effect which we shall discuss
below. Similar differences visible at red edges between the F96
standard and USNO are fluctuations in coatings. The temperature effect
is absent for the red edge.  The curves for the PT (with newer
filters) falls between the F96 standard and the 2.5m reference curves
(except for the $g$ band which lies somewhat outside the F96 standard,
as a result of fluctuations in the coating process): this is as
designed when we made the new set of filters for PT.

Interference coatings from different
coating batches lead to some fluctuations in the cutoff wavelengths up to
30-40\AA, which are within our specifications given to the
vendor. Interference films even in the same batch of coating may also vary 
to the extent up to 15\AA~due to inhomogeneity in the coating vessel. 
The transmission of coloured glass also shows 
fluctuations of the same order from piece to piece. 
We did not reject the filters unless
the effective wavelengths differ substantially 
more than by 30\AA~ from the specification.

For the $z$ band (Figure 7) the large difference between the 2.5m
reference response curve and the others is due to the lower quantum
efficiency of the front-illuminated thick CCDs in the main imager,
compared with thinned CCD used for the other curves.  This causes an
appreciable difference in the effective wavelengths.

As a quantitative measure for the variation of response functions we use
the effective wavelength or the effective frequency defined by
\begin{eqnarray}
&&\lambda_{\rm eff}=c/\nu_{\rm eff},\cr
&&\nu_{\rm eff}={\int \nu R(\nu) d\nu/h\nu \over \int  R(\nu) d\nu/h\nu},
\end{eqnarray}
where $R(\nu)$ is the response function and the integrand is weighted
to give the photon number\footnote{There are several definitions for effective
wavelengths. The traditionally used is $\int \lambda d\lambda/\int
d\lambda$ for photomultiplier measurements. For other definitions, see F96}.  
Note that this definition
of effective frequency $\nu_{\rm eff}$
differs from the one used in Table 2 of F96, where
$1/h\nu$ is not included. This definition is the average
with respect to the photon {\it number}, not energy, which is the relevant
quantity for nearly all modern detectors, CCDs included.
Table 2 also gives  the effective wavelengths $\lambda_{\rm eff}$ and
the wavelengths at the 50\% yield of the maximum at both blue and red edges  
for four response functions (without atmospheric transmission)
shown in Fig. 3--7.
The PT system before the replacement of filters is the F96 standard,
ignoring temperature effects.  Note that the shape of the response
function is modified by atmospheric extinction in actual observations,
while we work with the intrinsic response function without atmospheric
extinction in laboratories, and hence in this paper up to some
exceptions when we refer to the observation, 
for which we include the effect due to atmospheric extinction.

The response curves should vary from
column to column even within the main imager,
although the 
interference film transmission varies less, because all film coatings 
used for the main imager are made
in one single batch for each colour.
Figure 8 (a)$-$(e) shows the response curve for each column of the imager for
the five colour bands to show the column to column variation. These data
are taken from the measurement in October/November 2004.
The difference of  $\lambda_{\rm eff}$ relative
to the 2.5m reference measured at 3 epochs is presented in Table 3,
the dispersions being 8.9\AA~for $u$,
 5.8\AA~for $g$, 6.5\AA~for $r$, 4.1\AA~for $i$, 27.5\AA~for $z$.
The larger variations for the $z$ band are
due to different CCD sensitivities near the red cutoff. 
We saw a very small variation (0.8\AA) in
the transmission curve of $z$ filters used for the main imager in
the measurement when the filters are produced. For other 
colours the dispersions of the difference
are consistent with those of the filters we measured at the
time of procurement, except for the $u$ band 
for which {\it in~situ} measurements in combination with CCD 
give a  larger dispersion if only by a factor of two.
Observed in the measurement is approximately
what is anticipated for $g$, $r$ and $i$.
The maximum deviation of $\lambda_{\rm eff}$
against the reference is
($+9,-16$)\AA~for $u$, ($+6,-15$)\AA~for $g$, ($+4,-16$)\AA~for $r$,
($+6,-6$)\AA~for $i$ and ($+46,-31$)\AA~for $z$.

Figures 9(a)$-$(e) show the results of our measurements for the $u - z$
bands at specified
columns and Figure 9 (f) is for redleak of the $u$ filter at Camcol 1
on April 2000, December 2000, September 2001, July 2004,
October/November 2004 and December 2006.  The 2.5m reference is also
drawn for comparison.  Measurements for other columns give similar
results.  The most conspicuous variation at
different epochs is visible for the $u$ band as we have already noted
above (Figure 9(a)).  The response functions, notably at shorter wave
lengths, diminish as time passes till July 2004: the sensitivity
diminished by $\sim30\%$ between the year 2000 and 2004.  
The characteristics has gradually stabilised, so that 
such a large
variation was not visible after July 2004.  We confirmed that no
appreciable aging effects are visible with the $u$ filters at the PT
and the spare $u$ filters retained at the laboratory. 
We also noticed that the illumination pattern visible on the CCD
chip in our measurement
in 2004 differs from the one in earlier measurements for the $u$ band. 
No such conspicuous
aging effects are visible in other colour bands. These
observations lead us to suspect that aging is basically due to that of
the sensitivity of the ultraviolet sensitive CCD itself, probably the
deterioration of the ultraviolet enhancing coating and/or the associated
manipulation of the surface potential profiles on the devices.

The time variations of the response curves in other passbands are
small. We have shown in Figure 10 the time dependence of the effective
wavelengths and the cut-off wavelengths as a function of the epoch of
the measurement. Except for the $u$ filters the time variation of
$\lambda_{\rm eff}$ is of the order of $\lesssim10$\AA, which is
mostly seasonal.  For instance, we observe the average difference $4.2\pm
2.5$\AA~ between July 2004
and November 2004 for the $g$ band. 
We made careful wavelength calibration work in the operations of
July and November 2004.
So we focus on the two measurements.
This seasonal variation is largely ascribed to the 
temperature effect of GG400 colour glass used to cut off the blue side. 
The air temperature decreases by 17
deg between the two epochs.  The measured shift of the blue edge
6.8$\pm$1.3 \AA~ is consistent with 7.2 \AA~ expected from the
laboratory experiment for the temperature effect on GG400
\footnote{The Schott document gives the temperature coefficient
0.7\AA/deg for GG400 between 10$-$90 deg C, but our measurement in the
laboratory (FS09) for this glass piece gives 0.43/deg at -10$-$20 deg C, the
temperature variation being smaller than the Schott value by a factor of
2.  We noted that the temperature dependence is not quite linear. We found that
the temperature dependence is mild at low temperatures and it becomes
sharper at higher temperatures.
The observed wavelength shift is consistent with our laboratory
measurement.  Similarly, we get 0.75 \AA/deg for OG530 compared with
the Schott value of 1.3 \AA/deg, and 0.55 \AA/deg for RG695 compared with
the Schott value of 1.8 \AA/deg. For RG830 our 1.1 \AA/deg is compared
with 2.3 \AA/deg by Schott.
The $u$ filters are cut by colour glasses for both sides: BG38 (and UG11)
for the blue side and UG11 for the red side. The temperature dependence
is 0.5 \AA/deg at the laboratory, while the numbers are not given in the
Schott catalogue. Unlike the coating surface, this does not depend on
whether the environment is air or vacuum (FS09)}.
The red side cutoff varies little,
$1.8\pm1.6$\AA, which is consistent with null within the error.
The red side cutoff wavelengths of 
interference coating films such as those on the new PT filters 
made with ion-assisted deposition are stable
against temperature, which we have confirmed in laboratory
experiments (FS09).

Similarly, we find a shift of 14.7$\pm0.5$\AA~ in the blue edge of the $r$
passband as compared with 12.8\AA~ expected from the temperature
dependence measurement for colour glass pieces.  
The same trend also applies to the $i$
filter.  The shift in the blue edge is measured to be 21.0$\pm0.8$\AA~ for the
$i$ filter, as compared with 9.2\AA~  (35\AA~ if the Schott catalogue
value is used) expected from our laboratory
experiments. The measured shifts of the red edge cutoff are
2.0$\pm0.8$\AA~, $-$2.0$\pm0.8$\AA~
 for the $r$ and $i$ passbands, respectively, which are again consistent
with null shifts when the systematic error of 3\AA~ in the wavelength
calibration is taken into account.
The temperature dependence is not clearly identified for the $u$
passband in our measurements.
For the $z$ filter the measured blue edge shift 51$\pm$21\AA~
appears larger than those expected from the temperature dependence of a
laboratory measurement 19\AA, but is consistent with the Schott value 40\AA.
In summary, the effective wavelength shift of
the filter is grossly consistent with the temperature
dependence of the colour glass cutoff. This change is smaller than $\sim$
20\AA~ for the $u$ to $i$ bands,
small enough to cause any significant effects on photometry at a 1\%
level, as we will see in the next section.
This is also true with the $z$ band, although the shift itself is larger.

The $z$ band response function curves appear to show the time variation
in their shapes. 
We infer that the variation is due to the variation of the operational 
temperature of the CCD, which 
may have not been controlled well.
We know that the quantum efficiency is sensitive to the temperature 
in reddest wavelengths. 

We remark that the accuracy of our earlier measurements was somewhat lower:
the sampling pitch in early
measurements is too coarse for an accurate
characterisation of the response function; a careful wavelength calibration
is made only after July 2004; positional control of the
illumination system could not be done very accurately in the beginning,
etc. The measurements were
significantly better controlled in the 2004 and later runs.
We show in Table 5 summary of various uncertainties in the effective
wavelength. 

Figure 9 (f) presents the redleak measurement for the $u$ response
function for Camcol 1.  The $u$ filter composed of UG11 (1 mm) and
BG38 (1 mm) colour glass causes redleak between 6600\AA~ to 8000\AA~,
which is suppressed by interference coating.  Our best design of
coating still produces a small red leak at around 7900$-$8100\AA~ with
an amplitude of $\sim 5\times 10^{-4}$.  In the figure, sharp peaks
are seen at around 7700$-$7800\AA~ in the amount of $\sim 4\times
10^{-4}$ (we subtract the floor of the peak as discussed in what
follows).  
The 100\AA~redward shift of the interference coating compared with the 
design is caused by the vacuum environment.  
This would degrades the suppression; we
were concerned about this, but we found that the redleak suppression
is actually still sufficiently strong.  
We also observe another bump
at 6800$-$7500\AA.  Our laboratory measurements for the $u$ filters show
that the suppression in this wavelength range is strong enough and
does not exhibit any evidence for the leak in this region.  
We suspect from the
wavelength interval that this leakage signal might be produced by a
cross talk and scattered light that goes through the $i$ filter
(6800$-$8200\AA) which is placed in the row next to the $u$ filter. This
interpretation also accounts for the shoulder seen in the red wing of
the $u$ filter leak at around $\sim8000$\AA. We interpret this to
be due to scattered light in the measurement and, hence, a fake, so
that this floor is subtracted to estimate the redleak for $u$ filters.
Redleak depends on the filter. We detect the
peak at about 7670\AA~for Camcol 2, 5 and 6 and 7830\AA~for Camcol 3 and 4
as shown in Figure 11.
We have not seen the aging and/or seasonal effects in the u-band redleak more
than 10\% (Fig.9(f)).

This in turn indicates the order of magnitude of the effect of
the scattered light and shows that it is small enough and does not
affect the measurement for the main band pass.  

For most purposes, we want to work with one
representative response function for the 2.5m Telescope imager for the
entire duration of the survey observations.  We take the measurement
of October-November 2004, for which we have nearly complete data sets
(one measurement for r1 is missing, and we adopt the data from
measurement at other times and correct for the empirically known 
temperature effect).
The season is between summer and winter, at temperatures close to the
mean over the operation.  This response function was that shown in
Figure 3$-$7 above, with numerics given in 
Table 4\footnote{Tables for each column are available from 
http://www.ioa.s.u-tokyo.ac.jp/~doi/sdss/SDSSresponse.html}.  The
reflectivity for two fresh aluminium surfaces in the telescope is
multiplied. These data are used to calculate $\lambda_{\rm
eff}$ in Table 2 above.
For the $u$ band response function we include a representative
redleak that is obtained by averaging over the column to column variation.
The floor which is likely to be ascribed to scattered light
is removed as seen in Figure 11.
The atmospheric transmission at 1.3 airmass which we use
in this work is also shown in the last column of Table 4. Here the transmission
at the Palomar Observatory (altitude 1700m) is converted to the one at the
APO (altitude 2788m) assuming the exponential scale 
height of the atmosphere of 7000m.

\section{Effects on photometry}

We discuss the effects expected on photometry due to the
small differences in response functions from column to column and
from their variations with the epoch during the survey.
We also consider the difference in
photometry due to the use of different
response functions among the 2.5m telescope
main imager, Photometric Telescope, and the
USNO detector system. 
Comparisons with the original F96 standard are also made.
The SDSS primary standard stars (Smith et al. 2002) were observed using the 
USNO system, and then those observations are tied to the 2.5-m system with 
the Photometric Telescope. Our results confirm that the colour 
effects between two systems are sufficiently small. 

Let us consider brightness of the fundamental standard
star, F subdwarf BD+17$^\circ$4708 (Oke
1990), by a synthetic analysis. The AB magnitude is
defined by
\begin{equation}
m=-2.5\log {\int d(\log\nu) f(\nu) R(\nu) \over \int d(\log\nu) R(\nu)}
 - 48.60
\end{equation}
with $f(\nu)$ the object flux.
The flux of F subdwarf BD+17$^\circ$4708
calculated with the various response function is given in magnitude 
in Table 6.
Note that the photometric system used in F96 (and also in the present
paper) is AB$_{95}$,
which differs slightly from the AB$_{79}$ system used by Oke \& Gunn 
(1983)\footnote{If one uses the  AB$_{79}$
Vega spectrum, the brightness of  BD+17$^\circ$4708 for the F96 responses
will be $u=10.498$, $g=9.631$, $r=9.329$, $i=9.219$ and $z=9.194$.
These differ from the numbers given in Table 6 (AB$_{95}$) which are about 
0.030$-$0.038 mag fainter.
If one adopts the spectrum by Bohlin \& Gilliland (2004), the brightness of  
BD+17$^\circ$4708 for the F96 responses will be
$u=10.563$, $g=9.616$, $r=9.343$, $i=9.254$ and $z=9.247$.}.
The numbers in the bottom row is the flux that would be obtained
with the use of the response functions presented in the SDSS web site
(SDSS public web site 2001). 
We note that the survey adopts the system normalised to
the response function expected at 
1.3 airmasses of atmospheric extinction, so that 
the atmospheric effect that modifies the response function
must be included
in $R(\nu)$ when one discusses the actual observation
(while the atmospheric attenuation of the flux is removed).
The fluxes (in magnitude) 
that would be obtained in various detection systems with 1.3 airmass 
response functions for BD+17$^\circ$4708 are
presented in Table 7. They differ from the numbers in Table 6 by smaller
than 0.01 mag except for the $u$ passband for which the difference amounts to
about 0.036$-$0.039 mag due to a large atmospheric effect. 
The effective wavelengths for this case are also given in upper rows
in this table. 

In practice, in the application to SDSS photometry 
all measured fluxes are adjusted to fit the standard
scale adopted from F96, and the variations seen in this
table are absorbed into the calibration.
In Table 7 we see that the difference among the different systems 
is at most 0.038 mag ($u$ band). In $g$ and $r$ it is 0.034 mag and 0.007 mag
respectively,  
and even smaller for redder colour bands.
This change would cause  
constant shifts in photometry, so that
relative magnitudes will be unchanged.
These variations, however, are flattened by the calibration process
when the fluxes are presented in the catalogue.  
We emphasise that the fluxes in Table 7 are not yet intended to give accurate
AB magnitudes taken as the standard, but are referred to here for the 
purpose of comparisons among different systems.

The flux defined by eq. (3) depends primarily on $\lambda_{\rm eff}$,
and the change of the sensitivity tends to cancel between the
numerator and the denominator. We find that the $\lambda_{\rm eff}$
dependence is roughly $\delta m_u\simeq -0.105 \Delta\lambda_{\rm eff}/(100\AA)$,
 $\delta m_g\simeq -0.077 \Delta\lambda_{\rm eff}/(100\AA)$,
 $\delta m_r\simeq -0.044 \Delta\lambda_{\rm eff}/(100\AA)$,
 $\delta m_i\simeq -0.032 \Delta\lambda_{\rm eff}/(100\AA)$,
 $\delta m_z\simeq -0.024 \Delta\lambda_{\rm eff}/(100\AA)$,
where $\Delta\lambda_{\rm eff}$ is the change in the effective wavelength
between the two systems considered, 
for the BD+17$^\circ$4708 spectrum.

The values presented in Table 6 and 7 ignore redleaks of the $u$ filters.
The contribution depends much on colour of stars. For stars with $g-r<0.6$
it is smaller than $-$0.01 mag, but increases to $-$0.05 mag for 
$g-r\approx 1$,
and it can be as large as $-$0.5 mag for very red stars such as 
$g-r\approx 1.6$. Therefore caution is necessary when one deals with
red stars.

The time variations of the main imager are shown in the lowest two
panels of Figure 10  (they are shown column by column).
The panels (v) represent brightness of
BD+17$^\circ$4708 with 1.3 airmass atmosphere.  
In order to show the time-dependent variation of the
sensitivity, we also show in panel (iv) the CCD output, $\sim -2.5\log
N_{p.e.}$, i.e., eq.(3) where the denominator is
replaced by unity, so that it represents the variation of the number
of photoelectrons in CCD (using the response function without atmosphere). 
With the brightness definition 
this variation is largely
cancelled by the denominator in eq.(3): brightness of objects (magnitude) 
is sensitive solely
to the shift of the effective wavelength.  The decline of the CCD
sensitivity in the $u$ band seen in Fig. 10 (iv) is as much as 30\%
(or $\approx$0.3 mag; somewhat smaller if 1.3 airmass atmosphere is excised), 
which agrees with the size we expect from the change of the response
function seen in Fig. 9 (a).  This is reduced to a $-$0.03 mag
(in the opposite sign) change in magnitude if the brightness definition
is used (Fig. 10 (v)), and
such a variation should further be reduced by calibration procedures.
For other colour bands, the change of the sensitivity is 0.01 mag for
$g$ and $r$, and 0.03 mag for $i$ and $z$. These variations are also
largely cancelled by the denominator in eq.(3), and the variation in
brightness is smaller than 0.01 mag, which could in principle be
further reduced by calibration procedures. Note that the numbers
given here are colour dependent and apply only to BD+17$^\circ$4708.
The calibration would reduce these variations by
enforcing measured brightness to the SDSS
reference magnitudes obtained at USNO.
When one is interested
in sub-percent accuracy photometry, separate treatments of Camcols 
are desired (Ivezi\'c et al. 2007).

Similar considerations apply to colours. Each detector has a slightly
different response function, which causes some tilts in colour
space. We estimate the magnitude difference at the column $A$ of the camera relative to
the 2.5m reference $\Delta m_A=m_A-m_{\rm ref}$ as a function of
colour in Figure 12 for the five colours, where the stars used are
taken from the spectrophotometric atlas by Gunn \& Stryker (1983).  The
variation is largest in the $u$ and $z$ bands. It can be as much as 0.02
magnitudes,  especially for red stars. The variation is also fairly large for
$g2$ (and $r3$). For others it is mostly smaller than 0.01 mag. We see
that $\Delta m$ nearly vanishes for $g-r\simeq -0.5$ for $g$, and for
$g-r\sim 0$ for $r$ and $i$. These variations are without calibration
and are reduced by photometric calibrations.

We then consider the difference of colours that depends on associated
telescope systems used.
Figure 13-15 shows the colour-colour plot written in the form
$\Delta (j-k)_{\rm AB}=(j-k)_{\rm system~A}-(j-k)_{\rm system~B}$
as a function of $(j-k)_{\rm standard}$ to enhance the
scale of the difference which is tiny. Here, $j,k$ stand
for $u$, $g$, $r$, $i$, and $A$ and $B$ are two different systems. 
Again, Gunn-Stryker's spectrophotometric
atlas is used for the sample.  
This represents the bare system-dependent colour term of each system.
Figure 13 is the colour term
of the USNO system against the F96 original standard, Figure 14 is that
of PT against USNO, and Figure 15 is 
the 2.5m telescope imager (reference) against 
the USNO.
Note that colour measured with PT, and hence with the 2.5m telescope
imager, is adjusted to that with the 
USNO system for some set of standard stars
in the calibration procedure (Smith et al. 2002; Tucker et al. 2006). 
The output of the 2.5m
telescope is calibrated with stars measured by PT in secondary
patches which are calibrated with some sets of the standard stars,
which are given fixed values throughout all
observations, and therefore variations are, in principle, largely cancelled
by this reset.
The SDSS photometry is published after these calibrations against
secondary standard stars.

To verify that the variations we have seen here have not
directly propagated into the final SDSS output photometry catalogues,
we show in
Figure 16(a) and (b) the time variation of $r$ and $u$ magnitudes
of stars in the SDSS catalogue
in a patch of 2.5 deg square (R.A.=5$^\circ$, dec=0$^\circ$)
taken from Stripe 82 from the epoch
July 1999 to January 2004. The difference from the mean is calculated for
600 stars with $r<18$ for $r$ and 200 stars with $u<18.5$ for $u$ 
by tracing photometry in time
taken from Data Release 6 Supplement.
The error bars indicate the variance in the sample. 
We see that both $r$ and $u$ magnitudes are very stable, the variation
being less than 0.01 mag. In particular, the large variation
expected for the $u$ band sensitivity ($\Delta u\simeq 0.3$ mag) 
between year 2000 and 2004 does not appear in Figure 16(b). 
We confirmed that this is also true with brightness in other colour bands.

To differentiate the aging effect possibly seen in the $u$ band, we  
consider the variation of $u-g$ colour, as a function of 
$g-i$ colour.
Most of the variation in the $u$ band is absorbed into 
a time-varying constant
adjusted in the calibration, but the colour effect is suspected to
remain in the final catalogue, since the photometric calibrations do not
take account of the colour term.
The time variation in $u-g$ colours 
seen in photometry of stars at stripe 82 (Ivezi\'c et al. 2007)
is plotted in Figure 17. The figure shows the mean differences of $u-g$
measured in the first four years of the survey (1998 -- 2002) and the later 
four years (2004 -- 2007) as a function of $g-i$ colour.
We note that it is so adjusted in the calibration 
that $\Delta (u-g)$ is zero as a matter of principle for stars of  
normal colour, which are used to determine 
the zero points in $u$ and $g$. 
$\Delta (u-g)$ is about -0.01 mag for blue stars ($g-i \sim -0.2$)
and is about -0.01mag for red stars ($g-i \sim 2$).
This difference can be larger, as much as $-0.02$ mag, 
for extremely red stars (we note that $g-r\approx (g-i)/1.2$).
This is compared with the plot (shown by crosses) that shows
the expected variation using spectral synthesis of 
Gunn-Stryker stars between 2001 and 2004.
We conclude that photometry is consistent with what is expected from 
the secular variation of the
response function, although the variation is small, of the
order of 0.01 mag except for extreme colours.
We expect a larger change for very blue stars, but we have no
appropriate sample of such stars in our data base. 
We confirmed that the variation was no more than 0.005mag
for $g-r, r-i$, and $i-z$.

Another way to find the instrumental variation in observational data,
e.g., such as variations due to aging of the $u$ band detectors, may
be to look at the nightly calibration monitor data for the so called
a-term, the difference in the nightly zero point between the
atmospheric extinction corrected instrumental magnitude and the
magnitude taken as the standard (Tucker et al. 2006). This may
directly measure the sensitivity of the system.  Figures 18(a) and (b)
show this term as a function of the epoch of observations.  In the u
band (Figure 18(b)), we observe a rapid change as much as 0.4 mag from
MJD51000 (5 July 1998) to MJD52200 (6 November 2001), and a gradual
decline then to MJD 53200 (13 July 2004). The trend and size are
consistent with what is anticipated from the aging effect.  We also
observe some change in the $r$ band (Figure 18 (a)), to at most 0.1
mag, the reason of which is not clearly identified.

\section{Summary}

We have described the {\it in~situ} measurement of the response
functions of the SDSS 2.5 telescope imager, which has been used to
produce the SDSS astronomical catalogues during the period of 2000 to
2007.  We then discussed the effect of variations of the response
function on photometry.  We also presented the outline of the
monochromatic illumination system constructed for this purpose.

We studied the variation of the response functions from column to
column and also in time over the duration of the survey, both
secularly (aging) and with environmental temperature.  We presented
the reference response function of the 2.5m telescope imager as a mean
over the 6 columns of the imager.  We showed the detection of a
significant aging effect for the $u$ band, especially in the short
wavelength side, which amounts to about a 30 percent decrease in the
sensitivity over the period of the survey.

We confirmed, however, that brightness appears to be invariable to
0.01 mag over the years, and the SDSS catalogues is likely to be
accurate to this level: we confirmed that the error caused by
variations of the response function is not a dominant component of the
error in photometry.  Time variability was also detected for the
response function for other colour bands, but it is small and well
within our tolerable errors.  The variation other than that for the
$u$ passband is ascribed mostly to the temperature dependence of
transmission properties of the colour glass that composes the filters.

We have studied the seasonal, secular and column to column 
variations of the response function and their effects 
on photometry. We also studied the effect of the use of the associated
systems that have slightly different response functions in the SDSS
photometric calibration procedure.  We have verified that the effect
of the variation of the response function, which may amount to 0.01
mag in $g$, $r$, $i$ and $z$ bands and, and variations among 6
different devices of the imager, which also amount to 0.01$-$0.02 mag,
are cancelled very well by calibration procedures and do not appear in
the final SDSS catalogues; the residual effects are not larger
than 0.01 mag for all passbands.  We
have expected sizable aging effects in the response function for the
$u$ band, but photometric calibrations absorbs most of variations
and the variation visible in the SDSS catalogue is small.

The variation of $u-g$ colours is of the order of
$\Delta(u-g) \sim 0.01$ mag except for stars with extreme colours
for which it can be $0.02$ mag.

\vskip10mm\noindent
{\bf Acknowledgement}

We thank Kazuhiro Shimasaku for his work for filter experiments to
characterise environmental effects, carried out with one of the
present authors (MF). We are grateful to Maki Sekiguchi for his advice
in the construction of the monochromatic illumination system, and Jeff
Pier for his arrangement that allowed our work to characterise the
USNO 1m detector system. We thank Advanced Technology Centre of NAOJ
at which our laboratory filter measurements were carried out.  We also
thank Michael A. Carr, Brian R. Elms, George A. Pauls, Robert
H. Lupton, Constance M. Rockosi, Craig Loomis, James Annis, Daniel
C. Long, and Shannon Watters for their helps in installation and
operation of the monochrometer at APO. We thank Tanner Nakagawara for
his preliminary analysis of catalogued data. We also thank John
Marriner, 
Naotaka Suzuki, and an anonymous referee for their 
very useful comments.  MF is supported by the Friends of
the Institute at Princeton, and received Grant in Aid of the Ministry
of Education (Japan) at Tokyo. MD and NY are also supported by Grants
in Aid of the Ministry of Education (Japan) and by a JSPS core-to-core
program, International Research Network for Dark Energy. MD and \v{Z}I
would like to thank the Aspen Center for Physics for their kind
support to carry out a part of this work.

Funding for the SDSS and SDSS-II has been provided by the Alfred
P. Sloan Foundation, the Participating Institutions, the National
Science Foundation, the U.S. Department of Energy, the National
Aeronautics and Space Administration, the Japanese Monbukagakusho, the
Max Planck Society, and the Higher Education Funding Council for
England. The SDSS Web Site is http://www.sdss.org/.

The SDSS is managed by the Astrophysical Research Consortium for the
Participating Institutions. The Participating Institutions are the
American Museum of Natural History, Astrophysical Institute Potsdam,
University of Basel, University of Cambridge, Case Western Reserve
University, University of Chicago, Drexel University, Fermilab, the
Institute for Advanced Study, the Japan Participation Group, Johns
Hopkins University, the Joint Institute for Nuclear Astrophysics, the
Kavli Institute for Particle Astrophysics and Cosmology, the Korean
Scientist Group, the Chinese Academy of Sciences (LAMOST), Los Alamos
National Laboratory, the Max-Planck-Institute for Astronomy (MPIA),
the Max-Planck-Institute for Astrophysics (MPA), New Mexico State
University, Ohio State University, University of Pittsburgh,
University of Portsmouth, Princeton University, the United States
Naval Observatory, and the University of Washington.


\newpage
\begin{table}
\caption{Journal of response function measurements. The number stands
for measured Camcol.} 
\begin{center}
\begin{tabular}{clllll}
\hline
Date       & $u$   & $g$   & $r$   & $i$    & $z$   \\
\hline
Jan. 2000  & 6     & 1     & 4     & 1      & 2--4,6\\
Apr. 2000  & 1     & 1     & 1,4   & 1      & 1--2  \\
Dec. 2000  & 1--6  & 1--6  & 1--6  & 1--6   & 1--6  \\
Sep. 2001  &       & 1--2  & 1--2  & 1--6   &       \\
Jul. 2004  & 1--2  & 1--6  & 1--4  & 1--3   & 1--2  \\
Oct. 2004  & 1--5  &       &       &        &       \\
Nov. 2004  & 6     & 1--6  & 2--6  & 1--6   & 1--6  \\
May  2005  & 2--4  &       & 1     &        &       \\
Jul. 2005  & 2     &       & 1     &        &       \\
Sep. 2006  & 5--6  &       &       &        &       \\
Dec. 2006  & 1--6  &       & 1--6  &        &       \\
May  2008  & 1--6  & 1--6  & 1--6  & 1--6   & 1--6  \\
\hline
\end{tabular}
\label{tab:qe_summary}
\end{center}
\end{table}

\begin{landscape}
\begin{table}
\caption{Filter characteristics \tablenotemark{a}}
\vspace{0.5cm}
\centering
\footnotesize
\begin{tabular}{l|ccc|ccc|ccc|ccc|ccc}
\hline\hline 
system  & $\lambda_{\rm eff}(u)$ & $\lambda_{\rm blue}(u)$ & $\lambda_{\rm red}(u)$ &  $\lambda_{\rm eff}(g)$ & $\lambda_{\rm blue}(g)$ & $\lambda_{\rm red}(g)$ &  $\lambda_{\rm eff}(r)$ & $\lambda_{\rm blue}(r)$ & $\lambda_{\rm red}(r)$ &  $\lambda_{\rm eff}(i)$ & $\lambda_{\rm blue}(i)$ & $\lambda_{\rm red}(i)$ &  $\lambda_{\rm eff}(z)$ & $\lambda_{\rm blue}(z)$ & $\lambda_{\rm red}(z)$ \\
\hline
F96 standard & 3499 &  3216 &  3849 &  4728 &  4076 &  5485 &  6200 &  5548 &  6934 &  7615 &  6939 & 
8474 &  9054 &  8334 &  9741\\
PT-new  & 3496 &  3207 &  3838 &  4714 &  4035 &  5506 &  6180 &  5559 &  6879 &  7593 &  6971 & 
8346 &  9054 &  8334 &  9741\\
USNO      & 3500 &  3211 &  3847 &  4704 &  4057 &  5476 &  6201 &  5551 &  6930 &  7606 &  6931 & 
8461 &  9051 &  8329 &  9739\\
2.5m reference & 3531 &  3259 &  3852 &  4627 &  4019 &  5330 &  6140 &  5589 &  6748 &  7467 &  6902 & 
8178 &  8887 &  8257 &  9347\\
SDSS ``public'' & 3498 &  3203 &  3844 &  4627 &  4022 &  5330 &  6139 &  5593 &  6748 &  7467 &  6903 & 
8171 &  8927 &  8271 &  9449\\
\hline
\end{tabular}
\tablenotetext{a}{
$\lambda_{\rm eff}$ is defined in eq. (2) in the text,
$\lambda_{\rm blue}$ and $\lambda_{\rm red}$ are defined as 
the wavelengths that give $50\%$ of the peak response function. 
PT-new is the PT system after the filter replacement. The
system before the filter replacement is the same as F96 standard.
SDSS ``public'' means the response function that is made public through
the world wide web (2001).
All characteristics are without atmosphere transmissions.}
\end{table}
\end{landscape}

\begin{table}
\caption{
Measured $\lambda_{\rm eff}$ differences from the 2.5m reference case in \AA.
}
\vskip0.5cm
\centering \footnotesize
\begin{tabular}{llrrrrr}
\hline\hline 
epoch     &  column  &$\lambda_{\rm eff}(u)$ &$\lambda_{\rm eff}(g)$ &$\lambda_{\rm eff}(r)$ &$\lambda_{\rm eff}(i)$ & $\lambda_{\rm eff}(z)$\\
\hline
2.5m reference & &      3531 &  4627 &  6140 &  7467 &  8887\\
\hline
2000 Dec.      & 1 &   $-$40 &  $-$8 &    +0 &  $-$8 &   +23\\
               & 2 &   $-$43 & $-$18 &    +3 &    +5 &   +89\\
               & 3 &   $-$30 &  $-$1 & $-$19 &    +6 &   +57\\
               & 4 &   $-$28 &  $-$2 &    +1 &    +0 &   +50\\
               & 5 &   $-$37 &    +4 & $-$10 &  $-$2 &   +16\\
               & 6 &   $-$34 &    +1 &    +1 &  $-$5 &  $-$2\\
\hline
2004 Jul.      & 1 &      +7 &  $-$1 &   +16 &    +8 &   +20\\
               & 2 &      +7 & $-$11 &   +14 &   +19 &   +77\\
               & 3 &     --- &   +10 &  $-$8 &   +19 &   ---\\
               & 4 &     --- &    +8 &   +11 &   --- &   ---\\
               & 5 &     --- &    +9 &   --- &   --- &   ---\\
               & 6 &     --- &    +9 &   --- &   --- &   ---\\
\hline
2004 Oct./Nov. & 1 &    $-$2 &  $-$5 &   --- &  $-$6 & $-$17\\
               & 2 &    $-$5 & $-$15 &    +4 &    +5 &   +46\\
               & 3 &      +9 &    +6 & $-$16 &    +6 &    +9\\
               & 4 &    $-$7 &    +4 &    +2 &  $-$1 &   +19\\
               & 5 &   $-$16 &    +4 &    +0 &  $-$2 & $-$14\\
               & 6 &      +5 &    +5 &    +2 &  $-$6 & $-$31\\
\hline
\end{tabular}

\end{table}

\begin{deluxetable}{rrrrrrr}
\tablecaption{
2.5m telescope reference response function and adopted atmospheric transmission for APO (airmass = 1.3)
}
\tabletypesize{\scriptsize}
\tablecolumns{8}
\tablewidth{0pt}
\tablehead{
\colhead{$\lambda$} & \colhead{$u$} & \colhead{$g$} & \colhead{$r$} & \colhead{$i$} & \colhead{$z$} & \colhead{$T_{1.3 airmass}$}
}
\startdata
2940 & 0.0001  &   ---   &   ---   &   ---   &   ---   & 0.0162  \\
2960 & 0.0003  &   ---   &   ---   &   ---   &   ---   & 0.0242  \\
2980 & 0.0005  &   ---   &   ---   &   ---   &   ---   & 0.0351  \\
3000 & 0.0009  &   ---   &   ---   &   ---   &   ---   & 0.0459  \\
3020 & 0.0014  &   ---   &   ---   &   ---   &   ---   & 0.0644  \\
3040 & 0.0029  &   ---   &   ---   &   ---   &   ---   & 0.0828  \\
3060 & 0.0050  &   ---   &   ---   &   ---   &   ---   & 0.1048  \\
3080 & 0.0077  &   ---   &   ---   &   ---   &   ---   & 0.1303  \\
3100 & 0.0136  &   ---   &   ---   &   ---   &   ---   & 0.1558  \\
3120 & 0.0194  &   ---   &   ---   &   ---   &   ---   & 0.1882  \\
3140 & 0.0276  &   ---   &   ---   &   ---   &   ---   & 0.2206  \\
3160 & 0.0367  &   ---   &   ---   &   ---   &   ---   & 0.2628  \\
3180 & 0.0460  &   ---   &   ---   &   ---   &   ---   & 0.2962  \\
3200 & 0.0558  &   ---   &   ---   &   ---   &   ---   & 0.3269  \\
3220 & 0.0656  &   ---   &   ---   &   ---   &   ---   & 0.3521  \\
3240 & 0.0743  &   ---   &   ---   &   ---   &   ---   & 0.3728  \\
3260 & 0.0826  &   ---   &   ---   &   ---   &   ---   & 0.3912  \\
3280 & 0.0905  &   ---   &   ---   &   ---   &   ---   & 0.4067  \\
3300 & 0.0969  &   ---   &   ---   &   ---   &   ---   & 0.4224  \\
3320 & 0.1033  &   ---   &   ---   &   ---   &   ---   & 0.4369  \\
3340 & 0.1098  &   ---   &   ---   &   ---   &   ---   & 0.4496  \\
3360 & 0.1162  &   ---   &   ---   &   ---   &   ---   & 0.4608  \\
3380 & 0.1223  &   ---   &   ---   &   ---   &   ---   & 0.4687  \\
3400 & 0.1268  &   ---   &   ---   &   ---   &   ---   & 0.4766  \\
3420 & 0.1314  &   ---   &   ---   &   ---   &   ---   & 0.4824  \\
3440 & 0.1353  &   ---   &   ---   &   ---   &   ---   & 0.4882  \\
3460 & 0.1390  &   ---   &   ---   &   ---   &   ---   & 0.4977  \\
3480 & 0.1427  &   ---   &   ---   &   ---   &   ---   & 0.5109  \\
3500 & 0.1460  &   ---   &   ---   &   ---   &   ---   & 0.5241  \\
3520 & 0.1494  &   ---   &   ---   &   ---   &   ---   & 0.5311  \\
3540 & 0.1518  &   ---   &   ---   &   ---   &   ---   & 0.5382  \\
3560 & 0.1537  &   ---   &   ---   &   ---   &   ---   & 0.5452  \\
3580 & 0.1555  &   ---   &   ---   &   ---   &   ---   & 0.5522  \\
3600 & 0.1570  &   ---   &   ---   &   ---   &   ---   & 0.5593  \\
3620 & 0.1586  & 0.0002  &   ---   &   ---   &   ---   & 0.5660  \\
3640 & 0.1604  & 0.0015  &   ---   &   ---   &   ---   & 0.5728  \\
3660 & 0.1625  & 0.0027  &   ---   &   ---   &   ---   & 0.5795  \\
3680 & 0.1638  & 0.0038  &   ---   &   ---   &   ---   & 0.5862  \\
3700 & 0.1626  & 0.0035  &   ---   &   ---   &   ---   & 0.5930  \\
3720 & 0.1614  & 0.0032  &   ---   &   ---   &   ---   & 0.5979  \\
3740 & 0.1563  & 0.0031  &   ---   &   ---   &   ---   & 0.6027  \\
3760 & 0.1497  & 0.0037  &   ---   &   ---   &   ---   & 0.6076  \\
3780 & 0.1416  & 0.0048  &   ---   &   ---   &   ---   & 0.6125  \\
3800 & 0.1282  & 0.0067  &   ---   &   ---   &   ---   & 0.6174  \\
3820 & 0.1148  & 0.0115  &   ---   &   ---   &   ---   & 0.6231  \\
3840 & 0.0956  & 0.0220  &   ---   &   ---   &   ---   & 0.6289  \\
3860 & 0.0744  & 0.0353  &   ---   &   ---   &   ---   & 0.6346  \\
3880 & 0.0549  & 0.0507  &   ---   &   ---   &   ---   & 0.6404  \\
3900 & 0.0407  & 0.0740  &   ---   &   ---   &   ---   & 0.6461  \\
3920 & 0.0265  & 0.0973  &   ---   &   ---   &   ---   & 0.6534  \\
3940 & 0.0177  & 0.1224  &   ---   &   ---   &   ---   & 0.6606  \\
3960 & 0.0107  & 0.1484  &   ---   &   ---   &   ---   & 0.6679  \\
3980 & 0.0050  & 0.1757  &   ---   &   ---   &   ---   & 0.6751  \\
4000 & 0.0032  & 0.2081  &   ---   &   ---   &   ---   & 0.6824  \\
4020 & 0.0015  & 0.2404  &   ---   &   ---   &   ---   & 0.6876  \\
4040 & 0.0008  & 0.2617  &   ---   &   ---   &   ---   & 0.6928  \\
4060 & 0.0005  & 0.2785  &   ---   &   ---   &   ---   & 0.6981  \\
4080 & 0.0003  & 0.2954  &   ---   &   ---   &   ---   & 0.7033  \\
4100 & 0.0003  & 0.3122  &   ---   &   ---   &   ---   & 0.7086  \\
4120 & 0.0003  & 0.3290  &   ---   &   ---   &   ---   & 0.7129  \\
4140 & 0.0002  & 0.3411  &   ---   &   ---   &   ---   & 0.7172  \\
4160 & 0.0001  & 0.3512  &   ---   &   ---   &   ---   & 0.7215  \\
4180 &   ---   & 0.3603  &   ---   &   ---   &   ---   & 0.7258  \\
4200 &   ---   & 0.3660  &   ---   &   ---   &   ---   & 0.7301  \\
4220 &   ---   & 0.3717  &   ---   &   ---   &   ---   & 0.7339  \\
4240 &   ---   & 0.3773  &   ---   &   ---   &   ---   & 0.7378  \\
4260 &   ---   & 0.3830  &   ---   &   ---   &   ---   & 0.7416  \\
4280 &   ---   & 0.3886  &   ---   &   ---   &   ---   & 0.7455  \\
4300 &   ---   & 0.3943  &   ---   &   ---   &   ---   & 0.7493  \\
4320 &   ---   & 0.3999  &   ---   &   ---   &   ---   & 0.7529  \\
4340 &   ---   & 0.4043  &   ---   &   ---   &   ---   & 0.7564  \\
4360 &   ---   & 0.4083  &   ---   &   ---   &   ---   & 0.7600  \\
4380 &   ---   & 0.4122  &   ---   &   ---   &   ---   & 0.7635  \\
4400 &   ---   & 0.4161  &   ---   &   ---   &   ---   & 0.7671  \\
4420 &   ---   & 0.4200  &   ---   &   ---   &   ---   & 0.7705  \\
4440 &   ---   & 0.4240  &   ---   &   ---   &   ---   & 0.7739  \\
4460 &   ---   & 0.4279  &   ---   &   ---   &   ---   & 0.7773  \\
4480 &   ---   & 0.4314  &   ---   &   ---   &   ---   & 0.7808  \\
4500 &   ---   & 0.4337  &   ---   &   ---   &   ---   & 0.7842  \\
4520 &   ---   & 0.4359  &   ---   &   ---   &   ---   & 0.7873  \\
4540 &   ---   & 0.4381  &   ---   &   ---   &   ---   & 0.7904  \\
4560 &   ---   & 0.4404  &   ---   &   ---   &   ---   & 0.7935  \\
4580 &   ---   & 0.4426  &   ---   &   ---   &   ---   & 0.7965  \\
4600 &   ---   & 0.4448  &   ---   &   ---   &   ---   & 0.7996  \\
4620 &   ---   & 0.4470  &   ---   &   ---   &   ---   & 0.8021  \\
4640 &   ---   & 0.4488  &   ---   &   ---   &   ---   & 0.8047  \\
4660 &   ---   & 0.4504  &   ---   &   ---   &   ---   & 0.8072  \\
4680 &   ---   & 0.4521  &   ---   &   ---   &   ---   & 0.8097  \\
4700 &   ---   & 0.4537  &   ---   &   ---   &   ---   & 0.8122  \\
4720 &   ---   & 0.4553  &   ---   &   ---   &   ---   & 0.8141  \\
4740 &   ---   & 0.4569  &   ---   &   ---   &   ---   & 0.8160  \\
4760 &   ---   & 0.4586  &   ---   &   ---   &   ---   & 0.8179  \\
4780 &   ---   & 0.4601  &   ---   &   ---   &   ---   & 0.8199  \\
4800 &   ---   & 0.4611  &   ---   &   ---   &   ---   & 0.8218  \\
4820 &   ---   & 0.4622  &   ---   &   ---   &   ---   & 0.8237  \\
4840 &   ---   & 0.4633  &   ---   &   ---   &   ---   & 0.8256  \\
4860 &   ---   & 0.4644  &   ---   &   ---   &   ---   & 0.8276  \\
4880 &   ---   & 0.4655  &   ---   &   ---   &   ---   & 0.8295  \\
4900 &   ---   & 0.4666  &   ---   &   ---   &   ---   & 0.8314  \\
4920 &   ---   & 0.4677  &   ---   &   ---   &   ---   & 0.8327  \\
4940 &   ---   & 0.4687  &   ---   &   ---   &   ---   & 0.8340  \\
4960 &   ---   & 0.4698  &   ---   &   ---   &   ---   & 0.8353  \\
4980 &   ---   & 0.4709  &   ---   &   ---   &   ---   & 0.8366  \\
5000 &   ---   & 0.4719  &   ---   &   ---   &   ---   & 0.8379  \\
5020 &   ---   & 0.4730  &   ---   &   ---   &   ---   & 0.8388  \\
5040 &   ---   & 0.4741  &   ---   &   ---   &   ---   & 0.8397  \\
5060 &   ---   & 0.4752  &   ---   &   ---   &   ---   & 0.8406  \\
5080 &   ---   & 0.4762  &   ---   &   ---   &   ---   & 0.8415  \\
5100 &   ---   & 0.4770  &   ---   &   ---   &   ---   & 0.8423  \\
5120 &   ---   & 0.4765  &   ---   &   ---   &   ---   & 0.8432  \\
5140 &   ---   & 0.4753  &   ---   &   ---   &   ---   & 0.8441  \\
5160 &   ---   & 0.4731  &   ---   &   ---   &   ---   & 0.8450  \\
5180 &   ---   & 0.4704  &   ---   &   ---   &   ---   & 0.8458  \\
5200 &   ---   & 0.4672  &   ---   &   ---   &   ---   & 0.8467  \\
5220 &   ---   & 0.4625  &   ---   &   ---   &   ---   & 0.8477  \\
5240 &   ---   & 0.4512  &   ---   &   ---   &   ---   & 0.8487  \\
5260 &   ---   & 0.4326  &   ---   &   ---   &   ---   & 0.8497  \\
5280 &   ---   & 0.3996  &   ---   &   ---   &   ---   & 0.8507  \\
5300 &   ---   & 0.3429  &   ---   &   ---   &   ---   & 0.8517  \\
5320 &   ---   & 0.2768  &   ---   &   ---   &   ---   & 0.8527  \\
5340 &   ---   & 0.2013  &   ---   &   ---   &   ---   & 0.8537  \\
5360 &   ---   & 0.1397  &   ---   &   ---   &   ---   & 0.8547  \\
5380 &   ---   & 0.0899  & 0.0002  &   ---   &   ---   & 0.8557  \\
5400 &   ---   & 0.0585  & 0.0018  &   ---   &   ---   & 0.8567  \\
5420 &   ---   & 0.0398  & 0.0050  &   ---   &   ---   & 0.8571  \\
5440 &   ---   & 0.0269  & 0.0105  &   ---   &   ---   & 0.8576  \\
5460 &   ---   & 0.0189  & 0.0225  &   ---   &   ---   & 0.8580  \\
5480 &   ---   & 0.0136  & 0.0452  &   ---   &   ---   & 0.8585  \\
5500 &   ---   & 0.0096  & 0.0751  &   ---   &   ---   & 0.8589  \\
5520 &   ---   & 0.0068  & 0.1175  &   ---   &   ---   & 0.8594  \\
5540 &   ---   & 0.0051  & 0.1641  &   ---   &   ---   & 0.8598  \\
5560 &   ---   & 0.0037  & 0.2118  &   ---   &   ---   & 0.8602  \\
5580 &   ---   & 0.0024  & 0.2567  &   ---   &   ---   & 0.8607  \\
5600 &   ---   & 0.0013  & 0.2979  &   ---   &   ---   & 0.8611  \\
5620 &   ---   & 0.0002  & 0.3368  &   ---   &   ---   & 0.8616  \\
5640 &   ---   &   ---   & 0.3724  &   ---   &   ---   & 0.8620  \\
5660 &   ---   &   ---   & 0.4042  &   ---   &   ---   & 0.8625  \\
5680 &   ---   &   ---   & 0.4327  &   ---   &   ---   & 0.8629  \\
5700 &   ---   &   ---   & 0.4531  &   ---   &   ---   & 0.8634  \\
5720 &   ---   &   ---   & 0.4667  &   ---   &   ---   & 0.8638  \\
5740 &   ---   &   ---   & 0.4774  &   ---   &   ---   & 0.8643  \\
5760 &   ---   &   ---   & 0.4868  &   ---   &   ---   & 0.8647  \\
5780 &   ---   &   ---   & 0.4949  &   ---   &   ---   & 0.8652  \\
5800 &   ---   &   ---   & 0.5019  &   ---   &   ---   & 0.8656  \\
5820 &   ---   &   ---   & 0.5068  &   ---   &   ---   & 0.8672  \\
5840 &   ---   &   ---   & 0.5118  &   ---   &   ---   & 0.8688  \\
5860 &   ---   &   ---   & 0.5162  &   ---   &   ---   & 0.8704  \\
5880 &   ---   &   ---   & 0.5187  &   ---   &   ---   & 0.8720  \\
5900 &   ---   &   ---   & 0.5212  &   ---   &   ---   & 0.8736  \\
5920 &   ---   &   ---   & 0.5237  &   ---   &   ---   & 0.8752  \\
5940 &   ---   &   ---   & 0.5262  &   ---   &   ---   & 0.8768  \\
5960 &   ---   &   ---   & 0.5287  &   ---   &   ---   & 0.8783  \\
5980 &   ---   &   ---   & 0.5309  &   ---   &   ---   & 0.8799  \\
6000 &   ---   &   ---   & 0.5323  &   ---   &   ---   & 0.8815  \\
6020 &   ---   &   ---   & 0.5336  &   ---   &   ---   & 0.8841  \\
6040 &   ---   &   ---   & 0.5349  &   ---   &   ---   & 0.8866  \\
6060 &   ---   &   ---   & 0.5362  &   ---   &   ---   & 0.8891  \\
6080 &   ---   &   ---   & 0.5370  &   ---   &   ---   & 0.8917  \\
6100 &   ---   &   ---   & 0.5361  &   ---   &   ---   & 0.8942  \\
6120 &   ---   &   ---   & 0.5352  &   ---   &   ---   & 0.8961  \\
6140 &   ---   &   ---   & 0.5342  &   ---   &   ---   & 0.8980  \\
6160 &   ---   &   ---   & 0.5333  &   ---   &   ---   & 0.8998  \\
6180 &   ---   &   ---   & 0.5333  &   ---   &   ---   & 0.9017  \\
6200 &   ---   &   ---   & 0.5358  &   ---   &   ---   & 0.9036  \\
6220 &   ---   &   ---   & 0.5383  &   ---   &   ---   & 0.9048  \\
6240 &   ---   &   ---   & 0.5409  &   ---   &   ---   & 0.9059  \\
6260 &   ---   &   ---   & 0.5434  &   ---   &   ---   & 0.9071  \\
6280 &   ---   &   ---   & 0.5454  &   ---   &   ---   & 0.9083  \\
6300 &   ---   &   ---   & 0.5462  &   ---   &   ---   & 0.9095  \\
6320 &   ---   &   ---   & 0.5470  &   ---   &   ---   & 0.9104  \\
6340 &   ---   &   ---   & 0.5477  &   ---   &   ---   & 0.9114  \\
6360 &   ---   &   ---   & 0.5485  &   ---   &   ---   & 0.9123  \\
6380 &   ---   &   ---   & 0.5488  &   ---   &   ---   & 0.9133  \\
6400 &   ---   &   ---   & 0.5480  &   ---   &   ---   & 0.9142  \\
6420 &   ---   &   ---   & 0.5472  &   ---   &   ---   & 0.9152  \\
6440 &   ---   &   ---   & 0.5464  &   ---   &   ---   & 0.9161  \\
6460 &   ---   &   ---   & 0.5455  &   ---   &   ---   & 0.9171  \\
6480 &   ---   &   ---   & 0.5449  &   ---   &   ---   & 0.9180  \\
6500 &   ---   &   ---   & 0.5450  &   ---   &   ---   & 0.9190  \\
6520 &   ---   &   ---   & 0.5450  &   ---   &   ---   & 0.9194  \\
6540 &   ---   &   ---   & 0.5447  &   ---   &   ---   & 0.9198  \\
6560 &   ---   &   ---   & 0.5435  &   ---   &   ---   & 0.9202  \\
6580 &   ---   &   ---   & 0.5423  &   ---   &   ---   & 0.9206  \\
6600 &   ---   &   ---   & 0.5394  & 0.0002  &   ---   & 0.9210  \\
6620 &   ---   &   ---   & 0.5324  & 0.0008  &   ---   & 0.9214  \\
6640 &   ---   &   ---   & 0.5190  & 0.0012  &   ---   & 0.9217  \\
6660 &   ---   &   ---   & 0.4992  & 0.0017  &   ---   & 0.9221  \\
6680 &   ---   &   ---   & 0.4683  & 0.0026  &   ---   & 0.9225  \\
6700 &   ---   &   ---   & 0.4230  & 0.0046  &   ---   & 0.9228  \\
6720 &   ---   &   ---   & 0.3685  & 0.0080  &   ---   & 0.9232  \\
6740 &   ---   &   ---   & 0.3030  & 0.0131  &   ---   & 0.9236  \\
6760 &   ---   &   ---   & 0.2344  & 0.0226  &   ---   & 0.9239  \\
6780 &   ---   &   ---   & 0.1724  & 0.0365  &   ---   & 0.9243  \\
6800 &   ---   &   ---   & 0.1212  & 0.0560  &   ---   & 0.9246  \\
6820 &   ---   &   ---   & 0.0842  & 0.0834  &   ---   & 0.9250  \\
6840 &   ---   &   ---   & 0.0556  & 0.1162  &   ---   & 0.9027  \\
6860 &   ---   &   ---   & 0.0370  & 0.1553  &   ---   & 0.8804  \\
6880 &   ---   &   ---   & 0.0273  & 0.1952  &   ---   & 0.8581  \\
6900 &   ---   &   ---   & 0.0201  & 0.2377  &   ---   & 0.8722  \\
6920 &   ---   &   ---   & 0.0130  & 0.2839  &   ---   & 0.8863  \\
6940 &   ---   &   ---   & 0.0097  & 0.3222  &   ---   & 0.9004  \\
6960 &   ---   &   ---   & 0.0076  & 0.3565  &   ---   & 0.9145  \\
6980 &   ---   &   ---   & 0.0054  & 0.3869  &   ---   & 0.9286  \\
7000 &   ---   &   ---   & 0.0036  & 0.4104  &   ---   & 0.9298  \\
7020 &   ---   &   ---   & 0.0019  & 0.4301  &   ---   & 0.9301  \\
7040 &   ---   &   ---   & 0.0003  & 0.4458  &   ---   & 0.9305  \\
7060 &   ---   &   ---   &   ---   & 0.4565  &   ---   & 0.9308  \\
7080 &   ---   &   ---   &   ---   & 0.4648  &   ---   & 0.9312  \\
7100 &   ---   &   ---   &   ---   & 0.4706  &   ---   & 0.9315  \\
7120 &   ---   &   ---   &   ---   & 0.4764  &   ---   & 0.9319  \\
7140 &   ---   &   ---   &   ---   & 0.4791  &   ---   & 0.9322  \\
7160 &   ---   &   ---   &   ---   & 0.4814  &   ---   & 0.8928  \\
7180 &   ---   &   ---   &   ---   & 0.4823  &   ---   & 0.8533  \\
7200 &   ---   &   ---   &   ---   & 0.4815  &   ---   & 0.8703  \\
7220 &   ---   &   ---   &   ---   & 0.4806  &   ---   & 0.8873  \\
7240 &   ---   &   ---   &   ---   & 0.4771  &   ---   & 0.8896  \\
7260 &   ---   &   ---   &   ---   & 0.4732  &   ---   & 0.8919  \\
7280 &   ---   &   ---   &   ---   & 0.4694  &   ---   & 0.8942  \\
7300 &   ---   &   ---   &   ---   & 0.4655  &   ---   & 0.8966  \\
7320 &   ---   &   ---   &   ---   & 0.4617  &   ---   & 0.9156  \\
7340 &   ---   &   ---   &   ---   & 0.4578  &   ---   & 0.9346  \\
7360 &   ---   &   ---   &   ---   & 0.4539  &   ---   & 0.9358  \\
7380 &   ---   &   ---   &   ---   & 0.4505  &   ---   & 0.9365  \\
7400 &   ---   &   ---   &   ---   & 0.4477  &   ---   & 0.9371  \\
7420 &   ---   &   ---   &   ---   & 0.4449  &   ---   & 0.9371  \\
7440 &   ---   &   ---   &   ---   & 0.4421  &   ---   & 0.9371  \\
7460 &   ---   &   ---   &   ---   & 0.4393  &   ---   & 0.9371  \\
7480 &   ---   &   ---   &   ---   & 0.4364  &   ---   & 0.9371  \\
7500 &   ---   &   ---   &   ---   & 0.4335  &   ---   & 0.9371  \\
7520 &   ---   &   ---   &   ---   & 0.4306  &   ---   & 0.9371  \\
7540 &   ---   &   ---   &   ---   & 0.4264  &   ---   & 0.9371  \\
7560 &   ---   &   ---   &   ---   & 0.4220  &   ---   & 0.9371  \\
7580 &   ---   &   ---   &   ---   & 0.4176  &   ---   & 0.9209  \\
7600 &    ---    &   ---   &   ---   & 0.4132  &   ---   & 0.5647  \\
7620 & 0.000003  &   ---   &   ---   & 0.4088  &   ---   & 0.6334  \\
7640 & 0.000044  &   ---   &   ---   & 0.4042  &   ---   & 0.6037  \\
7660 & 0.000149  &   ---   &   ---   & 0.3996  &   ---   & 0.7830  \\
7680 & 0.000258  &   ---   &   ---   & 0.3951  & 0.0000  & 0.9396  \\
7700 & 0.000397  &   ---   &   ---   & 0.3905  & 0.0000  & 0.9407  \\
7720 & 0.000553  &   ---   &   ---   & 0.3860  & 0.0001  & 0.9410  \\
7740 & 0.000676  &   ---   &   ---   & 0.3815  & 0.0001  & 0.9412  \\
7760 & 0.000675  &   ---   &   ---   & 0.3770  & 0.0001  & 0.9415  \\
7780 & 0.000551  &   ---   &   ---   & 0.3725  & 0.0001  & 0.9417  \\
7800 & 0.000403  &   ---   &   ---   & 0.3680  & 0.0002  & 0.9420  \\
7820 & 0.000276  &   ---   &   ---   & 0.3636  & 0.0002  & 0.9422  \\
7840 & 0.000179  &   ---   &   ---   & 0.3610  & 0.0002  & 0.9424  \\
7860 & 0.000093  &   ---   &   ---   & 0.3586  & 0.0003  & 0.9427  \\
7880 & 0.000044  &   ---   &   ---   & 0.3562  & 0.0004  & 0.9429  \\
7900 & 0.000026  &   ---   &   ---   & 0.3539  & 0.0006  & 0.9432  \\
7920 & 0.000011  &   ---   &   ---   & 0.3515  & 0.0008  & 0.9433  \\
7940 & 0.000007  &   ---   &   ---   & 0.3492  & 0.0010  & 0.9434  \\
7960 &    ---    &   ---   &   ---   & 0.3469  & 0.0012  & 0.9435  \\
7980 &    ---    &   ---   &   ---   & 0.3449  & 0.0016  & 0.9437  \\
8000 &    ---    &   ---   &   ---   & 0.3432  & 0.0023  & 0.9438  \\
8020 &    ---    &   ---   &   ---   & 0.3411  & 0.0030  & 0.9439  \\
8040 &    ---    &   ---   &   ---   & 0.3388  & 0.0044  & 0.9440  \\
8060 &    ---    &   ---   &   ---   & 0.3362  & 0.0059  & 0.9442  \\
8080 &    ---    &   ---   &   ---   & 0.3328  & 0.0078  & 0.9443  \\
8100 &    ---    &   ---   &   ---   & 0.3279  & 0.0105  & 0.9444  \\
8120 &    ---    &   ---   &   ---   & 0.3215  & 0.0132  & 0.9205  \\
8140 &    ---    &   ---   &   ---   & 0.3043  & 0.0171  & 0.8966  \\
8160 &    ---    &   ---   &   ---   & 0.2763  & 0.0212  & 0.8966  \\
8180 &    ---    &   ---   &   ---   & 0.2379  & 0.0257  & 0.8966  \\
8200 &    ---    &   ---   &   ---   & 0.1857  & 0.0309  & 0.8966  \\
8220 &    ---    &   ---   &   ---   & 0.1355  & 0.0362  & 0.8966  \\
8240 &    ---    &   ---   &   ---   & 0.0874  & 0.0415  & 0.8966  \\
8260 &    ---    &   ---   &   ---   & 0.0578  & 0.0467  & 0.8966  \\
8280 &    ---    &   ---   &   ---   & 0.0360  & 0.0519  & 0.8966  \\
8300 &    ---    &   ---   &   ---   & 0.0212  & 0.0570  & 0.8966  \\
8320 &    ---    &   ---   &   ---   & 0.0144  & 0.0621  & 0.9167  \\
8340 &    ---    &   ---   &   ---   & 0.0094  & 0.0664  & 0.9368  \\
8360 &    ---    &   ---   &   ---   & 0.0061  & 0.0705  & 0.9469  \\
8380 &    ---    &   ---   &   ---   & 0.0020  & 0.0742  & 0.9470  \\
8400 &    ---    &   ---   &   ---   &   ---   & 0.0773  & 0.9470  \\
8420 &    ---    &   ---   &   ---   &   ---   & 0.0803  & 0.9471  \\
8440 &    ---    &   ---   &   ---   &   ---   & 0.0824  & 0.9472  \\
8460 &    ---    &   ---   &   ---   &   ---   & 0.0845  & 0.9473  \\
8480 &    ---    &   ---   &   ---   &   ---   & 0.0861  & 0.9474  \\
8500 &    ---    &   ---   &   ---   &   ---   & 0.0871  & 0.9475  \\
8520 &    ---    &   ---   &   ---   &   ---   & 0.0882  & 0.9476  \\
8540 &    ---    &   ---   &   ---   &   ---   & 0.0893  & 0.9477  \\
8560 &    ---    &   ---   &   ---   &   ---   & 0.0904  & 0.9478  \\
8580 &    ---    &   ---   &   ---   &   ---   & 0.0911  & 0.9479  \\
8600 &    ---    &   ---   &   ---   &   ---   & 0.0912  & 0.9480  \\
8620 &    ---    &   ---   &   ---   &   ---   & 0.0913  & 0.9481  \\
8640 &    ---    &   ---   &   ---   &   ---   & 0.0915  & 0.9482  \\
8660 &    ---    &   ---   &   ---   &   ---   & 0.0917  & 0.9483  \\
8680 &    ---    &   ---   &   ---   &   ---   & 0.0914  & 0.9484  \\
8700 &    ---    &   ---   &   ---   &   ---   & 0.0906  & 0.9484  \\
8720 &    ---    &   ---   &   ---   &   ---   & 0.0898  & 0.9485  \\
8740 &    ---    &   ---   &   ---   &   ---   & 0.0889  & 0.9486  \\
8760 &    ---    &   ---   &   ---   &   ---   & 0.0881  & 0.9487  \\
8780 &    ---    &   ---   &   ---   &   ---   & 0.0869  & 0.9488  \\
8800 &    ---    &   ---   &   ---   &   ---   & 0.0854  & 0.9489  \\
8820 &    ---    &   ---   &   ---   &   ---   & 0.0838  & 0.9490  \\
8840 &    ---    &   ---   &   ---   &   ---   & 0.0822  & 0.9491  \\
8860 &    ---    &   ---   &   ---   &   ---   & 0.0806  & 0.9492  \\
8880 &    ---    &   ---   &   ---   &   ---   & 0.0790  & 0.9493  \\
8900 &    ---    &   ---   &   ---   &   ---   & 0.0772  & 0.9493  \\
8920 &    ---    &   ---   &   ---   &   ---   & 0.0755  & 0.9190  \\
8940 &    ---    &   ---   &   ---   &   ---   & 0.0738  & 0.8888  \\
8960 &    ---    &   ---   &   ---   &   ---   & 0.0720  & 0.8585  \\
8980 &    ---    &   ---   &   ---   &   ---   & 0.0704  & 0.8282  \\
9000 &    ---    &   ---   &   ---   &   ---   & 0.0688  & 0.8387  \\
9020 &    ---    &   ---   &   ---   &   ---   & 0.0672  & 0.8492  \\
9040 &    ---    &   ---   &   ---   &   ---   & 0.0656  & 0.8597  \\
9060 &    ---    &   ---   &   ---   &   ---   & 0.0640  & 0.8701  \\
9080 &    ---    &   ---   &   ---   &   ---   & 0.0625  & 0.8701  \\
9100 &    ---    &   ---   &   ---   &   ---   & 0.0612  & 0.8701  \\
9120 &    ---    &   ---   &   ---   &   ---   & 0.0598  & 0.8701  \\
9140 &    ---    &   ---   &   ---   &   ---   & 0.0585  & 0.8701  \\
9160 &    ---    &   ---   &   ---   &   ---   & 0.0571  & 0.8701  \\
9180 &    ---    &   ---   &   ---   &   ---   & 0.0559  & 0.8701  \\
9200 &    ---    &   ---   &   ---   &   ---   & 0.0547  & 0.8701  \\
9220 &    ---    &   ---   &   ---   &   ---   & 0.0535  & 0.8701  \\
9240 &    ---    &   ---   &   ---   &   ---   & 0.0523  & 0.8701  \\
9260 &    ---    &   ---   &   ---   &   ---   & 0.0511  & 0.8701  \\
9280 &    ---    &   ---   &   ---   &   ---   & 0.0499  & 0.8043  \\
9300 &    ---    &   ---   &   ---   &   ---   & 0.0487  & 0.7385  \\
9320 &    ---    &   ---   &   ---   &   ---   & 0.0475  & 0.6727  \\
9340 &    ---    &   ---   &   ---   &   ---   & 0.0463  & 0.6069  \\
9360 &    ---    &   ---   &   ---   &   ---   & 0.0451  & 0.5861  \\
9380 &    ---    &   ---   &   ---   &   ---   & 0.0440  & 0.6104  \\
9400 &    ---    &   ---   &   ---   &   ---   & 0.0430  & 0.6346  \\
9420 &    ---    &   ---   &   ---   &   ---   & 0.0420  & 0.6588  \\
9440 &    ---    &   ---   &   ---   &   ---   & 0.0410  & 0.6365  \\
9460 &    ---    &   ---   &   ---   &   ---   & 0.0400  & 0.6142  \\
9480 &    ---    &   ---   &   ---   &   ---   & 0.0390  & 0.6331  \\
9500 &    ---    &   ---   &   ---   &   ---   & 0.0379  & 0.6520  \\
9520 &    ---    &   ---   &   ---   &   ---   & 0.0369  & 0.6491  \\
9540 &    ---    &   ---   &   ---   &   ---   & 0.0358  & 0.6461  \\
9560 &    ---    &   ---   &   ---   &   ---   & 0.0347  & 0.6728  \\
9580 &    ---    &   ---   &   ---   &   ---   & 0.0337  & 0.6994  \\
9600 &    ---    &   ---   &   ---   &   ---   & 0.0327  & 0.7322  \\
9620 &    ---    &   ---   &   ---   &   ---   & 0.0317  & 0.7651  \\
9640 &    ---    &   ---   &   ---   &   ---   & 0.0307  & 0.8003  \\
9660 &    ---    &   ---   &   ---   &   ---   & 0.0297  & 0.8355  \\
9680 &    ---    &   ---   &   ---   &   ---   & 0.0287  & 0.8707  \\
9700 &    ---    &   ---   &   ---   &   ---   & 0.0276  & 0.9059  \\
9720 &    ---    &   ---   &   ---   &   ---   & 0.0266  & 0.8880  \\
9740 &    ---    &   ---   &   ---   &   ---   & 0.0256  & 0.8701  \\
9760 &    ---    &   ---   &   ---   &   ---   & 0.0245  & 0.8846  \\
9780 &    ---    &   ---   &   ---   &   ---   & 0.0235  & 0.8990  \\
9800 &    ---    &   ---   &   ---   &   ---   & 0.0226  & 0.9135  \\
9820 &    ---    &   ---   &   ---   &   ---   & 0.0216  & 0.9279  \\
9840 &    ---    &   ---   &   ---   &   ---   & 0.0206  & 0.9423  \\
9860 &    ---    &   ---   &   ---   &   ---   & 0.0196  & 0.9568  \\
9880 &    ---    &   ---   &   ---   &   ---   & 0.0186  & 0.9568  \\
9900 &    ---    &   ---   &   ---   &   ---   & 0.0176  & 0.9568  \\
9920 &    ---    &   ---   &   ---   &   ---   & 0.0166  & 0.9568  \\
9940 &    ---    &   ---   &   ---   &   ---   & 0.0156  & 0.9568  \\
9960 &    ---    &   ---   &   ---   &   ---   & 0.0147  & 0.9568  \\
9980 &    ---    &   ---   &   ---   &   ---   & 0.0138  & 0.9569  \\
10000 &    ---    &   ---   &   ---   &   ---   & 0.0132  & 0.9569  \\
10020 &    ---    &   ---   &   ---   &   ---   & 0.0125  & 0.9569  \\
10040 &    ---    &   ---   &   ---   &   ---   & 0.0119  & 0.9569  \\
10060 &    ---    &   ---   &   ---   &   ---   & 0.0113  & 0.9569  \\
10080 &    ---    &   ---   &   ---   &   ---   & 0.0106  & 0.9570  \\
10100 &    ---    &   ---   &   ---   &   ---   & 0.0099  & 0.9570  \\
10120 &    ---    &   ---   &   ---   &   ---   & 0.0093  & 0.9570  \\
10140 &    ---    &   ---   &   ---   &   ---   & 0.0086  & 0.9570  \\
10160 &    ---    &   ---   &   ---   &   ---   & 0.0080  & 0.9570  \\
10180 &    ---    &   ---   &   ---   &   ---   & 0.0074  & 0.9571  \\
10200 &    ---    &   ---   &   ---   &   ---   & 0.0070  & 0.9571  \\
10220 &    ---    &   ---   &   ---   &   ---   & 0.0065  & 0.9571  \\
10240 &    ---    &   ---   &   ---   &   ---   & 0.0061  & 0.9571  \\
10260 &    ---    &   ---   &   ---   &   ---   & 0.0056  & 0.9571  \\
10280 &    ---    &   ---   &   ---   &   ---   & 0.0052  & 0.9571  \\
10300 &    ---    &   ---   &   ---   &   ---   & 0.0047  & 0.9572  \\
10320 &    ---    &   ---   &   ---   &   ---   & 0.0042  & 0.9572  \\
10340 &    ---    &   ---   &   ---   &   ---   & 0.0038  & 0.9572  \\
10360 &    ---    &   ---   &   ---   &   ---   & 0.0033  & 0.9572  \\
10380 &    ---    &   ---   &   ---   &   ---   & 0.0030  & 0.9572  \\
10400 &    ---    &   ---   &   ---   &   ---   & 0.0029  & 0.9573  \\
10420 &    ---    &   ---   &   ---   &   ---   & 0.0027  & 0.9573  \\
10440 &    ---    &   ---   &   ---   &   ---   & 0.0026  & 0.9573  \\
10460 &    ---    &   ---   &   ---   &   ---   & 0.0024  & 0.9573  \\
10480 &    ---    &   ---   &   ---   &   ---   & 0.0022  & 0.9573  \\
10500 &    ---    &   ---   &   ---   &   ---   & 0.0021  & 0.9574  \\
10520 &    ---    &   ---   &   ---   &   ---   & 0.0019  & 0.9574  \\
10540 &    ---    &   ---   &   ---   &   ---   & 0.0018  & 0.9574  \\
10560 &    ---    &   ---   &   ---   &   ---   & 0.0016  & 0.9574  \\
10580 &    ---    &   ---   &   ---   &   ---   & 0.0015  & 0.9574  \\
10600 &    ---    &   ---   &   ---   &   ---   & 0.0014  & 0.9574  \\
10620 &    ---    &   ---   &   ---   &   ---   & 0.0013  & 0.9575  \\
10640 &    ---    &   ---   &   ---   &   ---   & 0.0012  & 0.9575  \\
10660 &    ---    &   ---   &   ---   &   ---   & 0.0012  & 0.9575  \\
10680 &    ---    &   ---   &   ---   &   ---   & 0.0011  & 0.9575  \\
10700 &    ---    &   ---   &   ---   &   ---   & 0.0010  & 0.9575  \\
10720 &    ---    &   ---   &   ---   &   ---   & 0.0009  & 0.9576  \\
10740 &    ---    &   ---   &   ---   &   ---   & 0.0008  & 0.9576  \\
10760 &    ---    &   ---   &   ---   &   ---   & 0.0007  & 0.9576  \\
10780 &    ---    &   ---   &   ---   &   ---   & 0.0007  & 0.9576  \\
10800 &    ---    &   ---   &   ---   &   ---   & 0.0007  & 0.9576  \\
10820 &    ---    &   ---   &   ---   &   ---   & 0.0006  & 0.9577  \\
10840 &    ---    &   ---   &   ---   &   ---   & 0.0006  & 0.9577  \\
10860 &    ---    &   ---   &   ---   &   ---   & 0.0006  & 0.9577  \\
10880 &    ---    &   ---   &   ---   &   ---   & 0.0005  & 0.9577  \\
10900 &    ---    &   ---   &   ---   &   ---   & 0.0005  & 0.9577  \\
10920 &    ---    &   ---   &   ---   &   ---   & 0.0005  & 0.9577  \\
10940 &    ---    &   ---   &   ---   &   ---   & 0.0004  & 0.9578  \\
10960 &    ---    &   ---   &   ---   &   ---   & 0.0004  & 0.9578  \\
10980 &    ---    &   ---   &   ---   &   ---   & 0.0004  & 0.9578  \\
11000 &    ---    &   ---   &   ---   &   ---   & 0.0003  & 0.9578  \\
11020 &    ---    &   ---   &   ---   &   ---   & 0.0003  & 0.9578  \\
11040 &    ---    &   ---   &   ---   &   ---   & 0.0002  & 0.9579  \\
11060 &    ---    &   ---   &   ---   &   ---   & 0.0002  & 0.9579  \\
11080 &    ---    &   ---   &   ---   &   ---   & 0.0002  & 0.9579  \\
11100 &    ---    &   ---   &   ---   &   ---   & 0.0001  & 0.9579  \\
11120 &    ---    &   ---   &   ---   &   ---   & 0.0001  & 0.9579  \\
11140 &    ---    &   ---   &   ---   &   ---   & 0.0001  & 0.9580  \\
11160 &    ---    &   ---   &   ---   &   ---   & 0.0000  & 0.9580  \\
\enddata
\end{deluxetable}


\newpage
\begin{table}
\caption{\bf Summary of the uncertainties in the effective wavelength} 
\begin{center}
\begin{tabular}{llcl}
\hline
source      & uncertainty (\AA)   & band & comments   \\
\hline
monochromator repeatability  & $\pm 3 $ & all & r.m.s. \\
incidnent angle& -2 -- -3     & g, r, i & half of red cutoff shifts\\
filter temperature & 7, 15, 21, 51& g, r, i, z & peak-to-peak, color glass temperature effect\\
aging & $\approx -30$ & u & CCD coating? \\
CCD sensitivity? & $\approx \pm 30$ & z & temperature effect? \\ 
\hline
\end{tabular}
\label{tab:Table_new1}
\end{center}
\end{table}

\begin{table}
\caption{
Brightness of BD+17$^\circ$4708 in AB$_{95}$ 
without atmospheric extinction
}
\vskip0.5cm
\centering
\footnotesize
\begin{tabular}{lrrrrr}
\hline\hline 
             &$u$ &   $g$ &   $r$ &   $i$   & $z$\\
\hline
F96 standard   & 10.559 & 9.636 & 9.353 & 9.251 & 9.230\\
USNO 1m        & 10.558 & 9.644 & 9.353 & 9.252 & 9.230\\
PT-new         & 10.565 & 9.640 & 9.356 & 9.252 & 9.230\\
2.5m reference & 10.525 & 9.670 & 9.361 & 9.257 & 9.228\\
SDSS public web site (2001) & 10.560 & 9.671 & 9.361 & 9.257 & 9.229\\
\hline
\end{tabular}
\end{table}

\begin{table}
\caption{Characteristics of the response functions and
brightness of BD+17$^\circ$4708 with atmospheric transmission at 1.3 airmass.
Atmospheric extinction in flux itself is removed in brightness.
}
\vskip0.5cm
\centering
\footnotesize
\begin{tabular}{lrrrrr}
\hline\hline
                                                     &  $u$ &  $g$ &  $r$ &  $i$ &  $z$\\
\hline
F96 standard $\lambda_{\rm eff}$                     & 3537 & 4753 & 6209 & 7619 & 9032\\
USNO 1m      $\lambda_{\rm eff}$                     & 3539 & 4731 & 6210 & 7609 & 9030\\
PT-new       $\lambda_{\rm eff}$                     & 3534 & 4742 & 6189 & 7595 & 9032\\
2.5m reference $\lambda_{\rm eff}$                   & 3568 & 4653 & 6148 & 7468 & 8863\\
\hline
BD+17$^\circ$4708 with F96 standard (1.3 airmass)   & 10.520 & 9.628 & 9.353 & 9.251 & 9.231\\
BD+17$^\circ$4708 with USNO 1m (1.3 airmass)        & 10.519 & 9.635 & 9.353 & 9.252 & 9.231\\
BD+17$^\circ$4708 with PT-new (1.3 airmass)         & 10.527 & 9.632 & 9.355 & 9.252 & 9.231\\
BD+17$^\circ$4708 with 2.5m reference (1.3 airmass) & 10.489 & 9.662 & 9.360 & 9.257 & 9.228\\
BD+17$^\circ$4708 with SDSS ``public'' web (1.3 airmass) & 10.518 & 9.662 & 9.360 & 9.257 & 9.229\\
\hline

\end{tabular}
\end{table}

\clearpage

\begin{figure}[pt]
\begin{center}  
\includegraphics[width=12cm,bb=0 0 747 522]{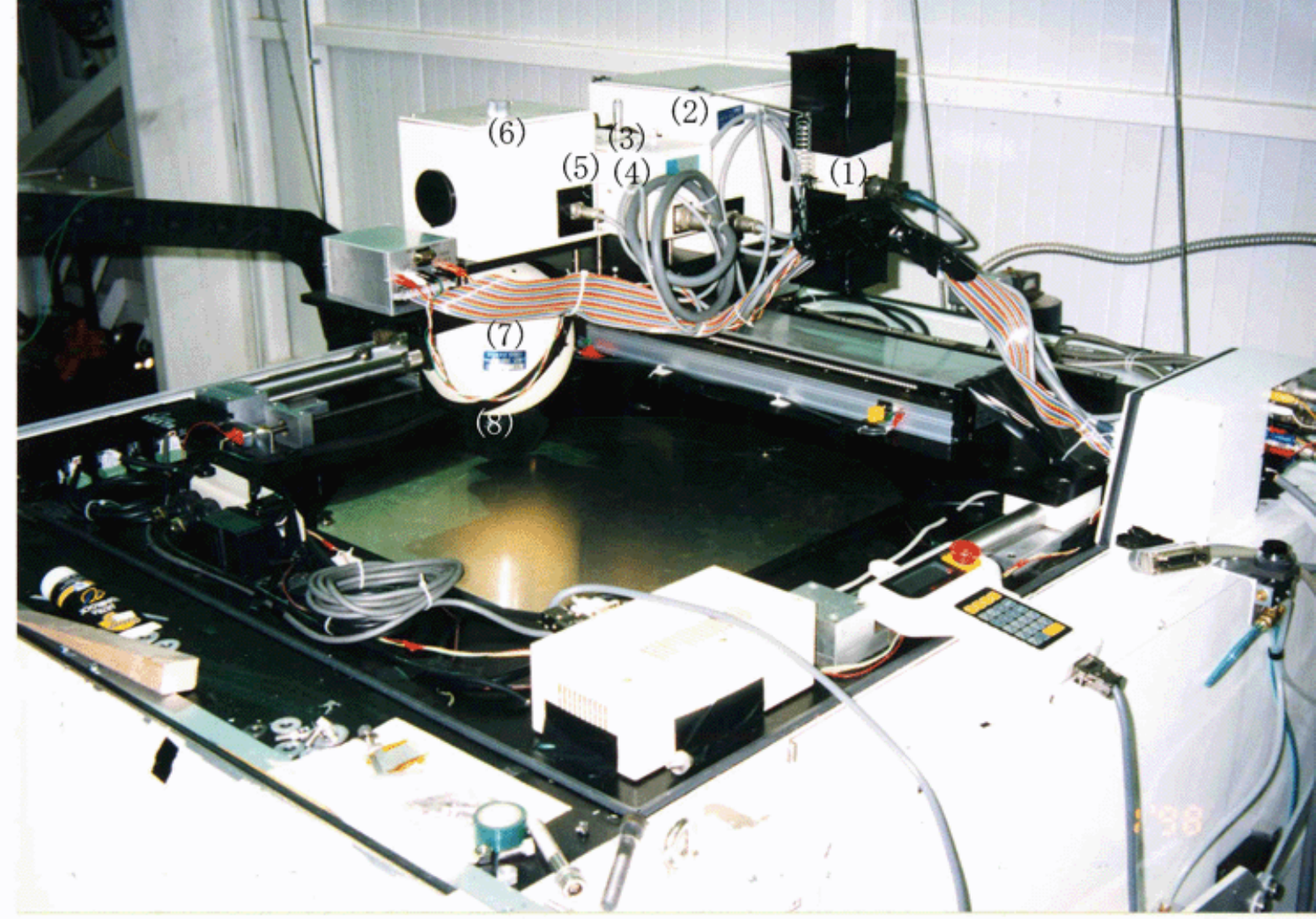}
\caption{Monochromatic illumination system overview. (1) lamp; (2)
light condensation mirror system; (3)  input slit;  
(4) monochrometer;  (5)  shutter;
(6) mirror switch to change the output direction;
(7) integration sphere;
(8) photodiode.
}
\label{fig:MSP}
\end{center}
\end{figure}

\begin{figure}[pt]
\begin{center} 
\includegraphics[height=10cm,bb=0 0 574 574]{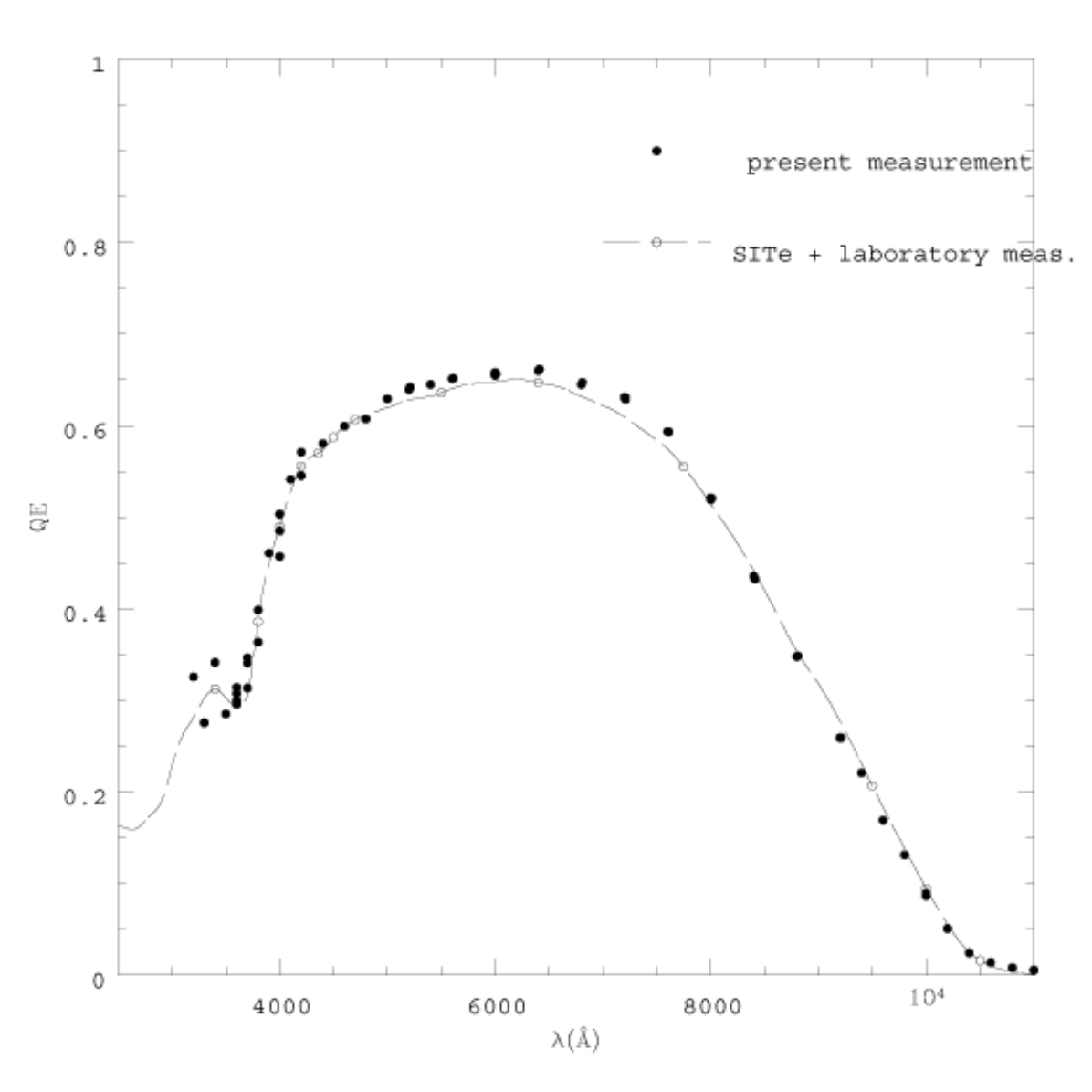}
\caption{Response function (quantum efficiency) of the ultraviolet-enhancing 
coated thinned back-illuminated CCD used at the Photometric
Telescope. The solid points are the measurement with the present
system, and the 
dashed curve is the quantum efficiency for the same CCD measured
at SITe at a room temperature, but tilted using a laboratory measurement
(open points) at a  cooled, operating  temperature, used in F96.
}
\label{fig:MTQE}
\end{center}
\end{figure}

\clearpage
\begin{figure}[pt]
\begin{center} 
\includegraphics[width=10cm,bb=100 0 565 354]{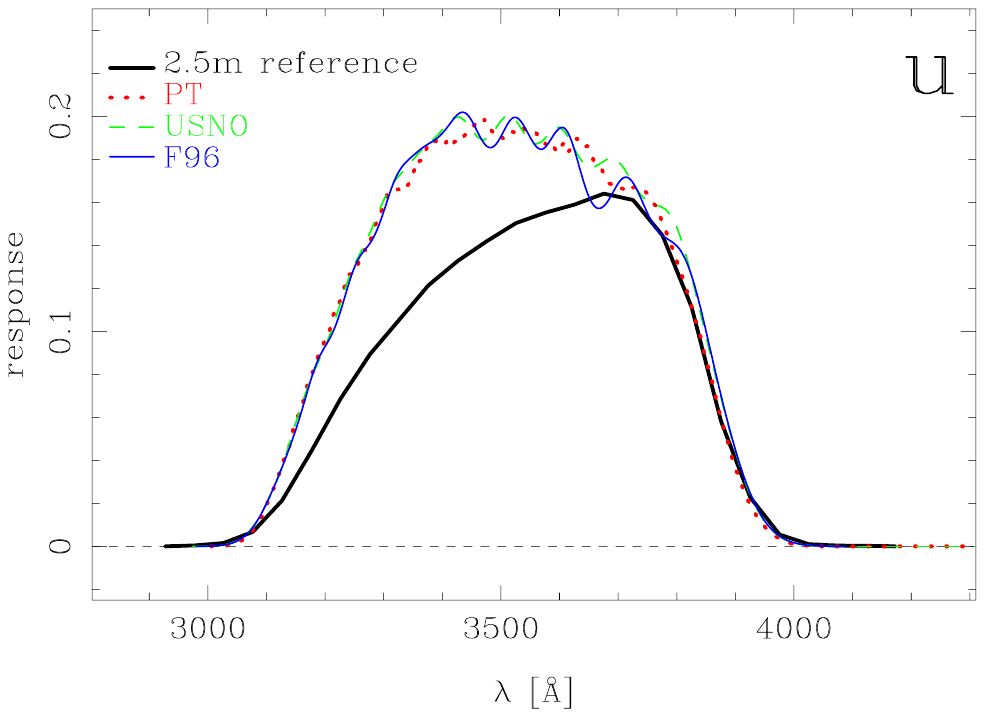}
\caption{Response functions for the $u$ band.
The thick solid curve is the reference response function for the 2.5 m
telescope imager defined in the text, 
thin solid curve is the F96 standard, the dashed curve
is for the USNO 1m telescope system, and the dotted curve is for the SDSS
Photometric Telescope (after the replacement of filters: see text)
}
\label{fig:u_refs}
\end{center}
\end{figure}

\begin{figure}[pt]
\begin{center} 
\includegraphics[width=10cm,bb=100 0 565 336]{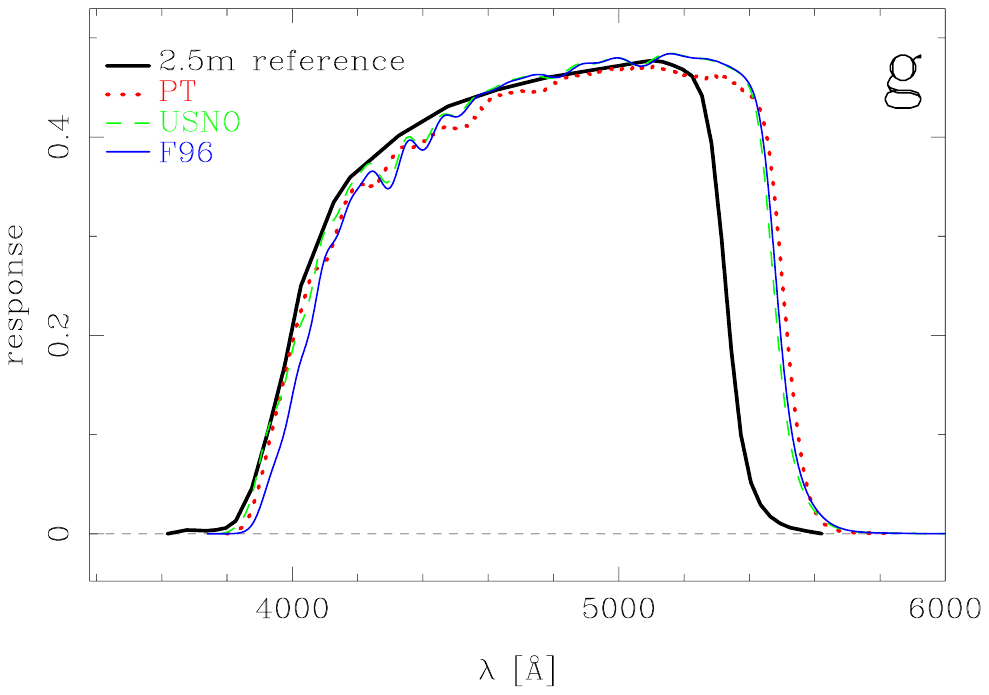}
\caption{Same as Figure 2, but for the $g$ band.
}
\label{fig:g_refs}
\end{center}
\end{figure}

\begin{figure}[pt]
\begin{center} 
\includegraphics[width=10cm,bb=100 0 565 360]{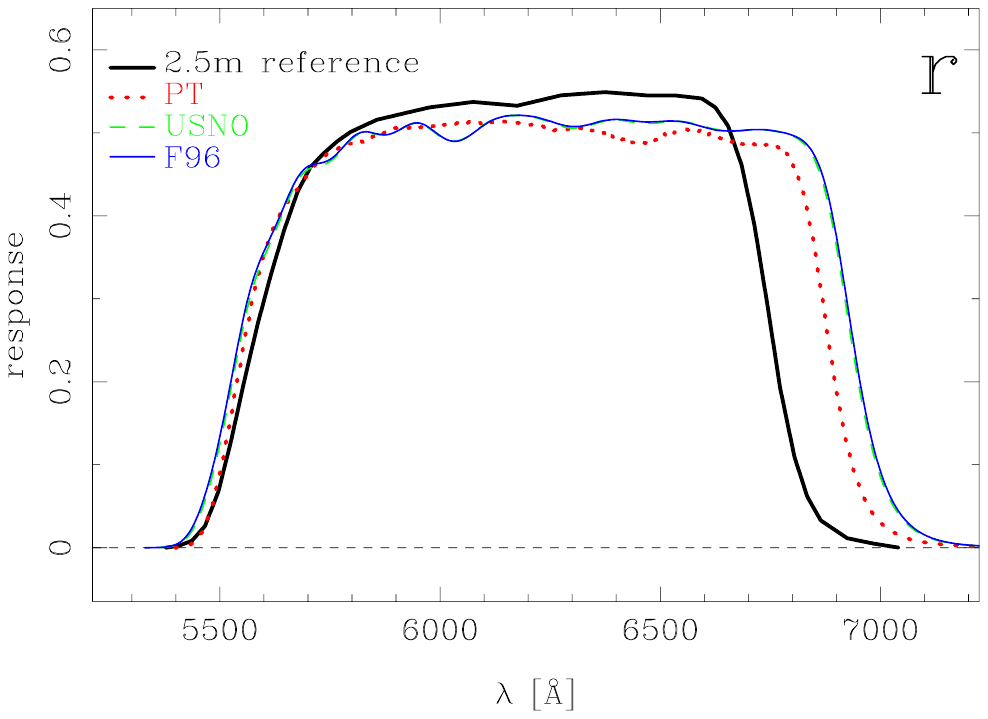}
\caption{Same as Figure 2, but for the $r$ band.
}
\label{fig:r_refs}
\end{center}
\end{figure}

\begin{figure}[pt]
\begin{center} 
\includegraphics[width=10cm,bb=100 0 579 348]{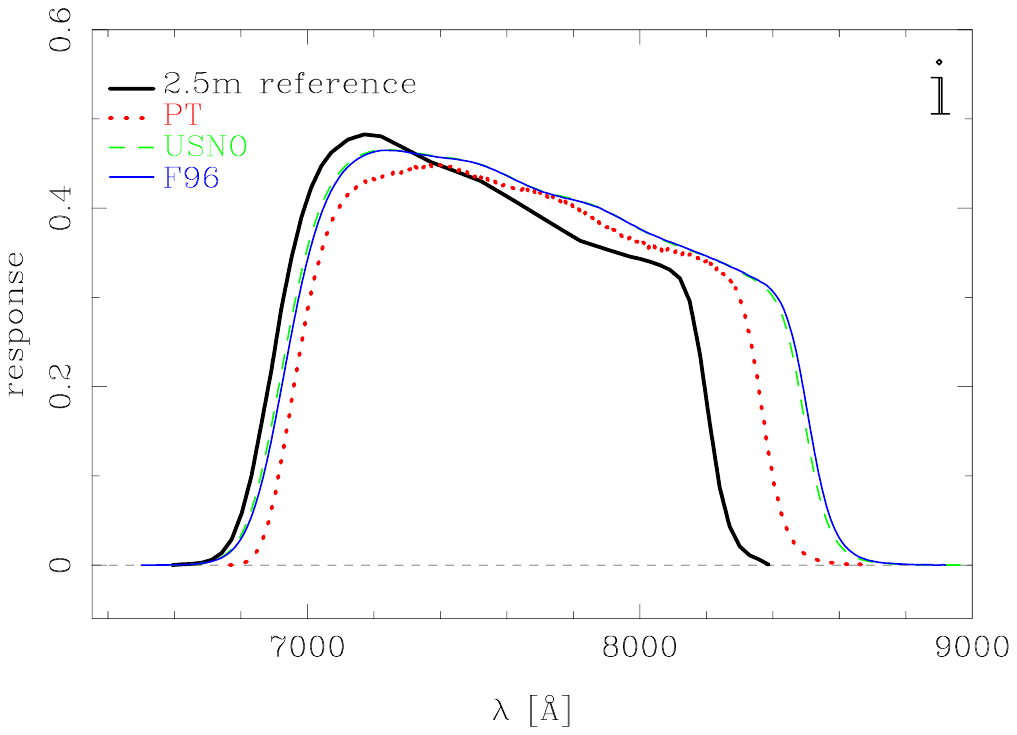} 
\caption{Same as Figure 2, but for the $i$ band.
}
\label{fig:i_refs}
\end{center}
\end{figure}

\begin{figure}[pt]
\begin{center} 
\includegraphics[width=10cm,bb=100 0 579 369]{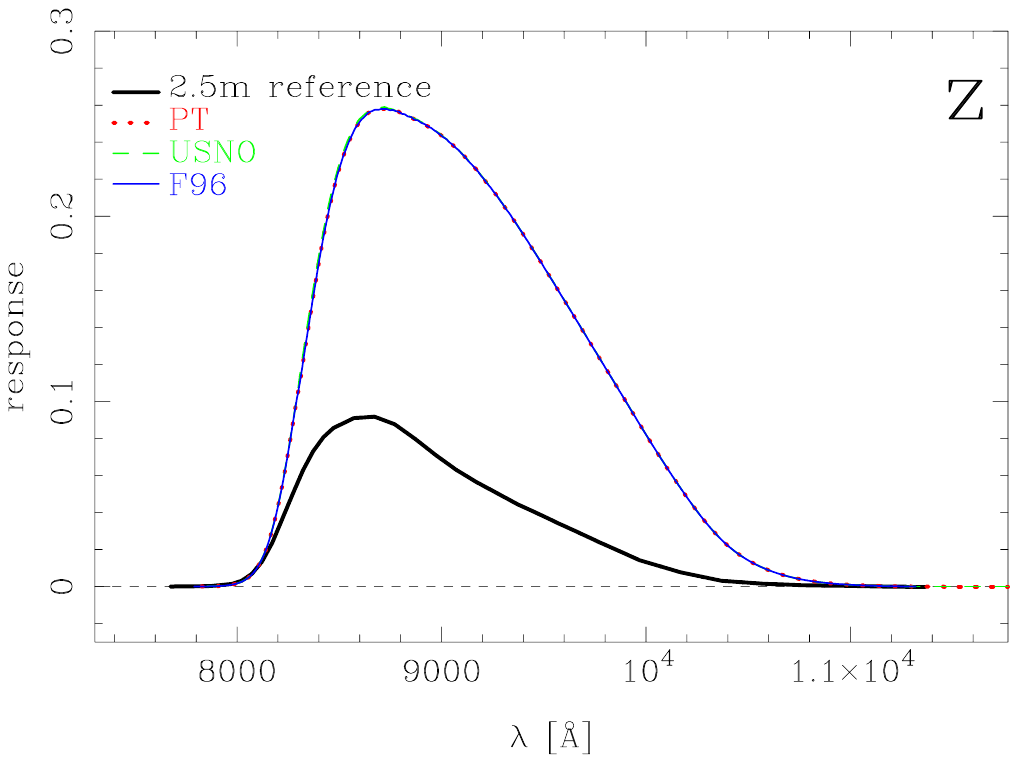}
\caption{Same as Figure 2, but for the $z$ band.}
\label{fig:z_refs}
\end{center}
\end{figure}

\clearpage
\begin{subfigures}
\begin{figure}[pt]
\begin{center} 
\includegraphics[width=10cm,bb=100 0 565 363]{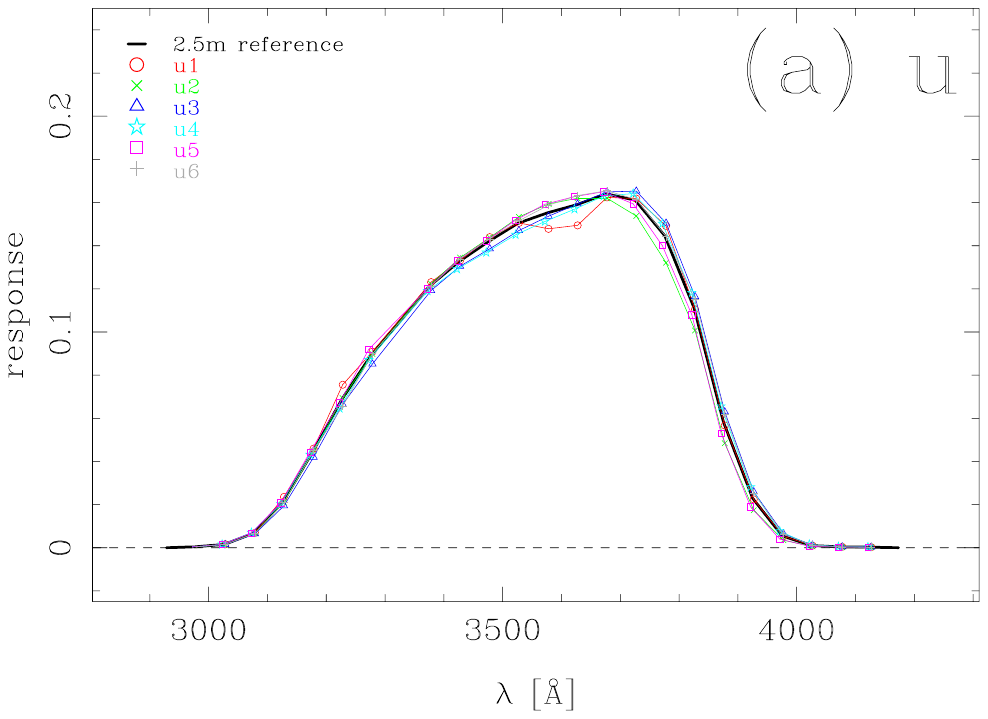}
\caption{Response function for each column of the detector 
in the main camera.
(a) $u$ band, (b) $g$ band, (c) $r$ band, (d) $i$ band,
(e) $z$ band.
}
\label{fig:uc}
\end{center}
\end{figure}

\begin{figure}[pt]
\begin{center} 
\includegraphics[width=10cm,bb=100 0 565 363]{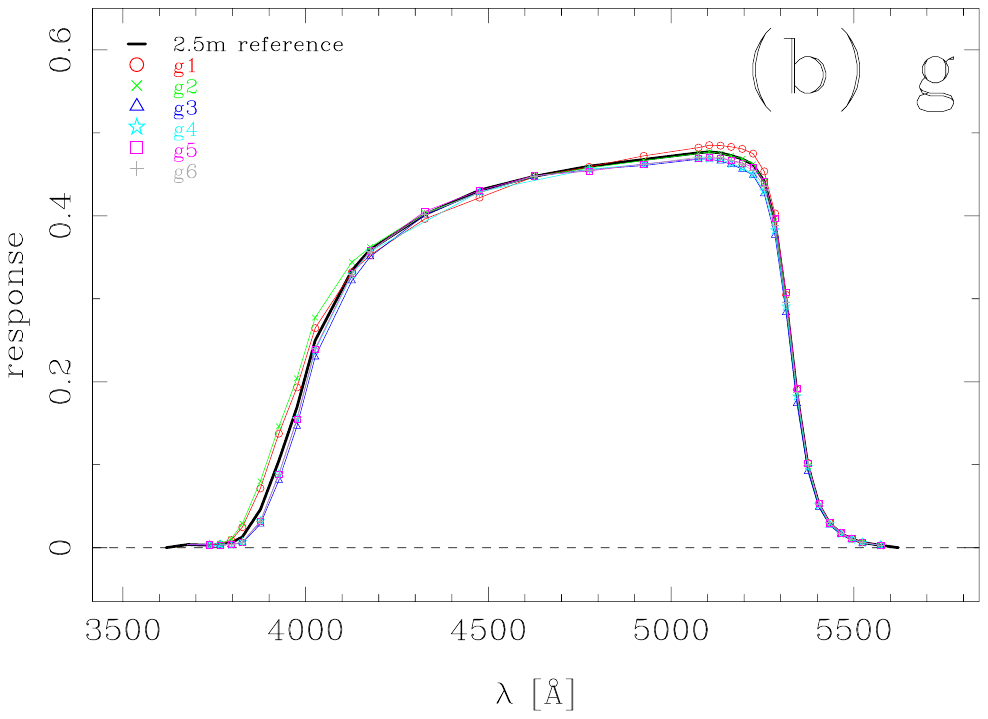}
\caption{}
\label{fig:gc}
\end{center}
\end{figure}

\begin{figure}[pt]
\begin{center} 
\includegraphics[width=10cm,bb=100 0 565 363]{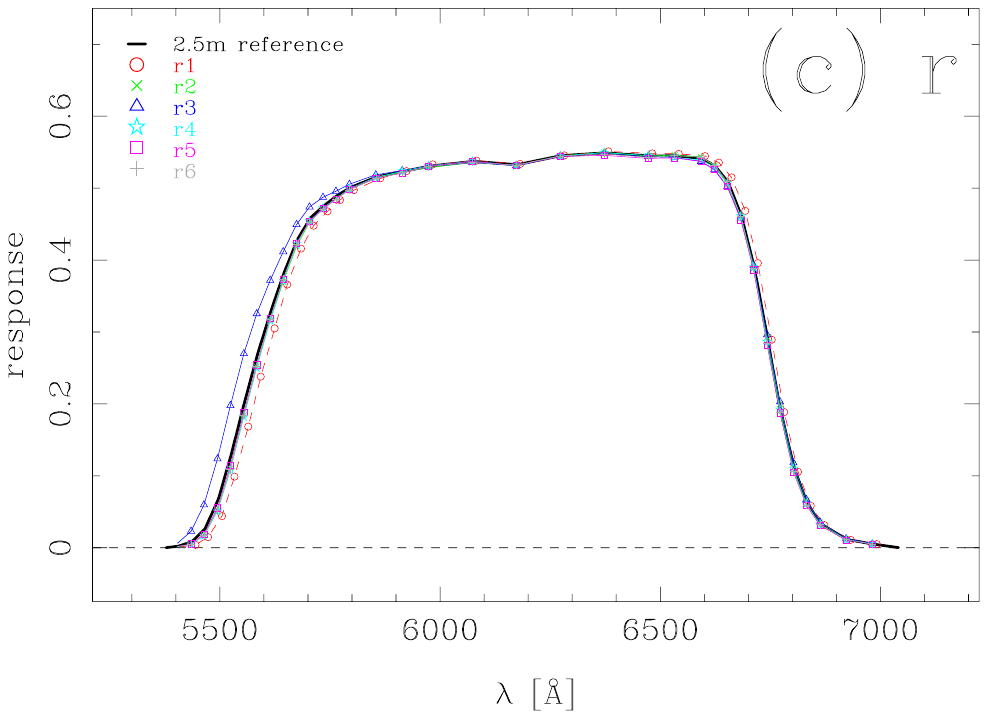}
\caption{}
\label{fig:rc}
\end{center}
\end{figure}

\begin{figure}[pt]
\begin{center} 
\includegraphics[width=10cm,bb=100 0 565 363]{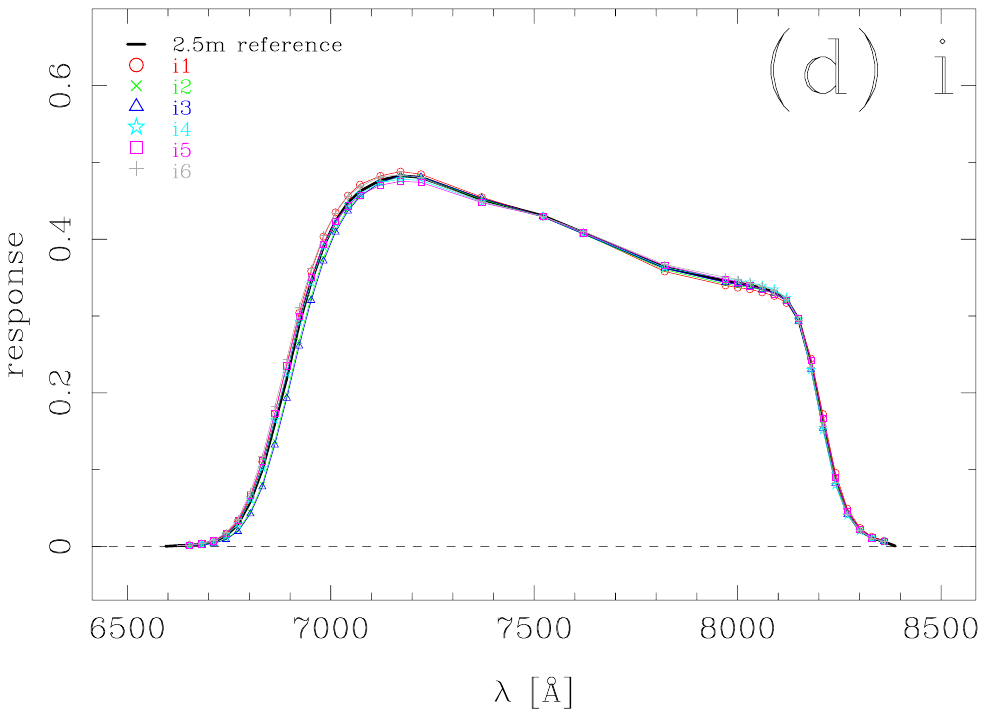}
\caption{}
\label{fig:ic}
\end{center}
\end{figure}

\begin{figure}[pt]
\begin{center} 
\includegraphics[width=10cm,bb=100 0 565 363]{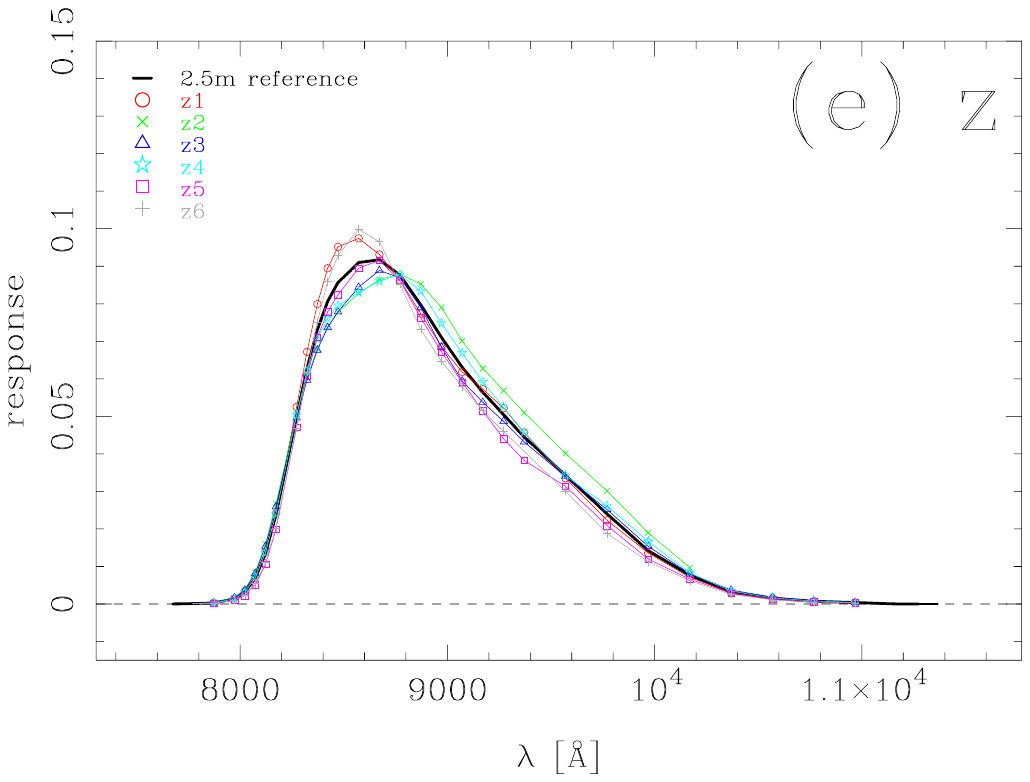}
\caption{}
\label{fig:zc}
\end{center}
\end{figure}
\end{subfigures}

\clearpage
\begin{subfigures}
\begin{figure}[pt]
\begin{center} 
\includegraphics[width=10cm,bb=100 0 579 383]{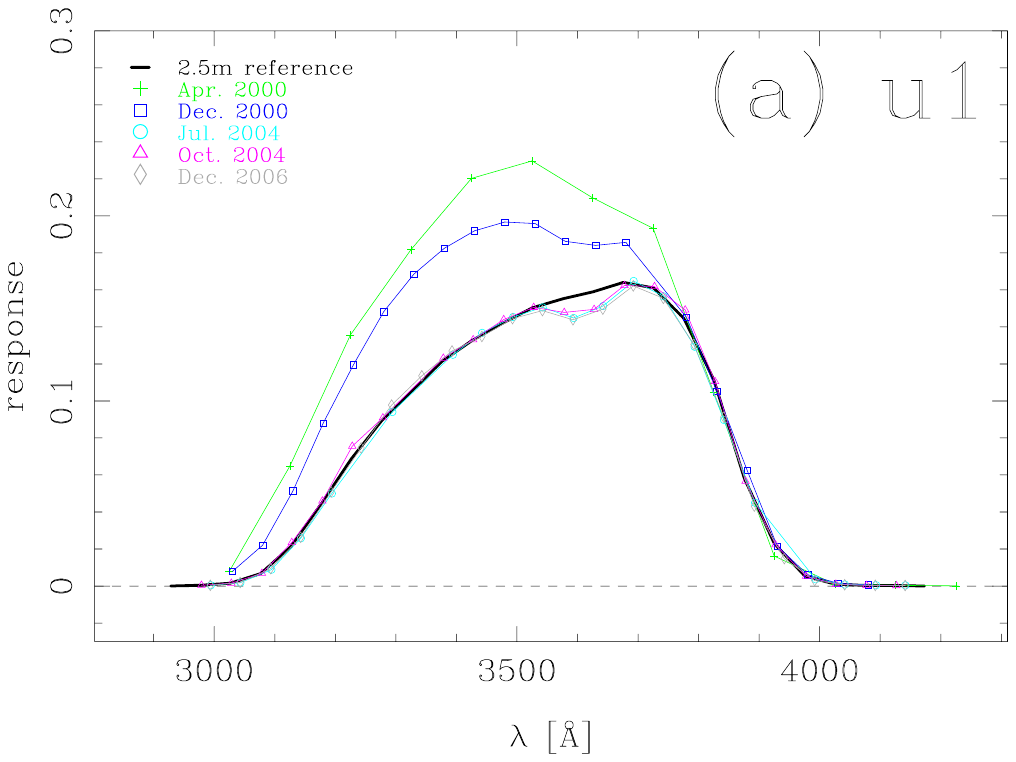}
\caption{Response functions at various epochs of measurements for 
the imager in the main camera.
 (a) $u$ band, Camcol 1, (b) $g$ band, Camcol 1, (c) $r$ band Camcol 1, 
(d) $i$ band, Camcol 1,  (e) $z$ band Camcol 1, and (f) red leak
for the $u$ band, Camcol 1.
}
\label{fig:qe_u1}
\end{center}
\end{figure}

\begin{figure}[pt]
\begin{center} 
\includegraphics[width=10cm,bb=100 0 579 383]{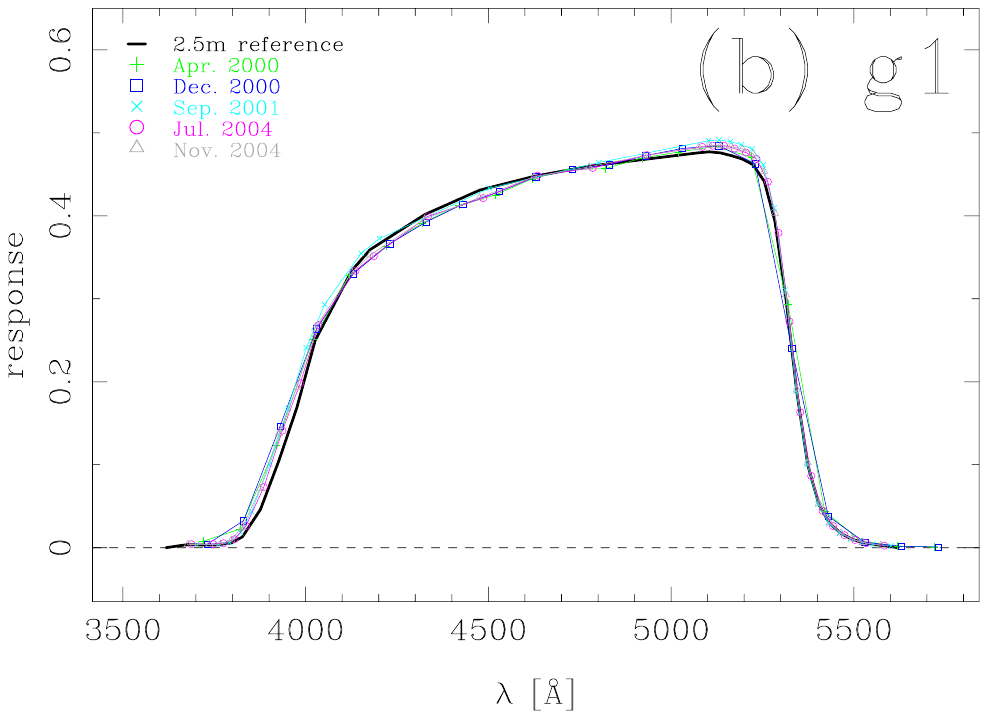}
\caption{}
\label{fig:qe_g1}
\end{center}
\end{figure}

\begin{figure}[pt]
\begin{center} 
\includegraphics[width=10cm,bb=100 0 579 383]{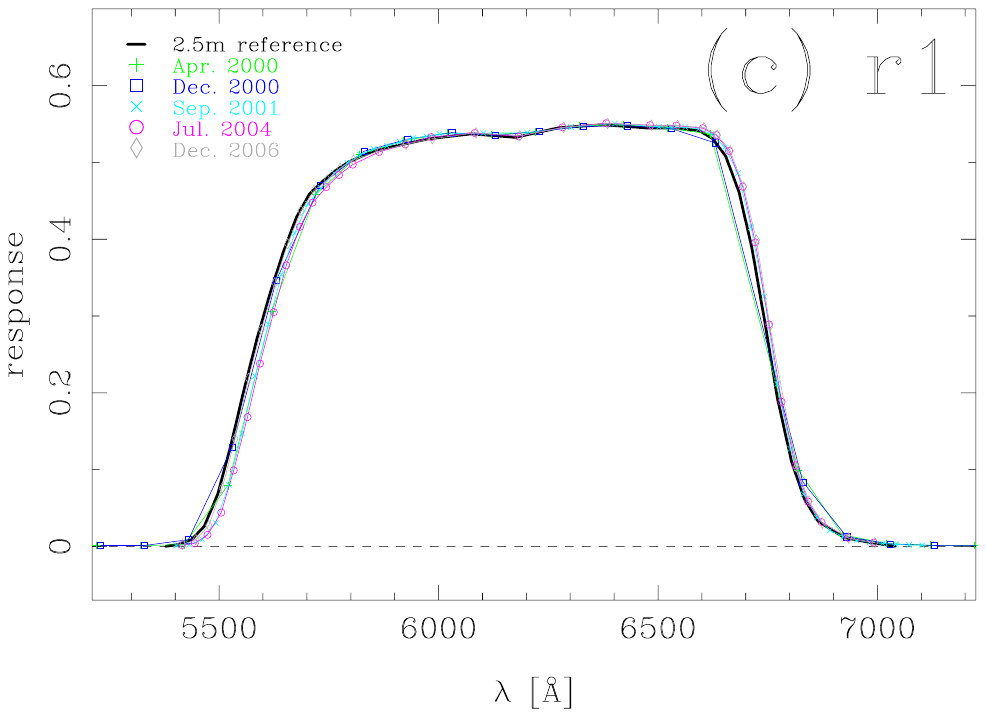}
\caption{}
\label{fig:qe_r1}
\end{center}
\end{figure}

\begin{figure}[pt]
\begin{center} 
\includegraphics[width=10cm,bb=100 0 579 383]{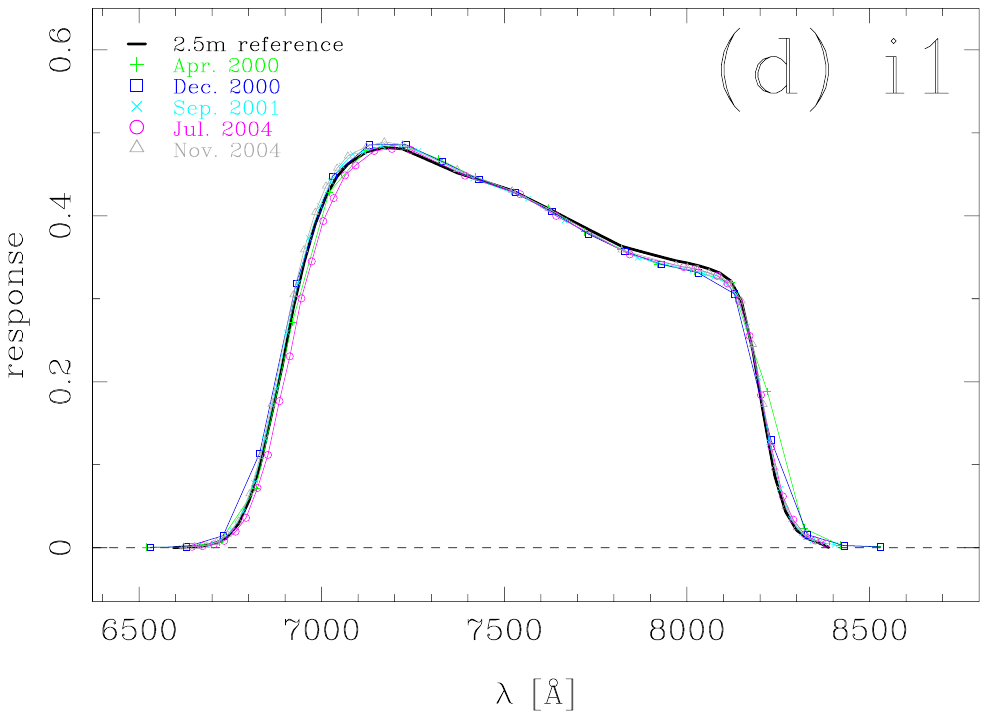}  
\caption{}
\label{fig:qe_i1}
\end{center}
\end{figure}

\begin{figure}[pt]
\begin{center} 
\includegraphics[width=10cm,bb=100 0 579 383]{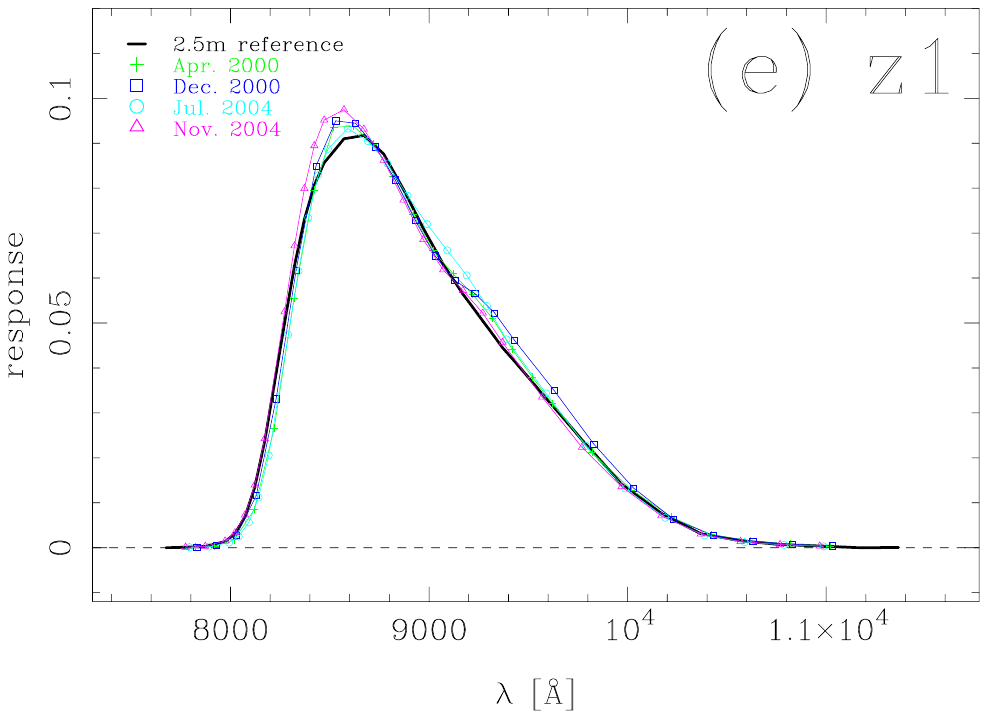} 
\caption{}
\label{fig:qe_z1}
\end{center}
\end{figure}

\begin{figure}[pt]
\begin{center} 
\includegraphics[width=10cm,bb=100 0 579 383]{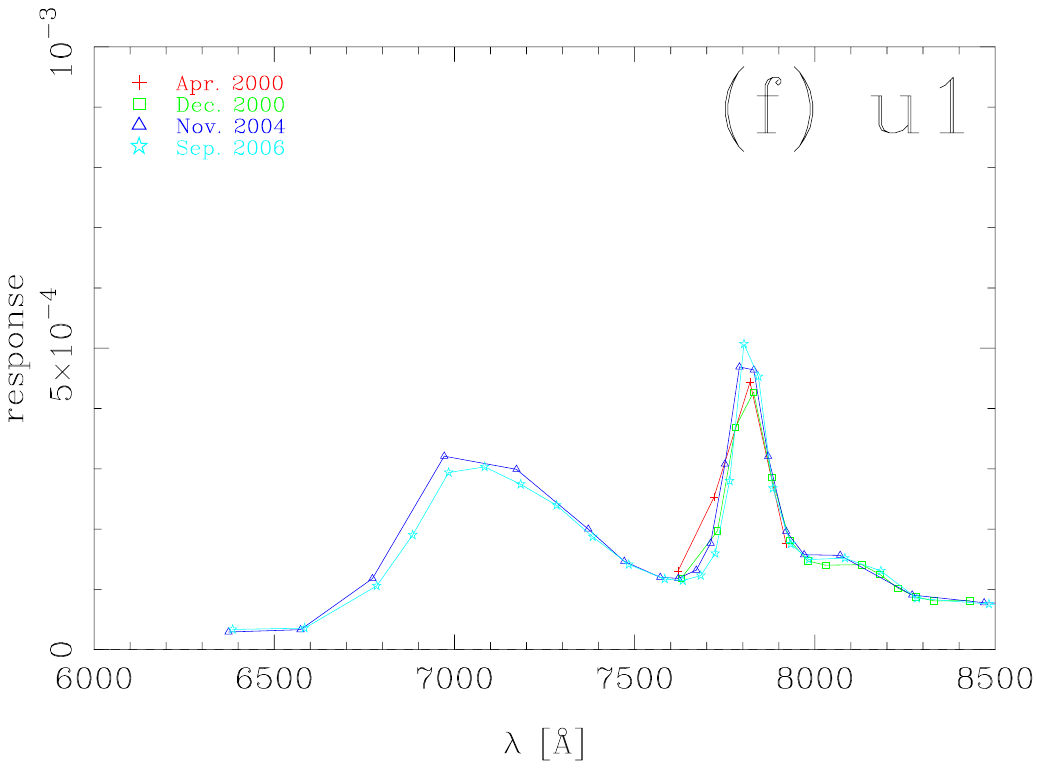}
\caption{}
\label{fig:qe_uleak}
\end{center}
\end{figure}
\end{subfigures}

\clearpage
\begin{subfigures}
\begin{figure}[pt]
\begin{center} 
\includegraphics[height=10cm,bb=0 0 458 396]{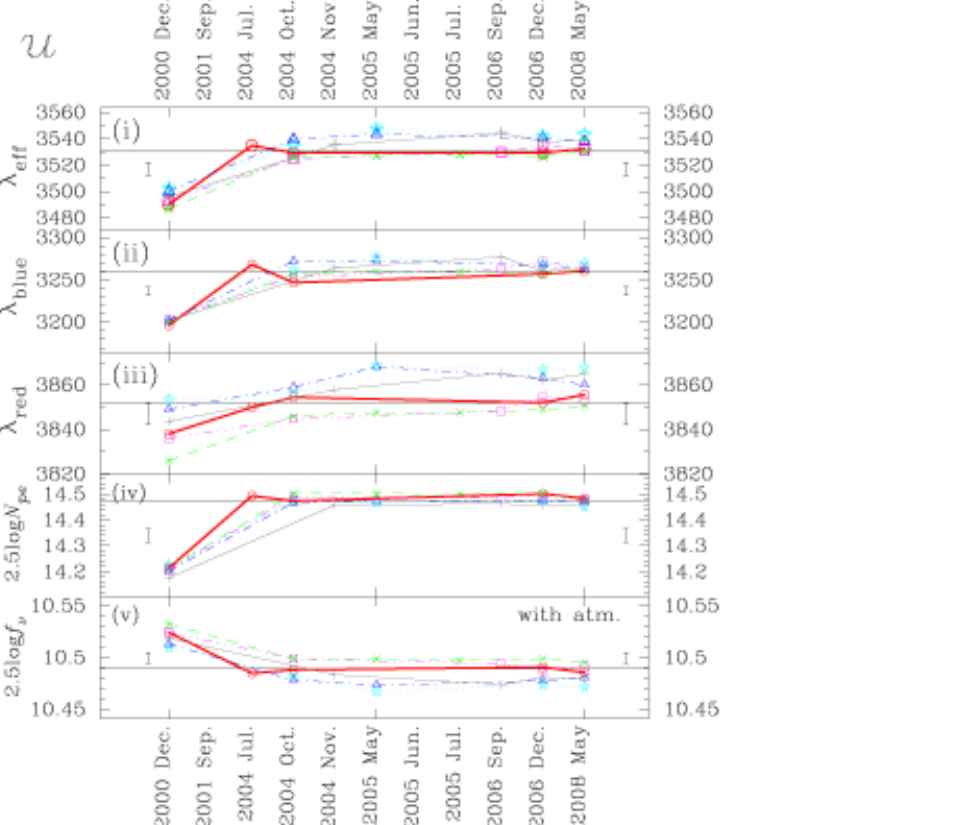}
\caption{Time variations of the response functions and their effects on
photometry. From the top to the bottom:
(i) the effective wavelength, $\lambda_{\rm eff}$,
(ii) 50\%-response wavelengths of the blue edges, (iii) 50\%-response 
wavelengths of the red edges,
(iv) the detector sensitivity  i.e, the quantity proportional to
$-2.5\log N_{p.e.}$ (arbitrary unit) where $N_{p.e.}$
is the number of photoelectrons in the detector, or brightness 
defined by eq. (3)
but with the denominator set equal to unity. The star is BD+17$^\circ$4708; 
(v) brightness of BD+17$^\circ$4708 from eq. (3). The response function
includes the effect of atmosphere at 1.3 airmass.
The symbols are the same as in Fig. 8 (circle, cross, triangle,
star, square, and plus present column 1-6, respectively).
The horizontal line in each panel is for the reference response.
In the panels (i), (ii), and (iii), the error bars show 10$\AA$,
and in (iv) and (v), they show 0.05 mag and 0.01 mag, respectively.
}
\label{fig:u_evol}
\end{center}
\end{figure}

\begin{figure}[pt]
\begin{center} 
\includegraphics[height=10cm,bb=0 0 458 396]{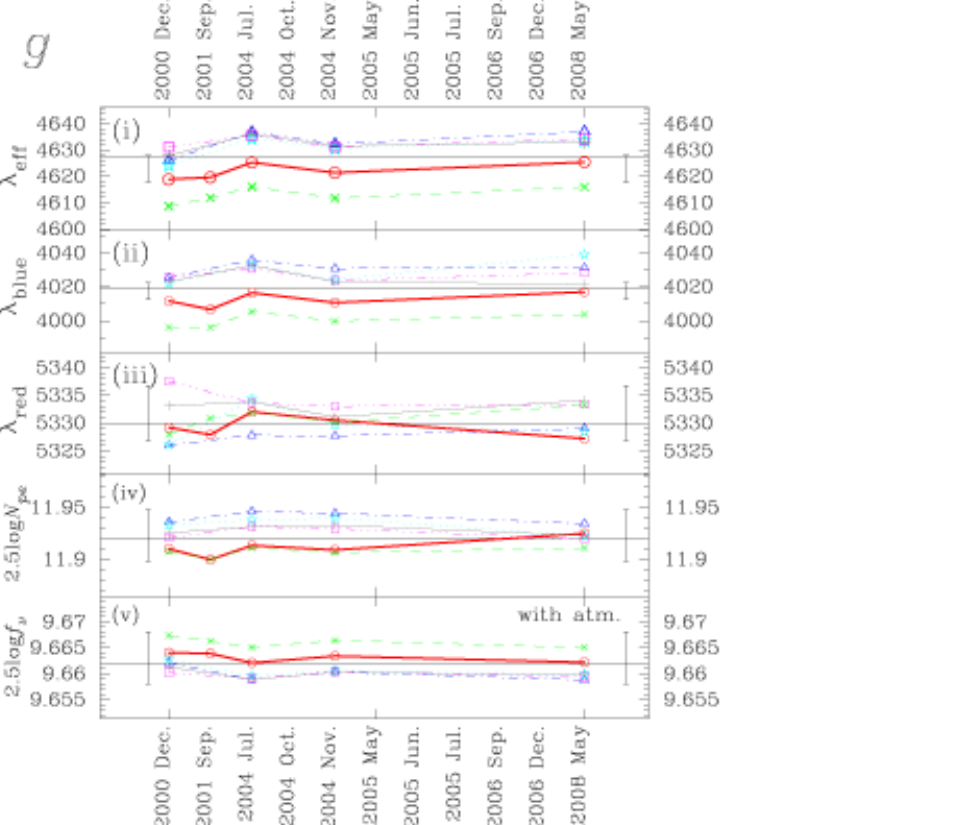}
\caption{}
\label{fig:g_evol}
\end{center}
\end{figure}

\begin{figure}[pt]
\begin{center} 
\includegraphics[height=10cm,bb=0 0 458 396]{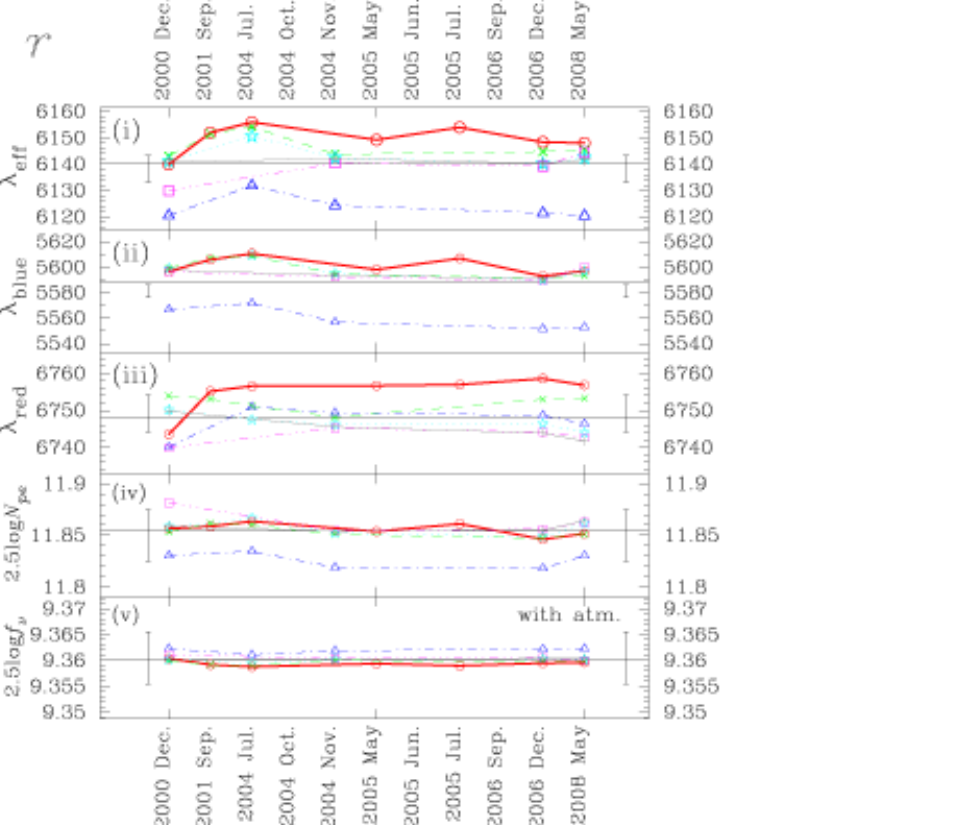}
\caption{}
\label{fig:r_evol}
\end{center}
\end{figure}

\begin{figure}[pt]
\begin{center} 
\includegraphics[height=10cm,bb=0 0 458 396]{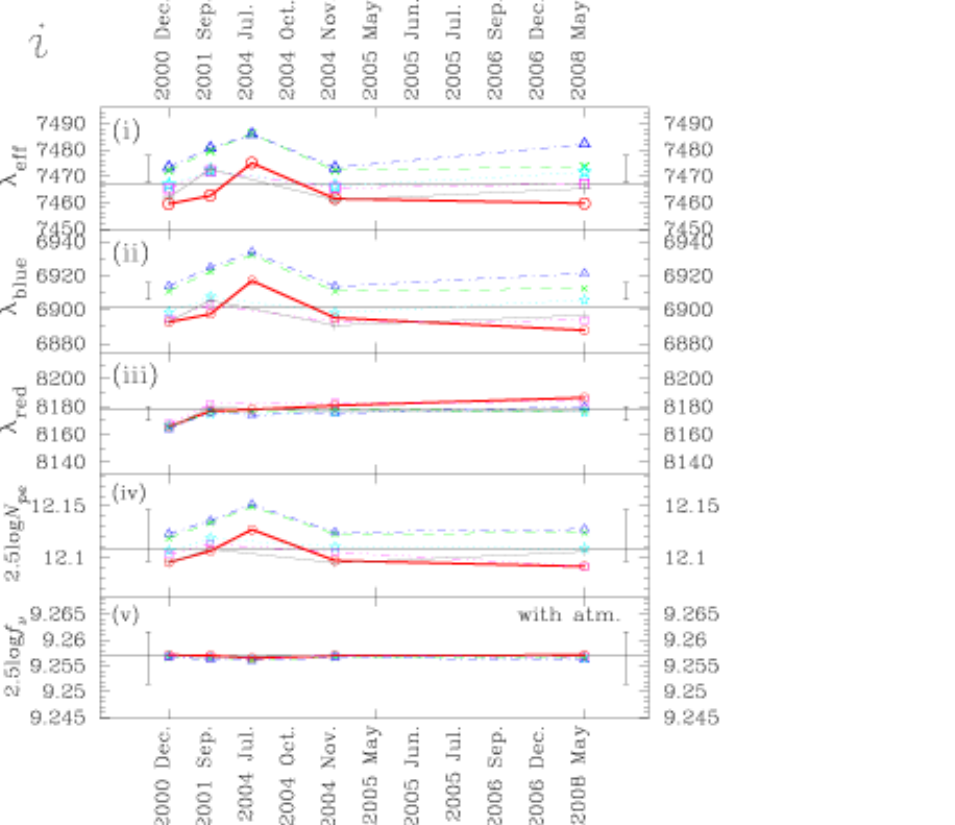}  
\caption{}
\label{fig:i_evol}
\end{center}
\end{figure}

\begin{figure}[pt]
\begin{center} 
\includegraphics[height=10cm,bb=0 0 458 396]{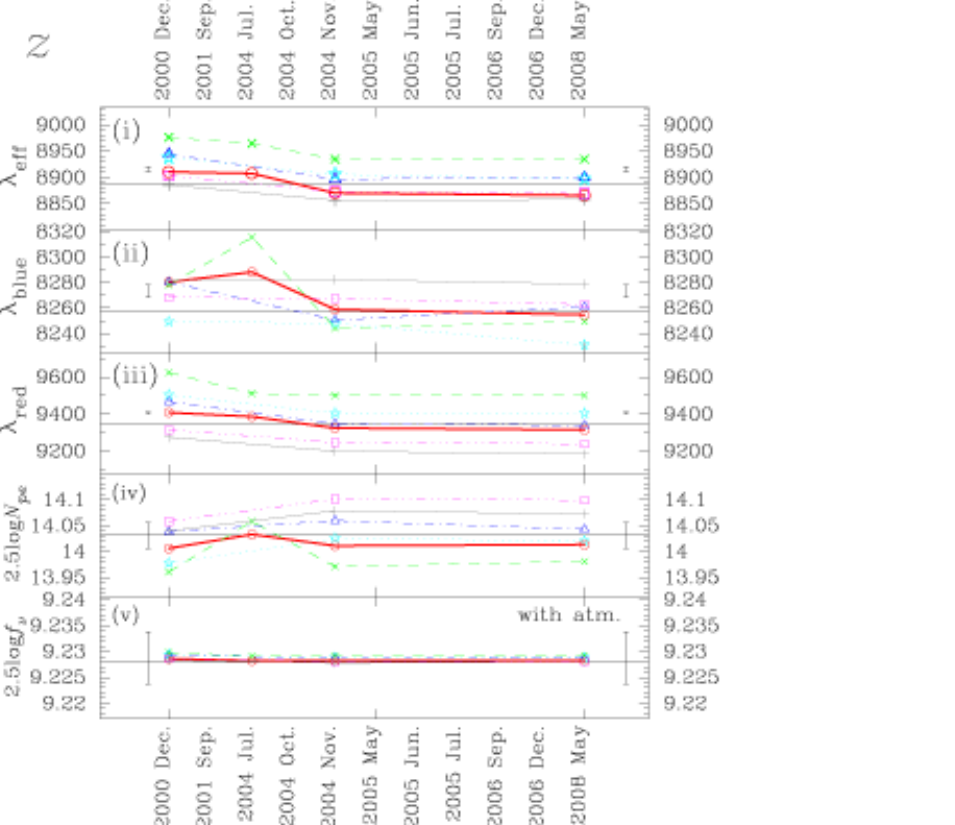} 
\caption{}
\label{fig:z_evol}
\end{center}
\end{figure}
\end{subfigures}

\begin{figure}[pt]
\begin{center} 
\includegraphics[height=10cm,bb=50 0 553 411]{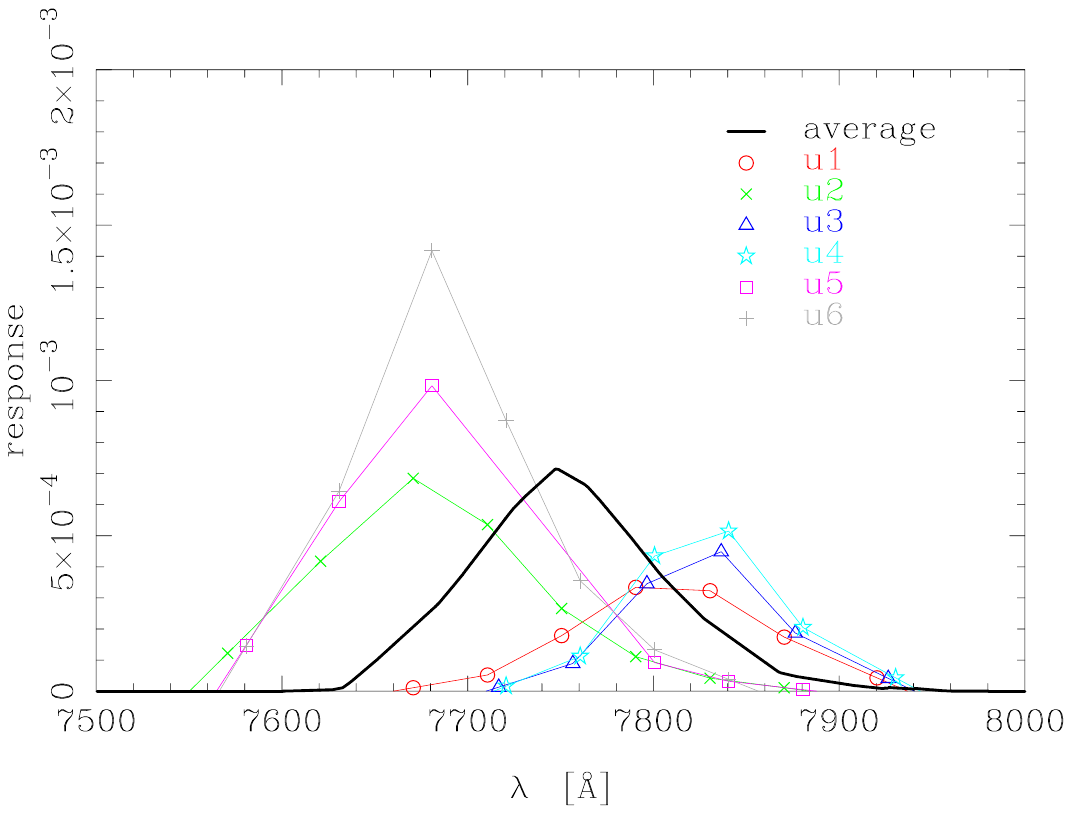}
\caption{The u-band redleak response after subtraction of the
possible scattered light. The thick line shows the average
of all six columns.}
\label{fig:qe_uleak}
\end{center}
\end{figure}

\clearpage
\begin{subfigures}
\begin{figure}[pt]
\begin{center} 
\includegraphics[height=8cm,bb=100 0 512 393]{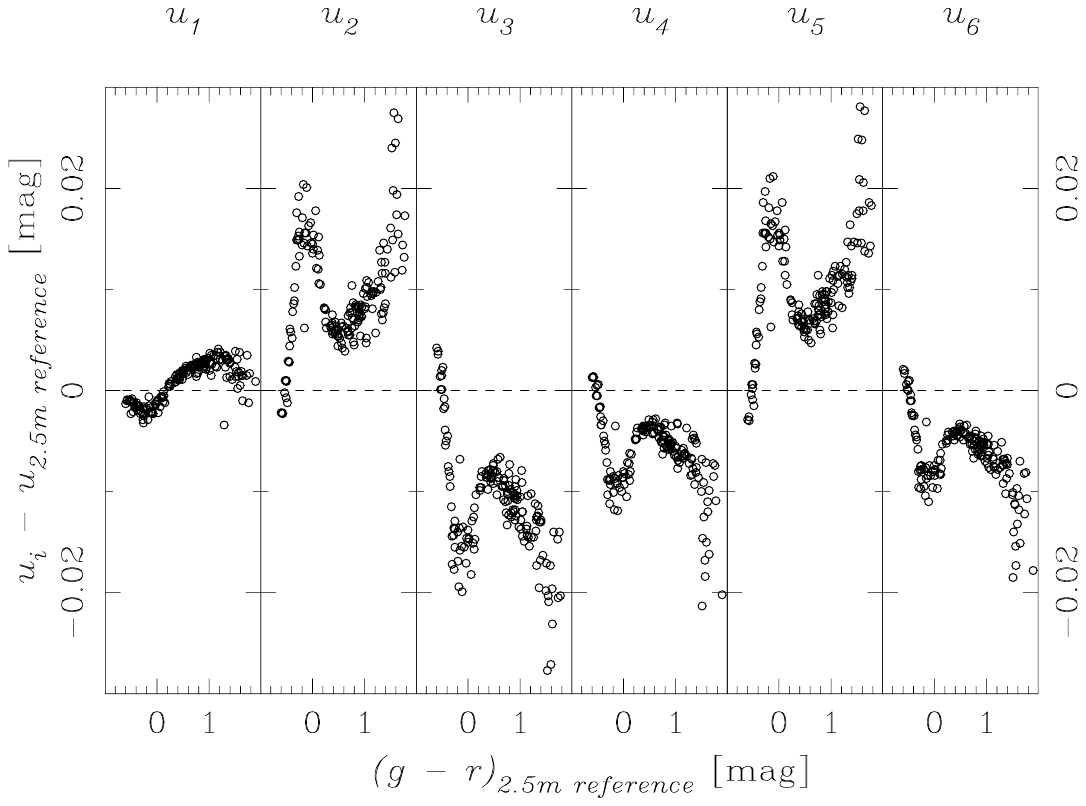}
\caption{Column to column colour variations expected in the $u$ band as
a function of $g-r$ colour of stars. The response function involves
atmospheric extinction at 1.3 airmass.
}
\label{fig:u_column_vs_ref}
\end{center}
\end{figure}

\begin{figure}[pt]
\begin{center} 
\includegraphics[height=8cm,bb=100 0 512 393]{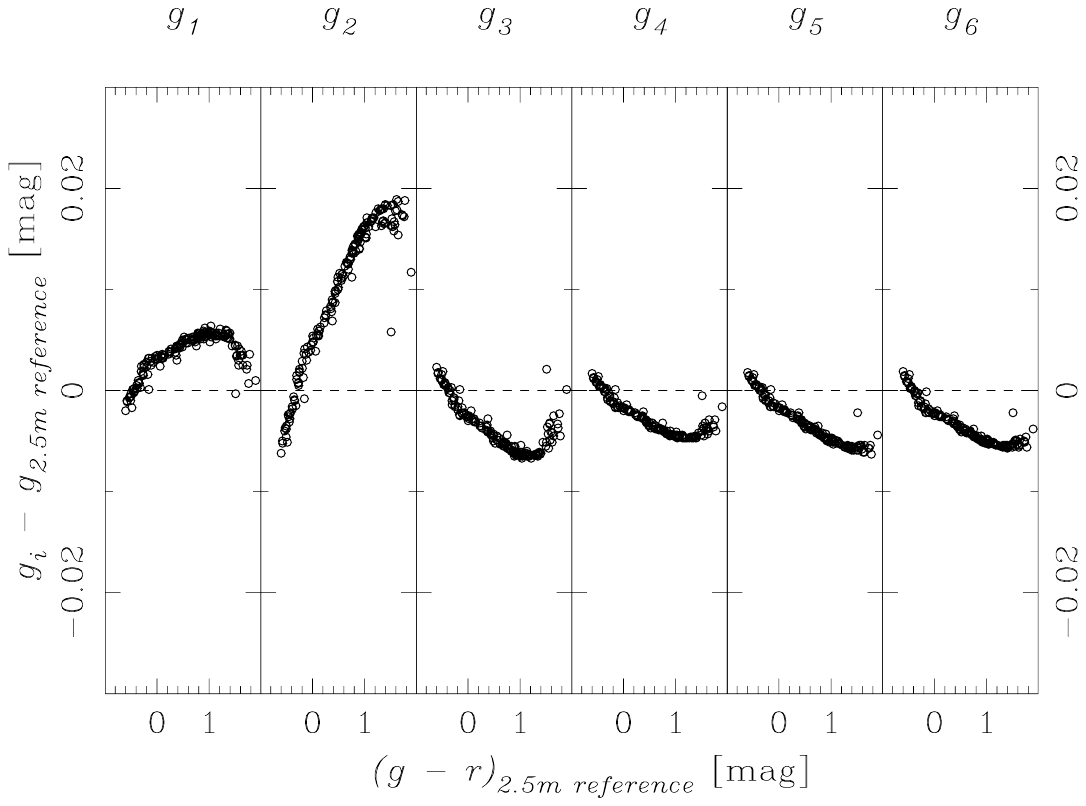}
\caption{The same as Fig.11(a) but in the $g$ band.}
\label{fig:g_column_vs_ref}
\end{center}
\end{figure}

\begin{figure}[pt]
\begin{center} 
\includegraphics[height=8cm,bb=100 0 512 393]{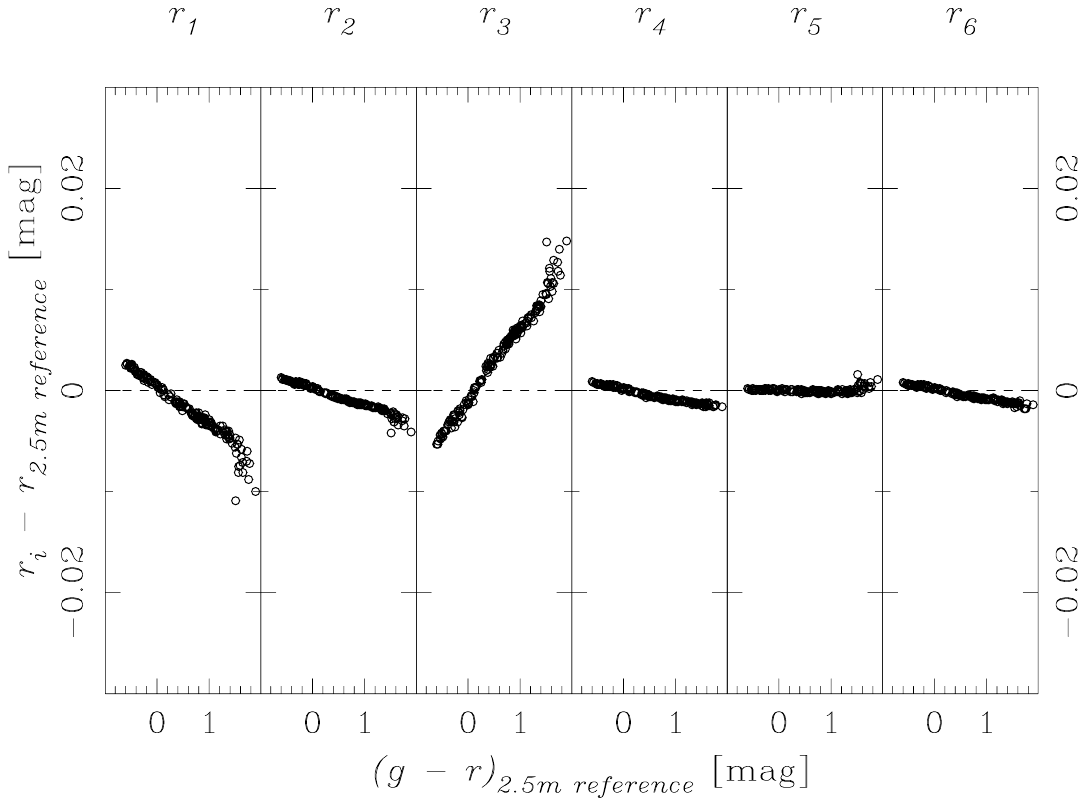} 
\caption{The same as Fig.11(a) but in the $r$ band.}
\label{fig:r_column_vs_ref}
\end{center}
\end{figure}

\begin{figure}[pt]
\begin{center} 
\includegraphics[height=8cm,bb=100 0 512 393]{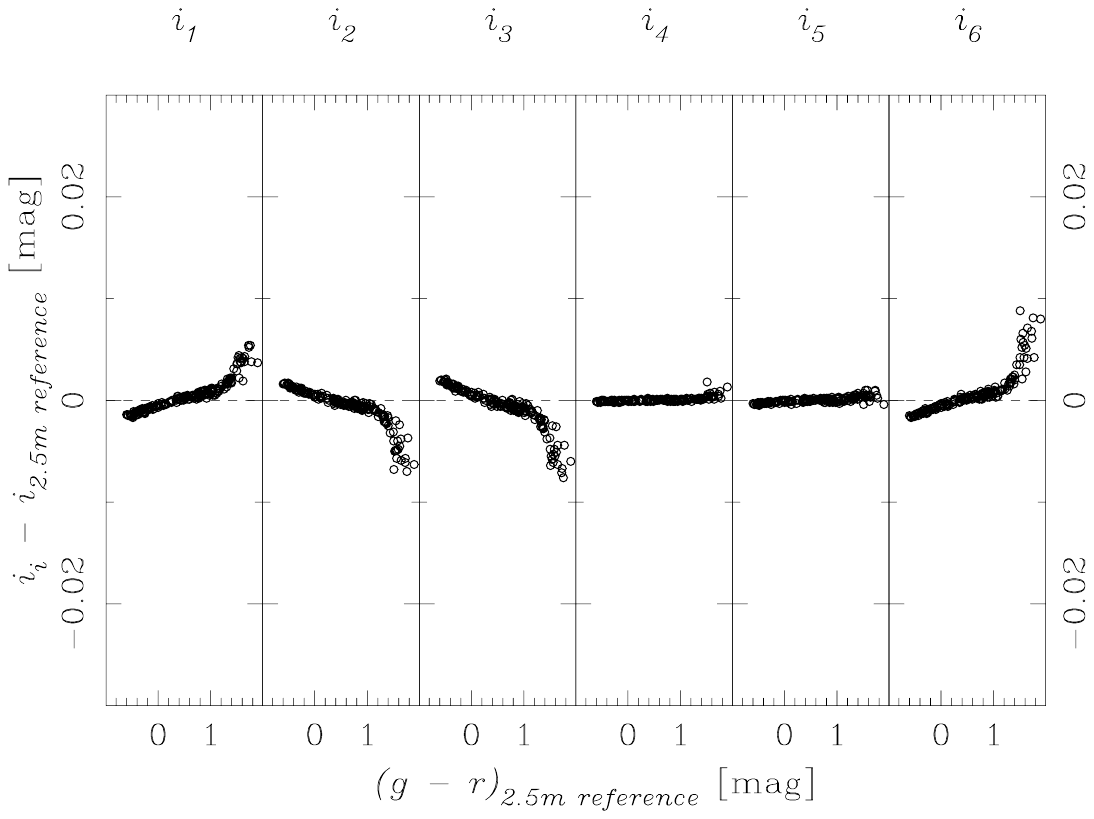} 
\caption{The same as Fig.11(a) but in the $i$ band.}
\label{fig:i_column_vs_ref}
\end{center}
\end{figure}

\begin{figure}[pt]
\begin{center} 
\includegraphics[height=8cm,bb=100 0 512 393]{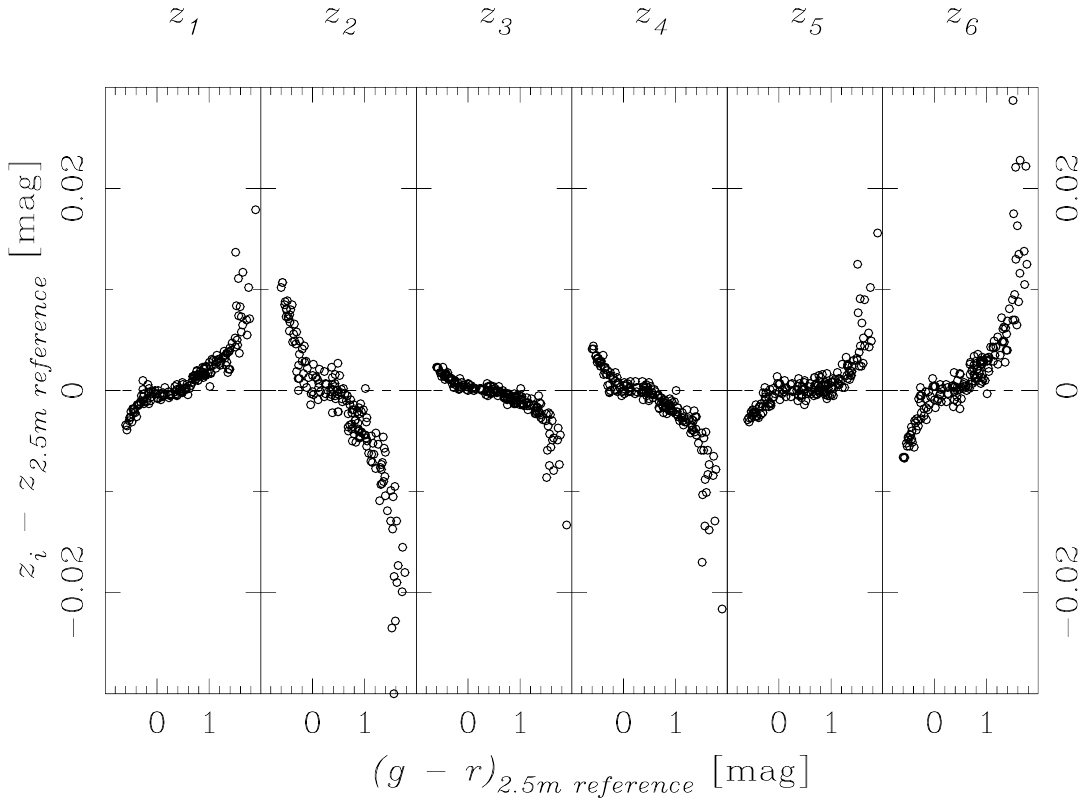} 
\caption{The same as Fig.11(a) but in the $z$ band.}
\label{fig:z_column_vs_ref}
\end{center}
\end{figure}
\end{subfigures}

\clearpage
\begin{subfigures}
\begin{figure}[pt]
\begin{center} 
\includegraphics[height=10cm,bb=0 0 483 393]{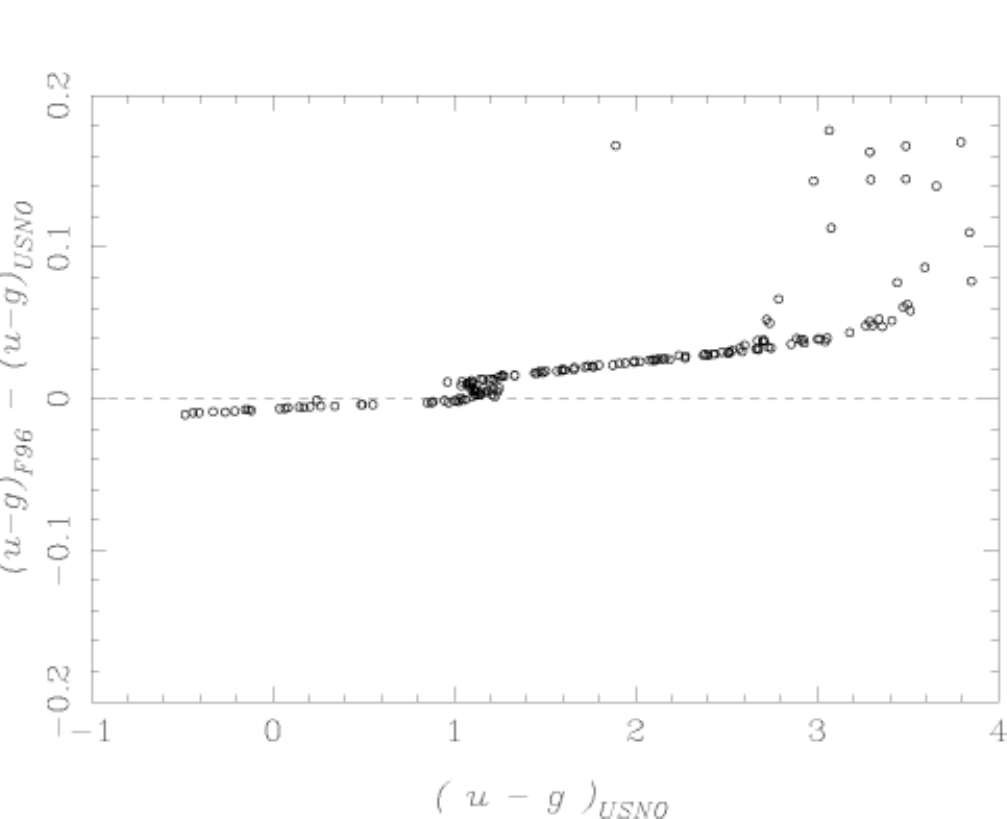} 
\caption{Expected offset of $u-g$ colour between the USNO photometric
system and the original F96 standard,
$\Delta (u-g)_{\rm AB}=(u-g)_{\rm USNO}-(u-g)_{\rm F96 standard}$
plotted as a function of $u-g$. The response function involves 
atmospheric extinction.
}
\label{fig:usno_f96_ug}
\end{center}
\end{figure}

\begin{figure}[pt]
\begin{center} 
\includegraphics[height=10cm,bb=0 0 483 393]{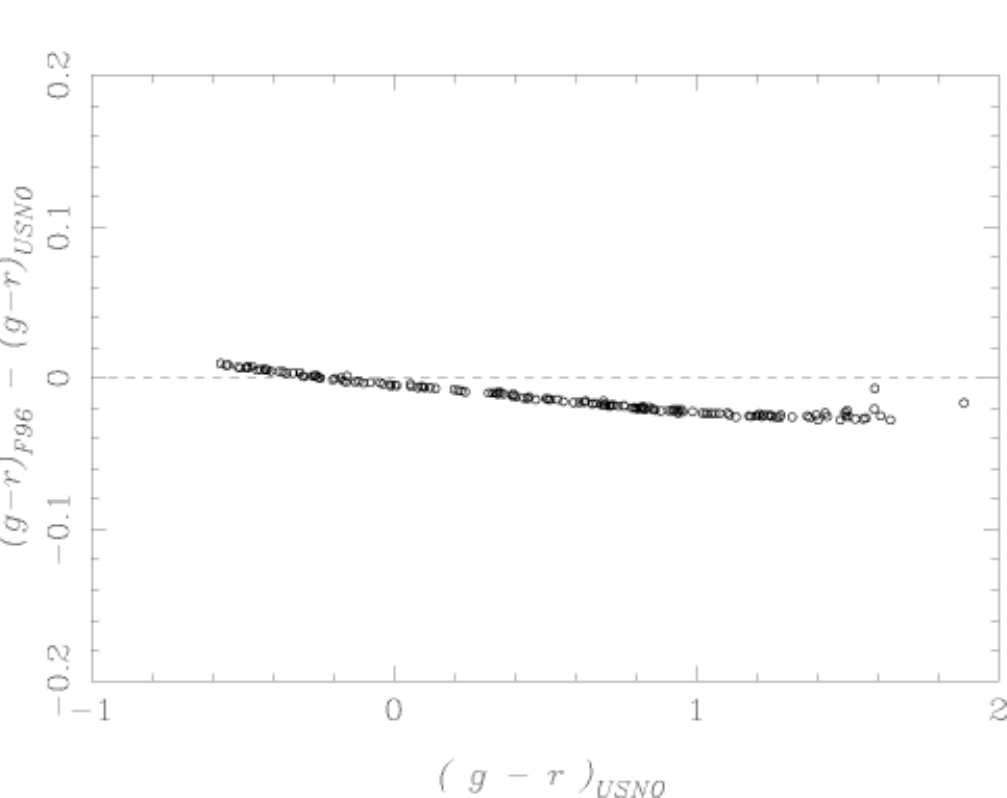} 
\caption{The same as Fig.12(a) but for $g-r$.}
\label{fig:usno_f96_gr}
\end{center}
\end{figure}

\begin{figure}[pt]
\begin{center} 
\includegraphics[height=10cm,bb=0 0 483 393]{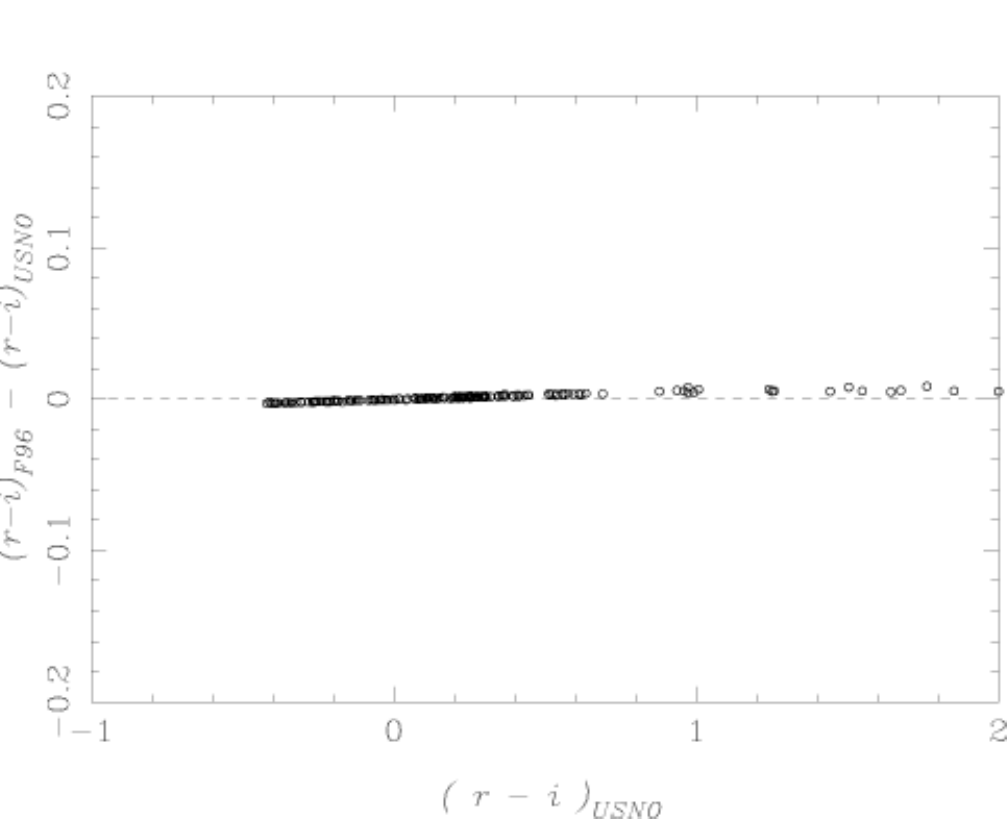} 
\caption{The same as Fig.12(a) but for $r-i$.}
\label{fig:usno_f96_ri}
\end{center}
\end{figure}
\end{subfigures}

\clearpage
\begin{subfigures}
\begin{figure}[pt]
\begin{center} 
\includegraphics[height=10cm,bb=0 0 483 393]{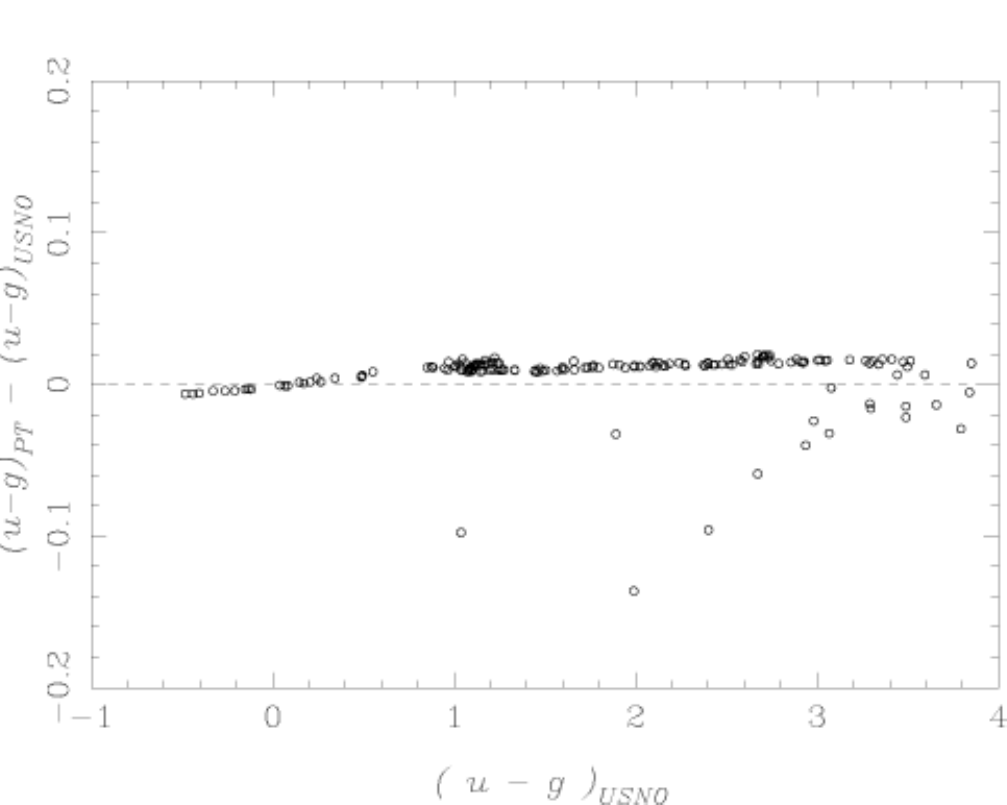} 
\caption{Expected offset of $u-g$ colour between the Photometric Telescope
PT-new and the USNO photometric
system,
$\Delta (u-g)_{\rm AB}=(u-g)_{\rm PT-new}-(u-g)_{\rm USNO}$
plotted as a function of $u-g$. 
}
\label{fig:pt_usno_ug}
\end{center}
\end{figure}

\begin{figure}[pt]
\begin{center} 
\includegraphics[height=10cm,bb=0 0 483 393]{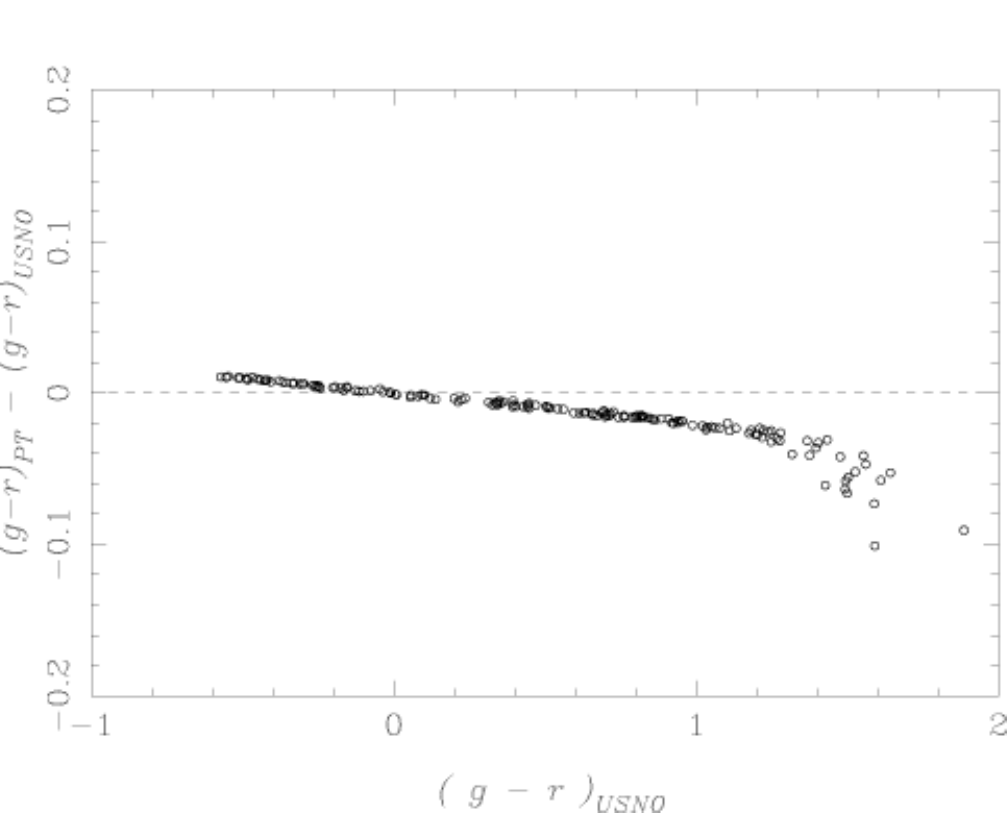} 
\caption{The same as Fig.13(a) but for $g-r$.}
\label{fig:pt_usno_gr}
\end{center}
\end{figure}

\begin{figure}[pt]
\begin{center} 
\includegraphics[height=10cm,bb=0 0 483 393]{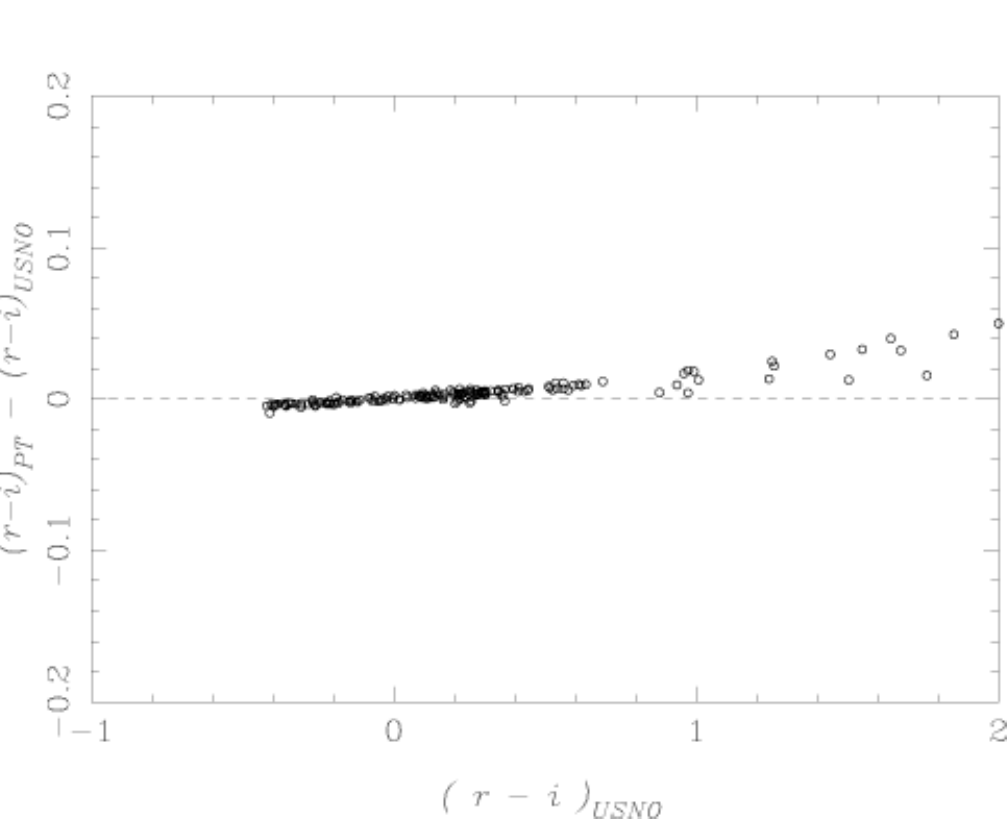} 
\caption{The same as Fig.13(a) but for $r-i$.}
\label{fig:pt_usno_ri}
\end{center}
\end{figure}
\end{subfigures}

\clearpage
\begin{subfigures}
\begin{figure}[pt]
\begin{center} 
\includegraphics[height=10cm,bb=0 0 483 393]{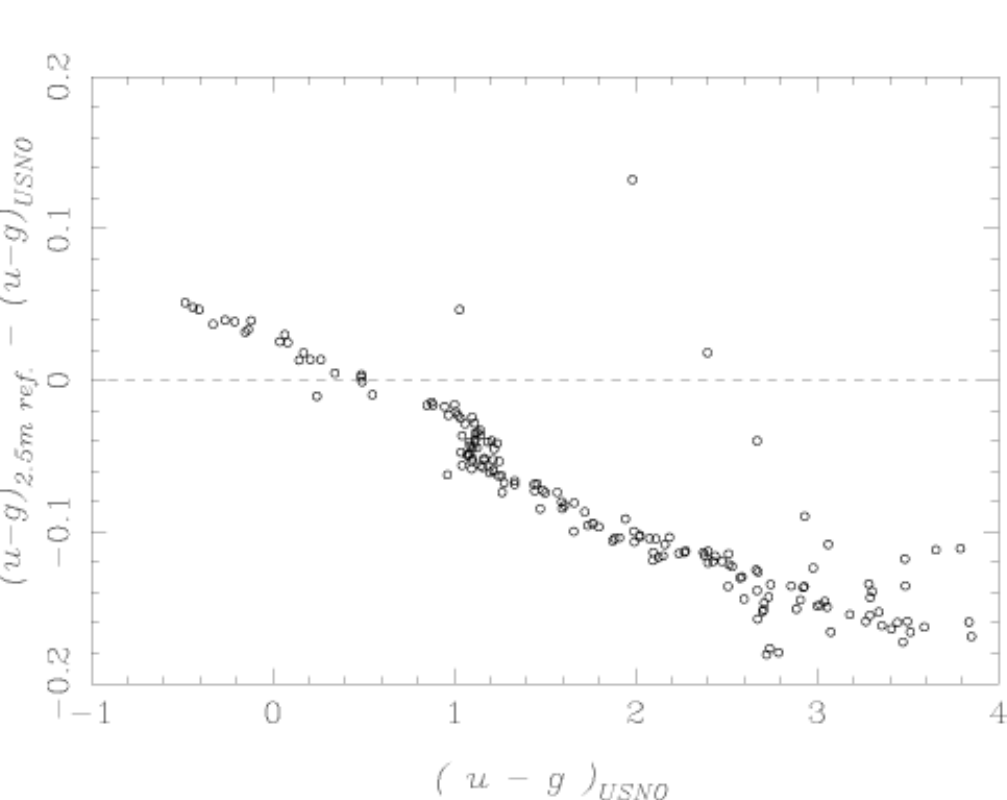}  
\caption{Expected offset of $u-g$ colour between the 2.5m reference 
 and the USNO photometric
system,
$\Delta (u-g)_{\rm AB}=(u-g)_{\rm 2.5m~reference}-(u-g)_{\rm USNO}$
plotted as a function of $u-g$. 
}
\label{fig:usno_ref_ug}
\end{center}
\end{figure}

\begin{figure}[pt]
\begin{center} 
\includegraphics[height=10cm,bb=0 0 483 393]{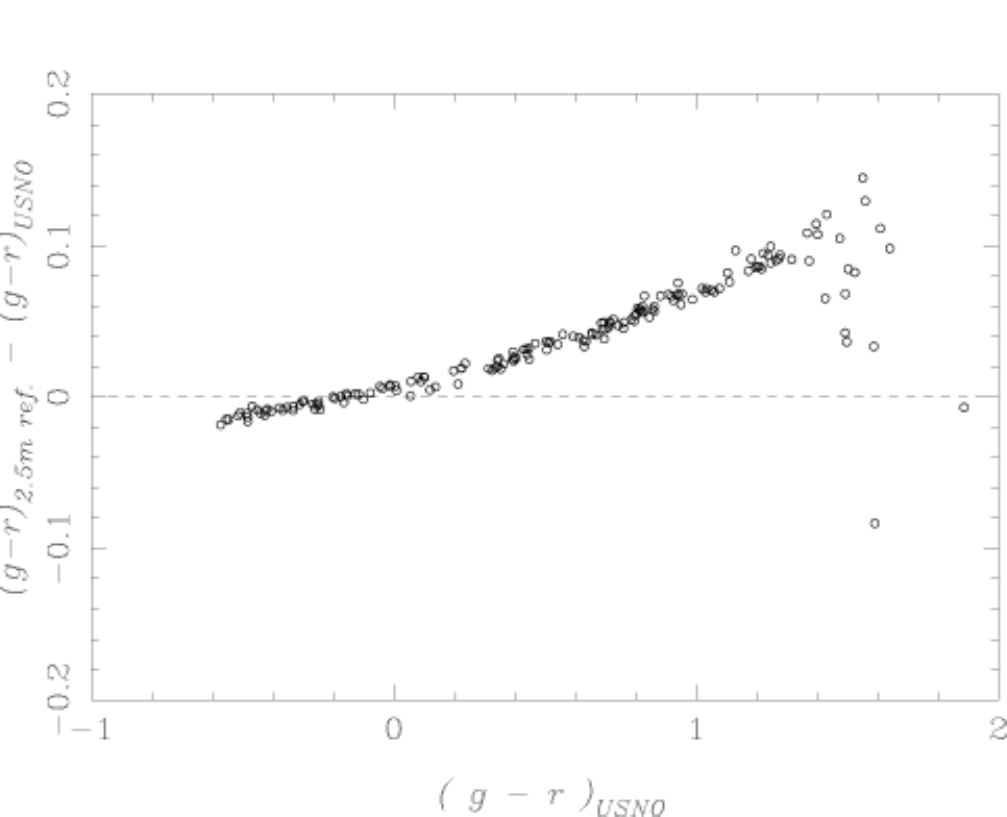}  
\caption{The same as Fig.14(a) but for $g-r$.}
\label{fig:usno_ref_gr}
\end{center}
\end{figure}

\begin{figure}[pt]
\begin{center} 
\includegraphics[height=10cm,bb=0 0 483 393]{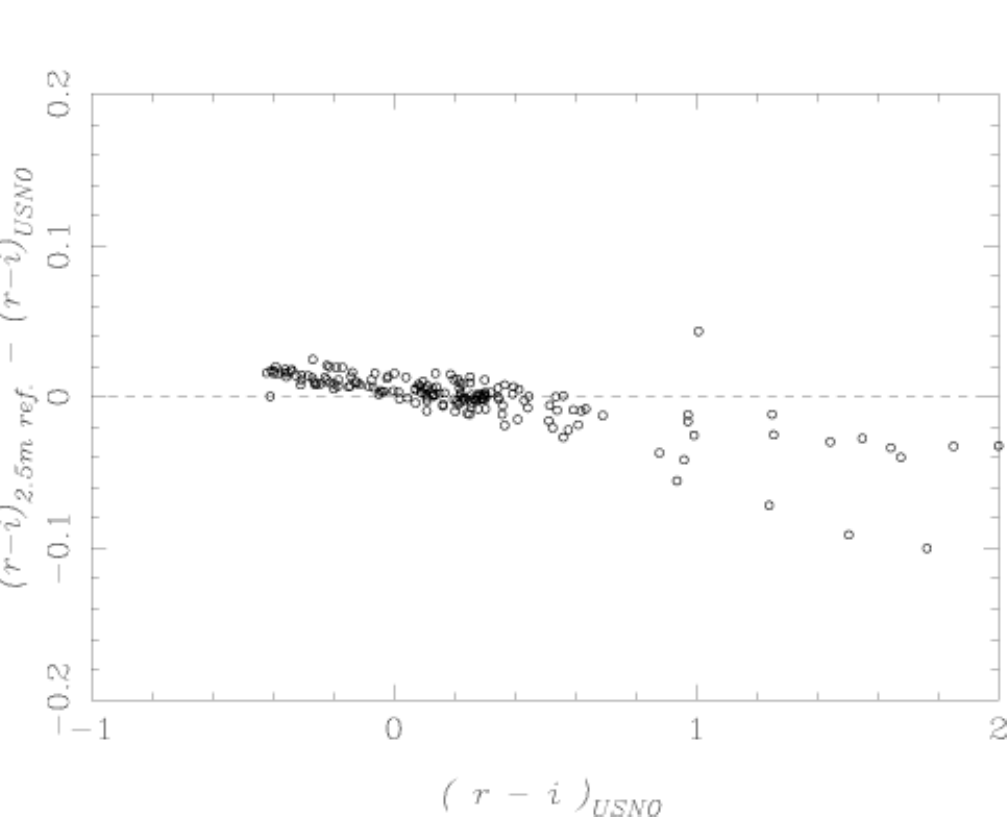}  
\caption{The same as Fig.14(a) but for $r-i$.}
\label{fig:usno_ref_ri}
\end{center}
\end{figure}
\end{subfigures}

\clearpage
\begin{subfigures}
\begin{figure}[pt]
\begin{center} 
\includegraphics[width=10cm,bb=0 0 574 574]{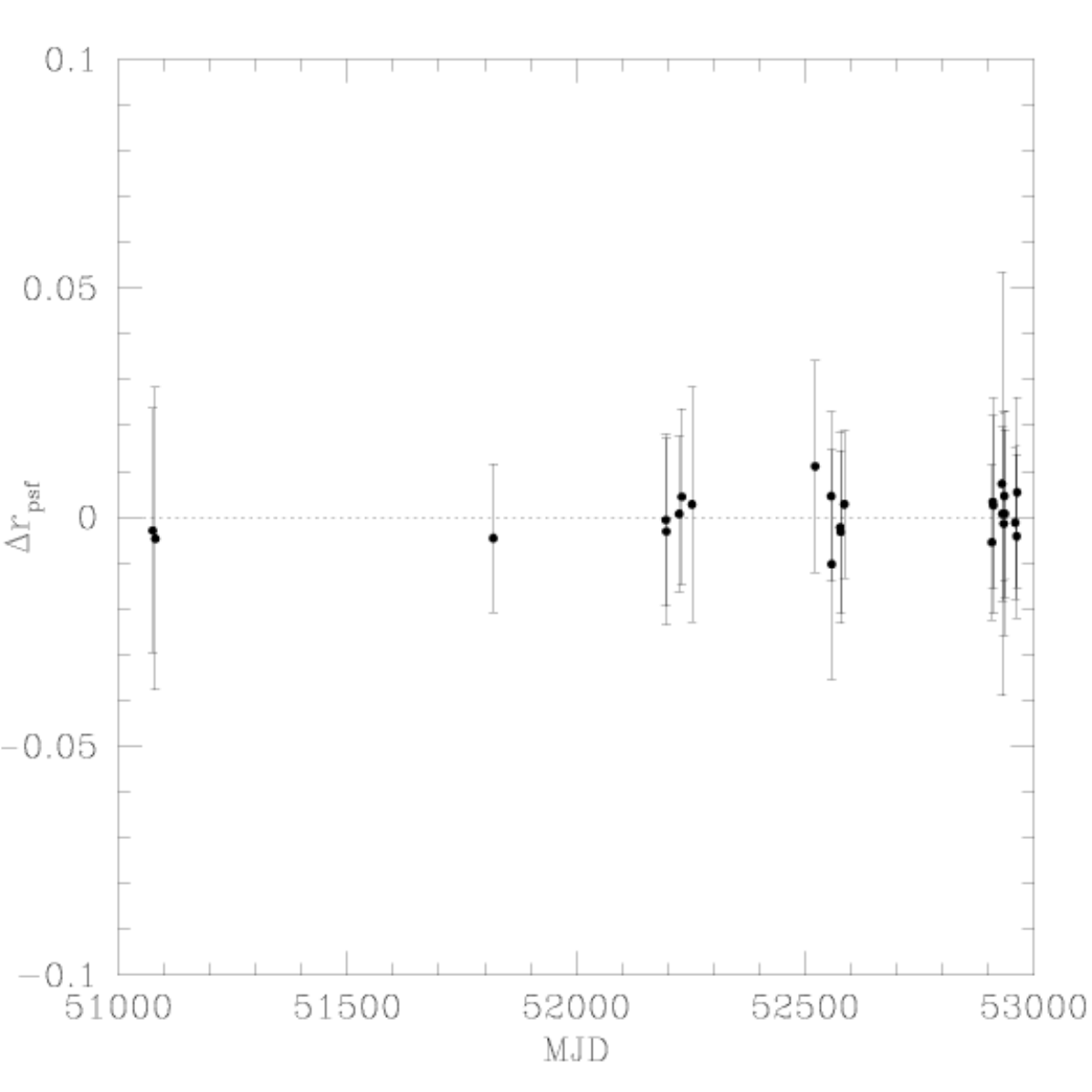}  
\caption{Mean of $r$ brightness of about 600 stars in a patch of 2.5 square
degree in Stripe 82 shown as a function of observation epochs from the 
SDSS catalogue. Error bars mean the variance of the sample.}
\label{fig:psfMag_r}
\end{center}
\end{figure}

\begin{figure}[pt]
\begin{center} 
\includegraphics[width=9cm,bb=0 0 574 574]{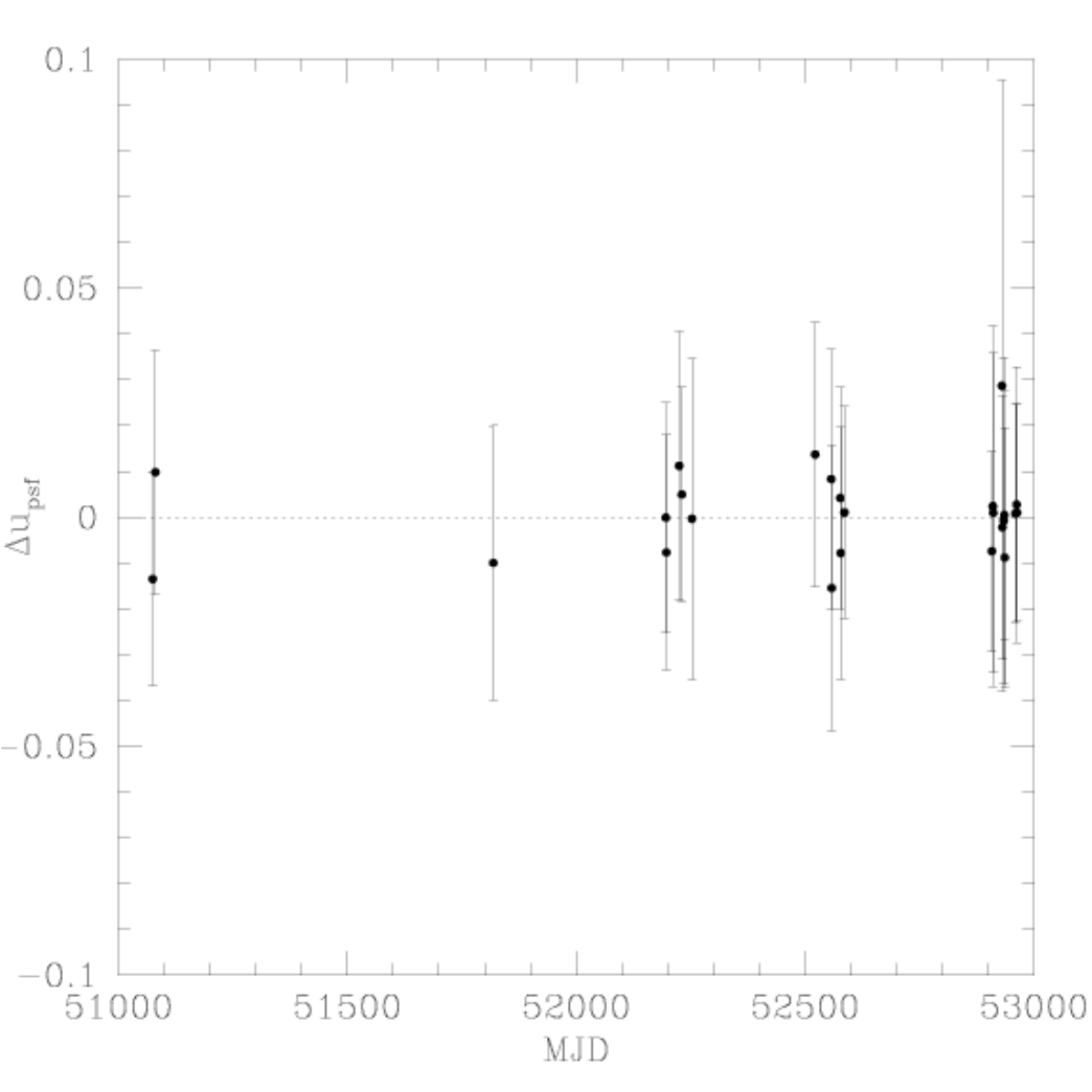} 
\caption{Mean of $u$ brightness using about 200 stars in a patch of 2.5 square
degree in Stripe 82 as a function of observation epochs. 
}
\label{fig:psfMag_u}
\end{center}
\end{figure}
\end{subfigures}

\clearpage
\begin{figure}
\begin{center} 
\includegraphics[width=9cm,bb=0 0 574 574]{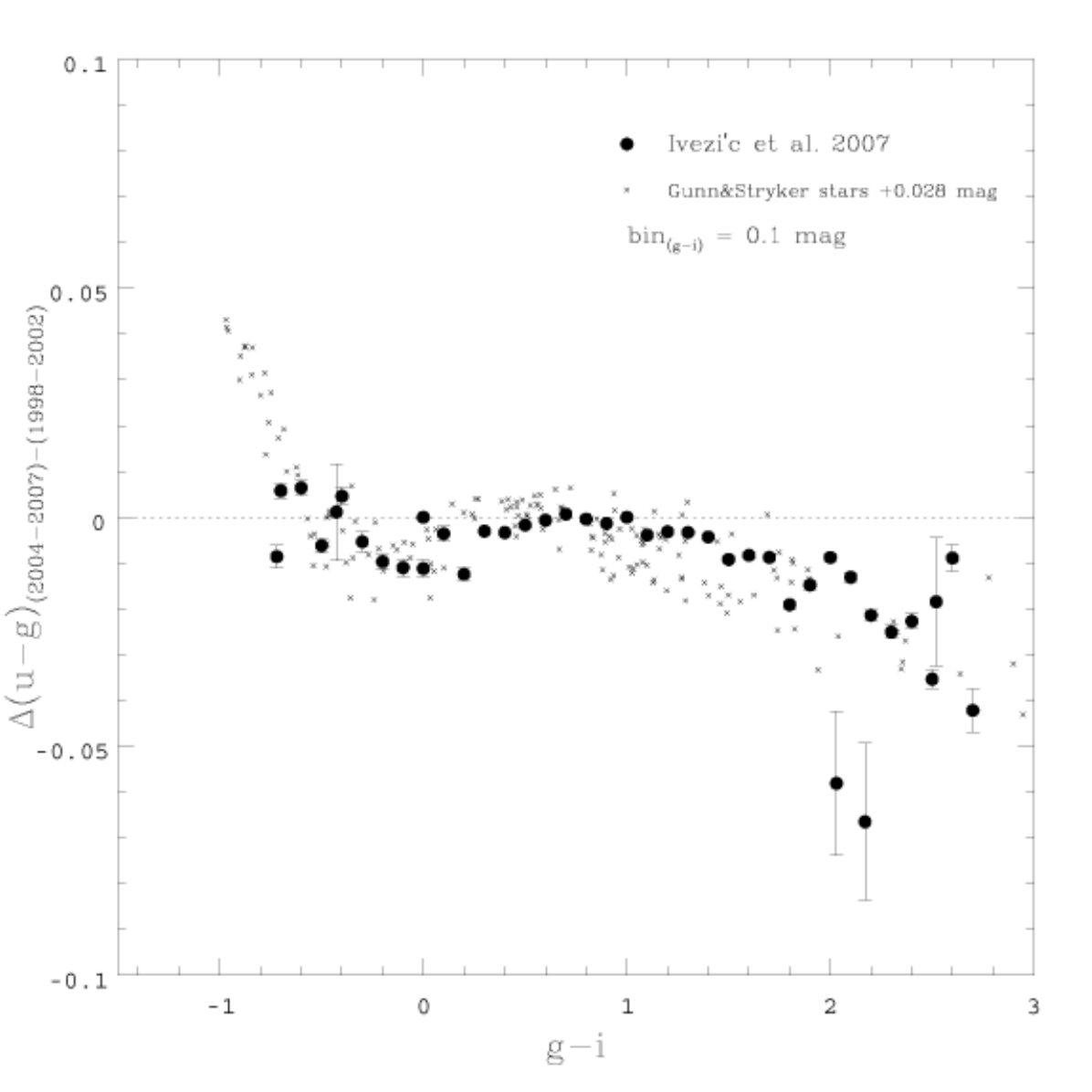} 
\caption{The time variation of $u-g$ colours 
in repeated photometry at stripe 82 (Ivezi\'c et al. 2007)
are shown by filled circles. All the non-variable stars with $u<21$ were used. 
Crosses show colour variation $\Delta(u-g)$ expected for Gunn-Stryker stars
between the 2001 (taken as the zero point) and the 2004 measurements
as a function of 
$g-i$. 
The Error bar is the mean and
variance of the data in each colour bin. }
\label{fig:uvars}
\end{center}
\end{figure}

\clearpage
\begin{subfigures}
\begin{figure}[pt]
\begin{center} 
\includegraphics[width=10cm,bb=0 0 574 574]{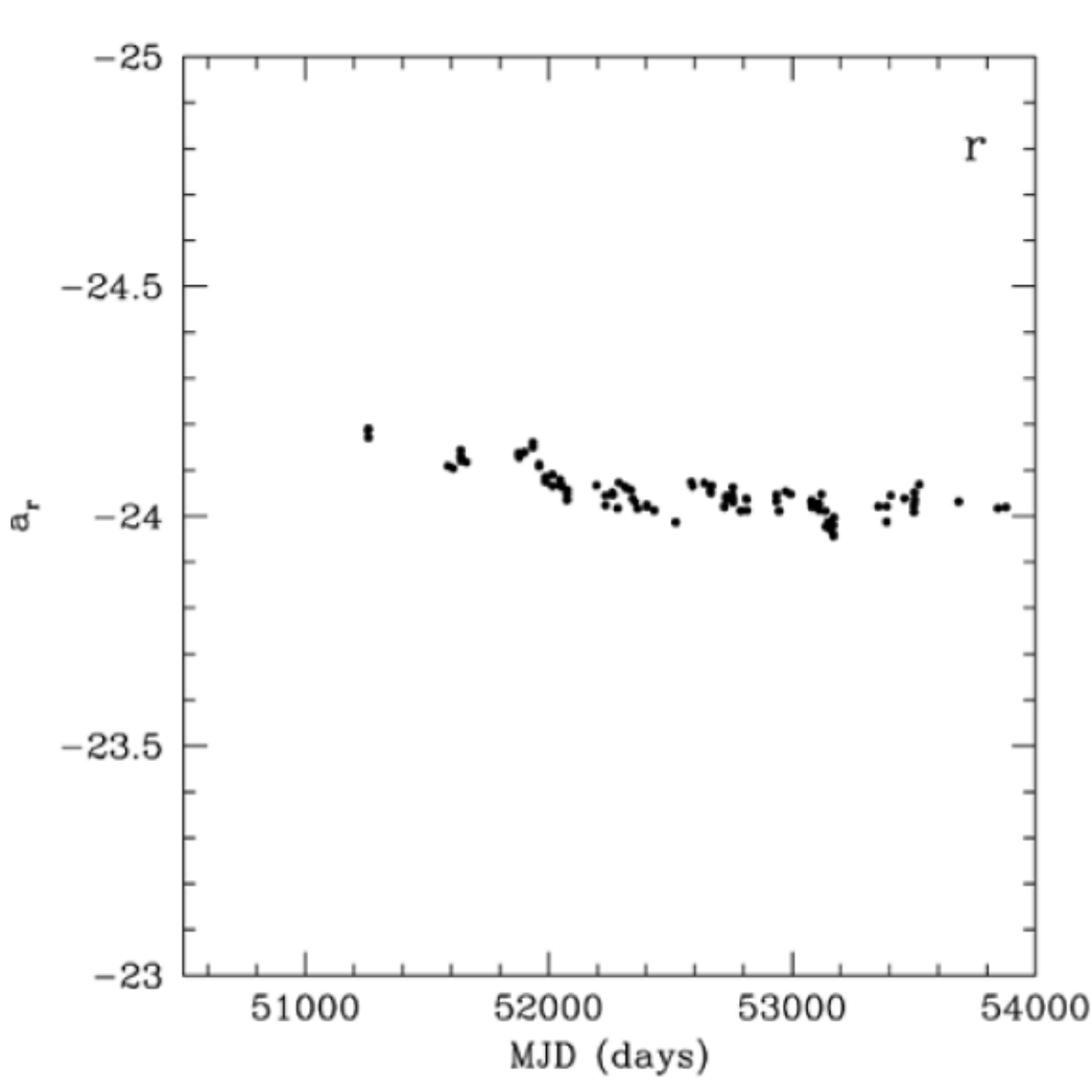}
\caption{Nightly variation of the instrumental photometric 
zero points of 2.5-m 
photometry in the $r$ band.}
\label{fig:ratm}
\end{center}
\end{figure}

\begin{figure}[pt]
\begin{center} 
\includegraphics[width=10cm,bb=0 0 574 574]{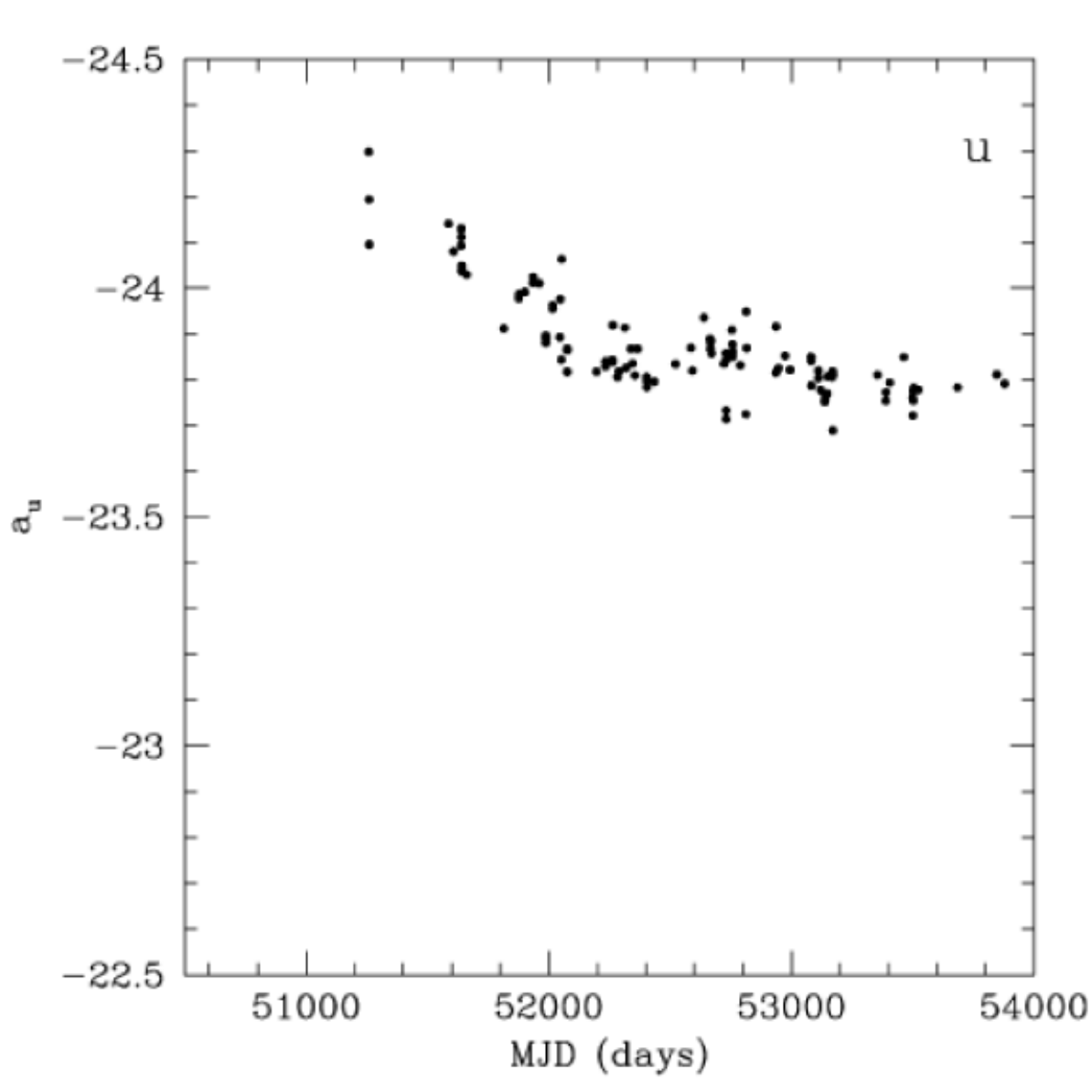}
\caption{Same as (a) but in the $u$ band.}
\label{fig:uatm}
\end{center}
\end{figure}
\end{subfigures}

\end{document}